\documentclass[showpacs,aps,prd,reprint,superscriptaddress,nofootinbib,longbibliography]{revtex4-1}
\usepackage[colorlinks=true, pdfstartview=FitV, linkcolor=magenta,citecolor=blue, urlcolor=blue,
bookmarks=true, bookmarksnumbered=true, breaklinks]{hyperref}
\usepackage[dvipdfmx]{graphicx}
\usepackage{amsmath,amssymb,bm,color,longtable,mathrsfs,slashed}
\usepackage{ulem}

\def\beqnn#1{\begin{eqnarray}#1\end{eqnarray}}

\newcommand{\Slash}[1]{{\ooalign{\hfil/\hfil\crcr$#1$}}}

\begin{document}

\begin{flushright}
\flushright{KEK-TH-2134, OU-HET-1019}
\end{flushright}

\title{$D$ mesons as a probe of Casimir effect for chiral symmetry breaking}

\author{Tsutomu~Ishikawa}
\email[]{{\tt tsuto@post.kek.jp}}
\affiliation{Graduate University for Advanced Studies (SOKENDAI), Tsukuba 305-0801, Japan}
\affiliation{KEK Theory Center, Institute of Particle and Nuclear
Studies, High Energy Accelerator Research Organization (KEK), Tsukuba 305-0801, Japan}

\author{Katsumasa~Nakayama}
\email[]{{\tt katumasa@post.kek.jp}}
\affiliation{Department of Physics, Osaka University, Toyonaka 560-0043, Japan}

\author{Daiki~Suenaga}
\email[]{\tt suenaga@mail.ccnu.edu.cn}
\affiliation{Key Laboratory of Quark and Lepton Physics (MOE) and Institute of Particle Physics, Central China Normal University, Wuhan 430079, China}

\author{Kei~Suzuki}
\email[]{{\tt k.suzuki.2010@th.phys.titech.ac.jp}}
\affiliation{Advanced Science Research Center, Japan Atomic Energy Agency (JAEA), Tokai 319-1195, Japan}

\date{\today}

\begin{abstract}
We propose $D$ mesons as probes to investigate finite-volume effects for chiral symmetry breaking at zero and finite temperatures.
By using the $2+1$-flavor linear sigma model with constituent light quarks, we analyze the Casimir effects for the $\sigma$ mean fields; the chiral symmetry is rapidly restored by the antiperiodic boundary for light quarks, and the chiral symmetry breaking is catalyzed by the periodic boundary.
We also show the phase diagram of the $\sigma$ mean fields on the volume and temperature plane.
For $D$ mesons, we employ an effective model based on the chiral-partner structure, in which the volume dependence of $D$ mesons is induced by the $\sigma$ mean fields.
We find that $D_s$ mesons are less sensitive to finite volume than $D$ mesons, which is caused by the insensitivity of $\sigma_s$ mean fields.
An anomalous mass shift of $D$ mesons at high temperature with the periodic boundary will be useful in examinations with lattice QCD simulations.
The dependence on the number of compactified spatial dimensions is also studied.
\end{abstract}


\maketitle

\section{Introduction}
Chiral symmetry breaking induced by the chiral condensate is a unique property in the low-energy QCD.
It is related to the properties of many hadrons, such as masses and decay constants.
Under an extreme environment such as high temperature/density, the chiral symmetry is partially restored by the external effects, and hadron properties are simultaneously modified. 
An intuitive scenario connecting the chiral symmetry to hadron properties is the {\it chiral-partner structure}, which means that the masses of partners become degenerate in the chiral-restored phase.

The chiral-partner structure for $D$ mesons was first suggested in Refs.~\cite{Nowak:1992um,Bardeen:1993ae}, and it was later extended to strange sectors such as $D_s$ mesons \cite{Bardeen:2003kt,Nowak:2003ra,Harada:2003kt}.\footnote{Note that this model is similar to the heavy-meson chiral perturbation theory (HM$\chi$PT) \cite{Wise:1992hn,Burdman:1992gh,Yan:1992gz} in the sense that both the models are motivated by the heavy-quark effective theory \cite{Eichten:1989zv,Georgi:1990um} based on the heavy-quark spin symmetry \cite{Isgur:1989vq,Isgur:1989ed}, but they differ in that the chiral-partner structure for heavy-light mesons is absent in the HM$\chi$PT.}
Since a heavy-light meson consists of one light ($u$, $d$, or $s$) quark and one heavy ($c$ or $b$) quark, the light quark can be a probe of the chiral symmetry breaking/restoration.
In particular, heavy-light mesons are slightly different from hadrons including only light quarks in the sense that the heavy quark cannot be a probe of the chiral condensate.
Thanks to this point, heavy-light meson chiral partners at chiral-restored environments could be a more visible probe than light-hadron partners.\footnote{In the language of QCD sum rules, the chiral partners of light hadrons such as $\rho$-$a_1$ mesons are significantly affected by the {\it four-quark} condensates, while the contribution from the {\it two-quark} chiral condensate is suppressed.
On the other hand, the contributions from four-quark condensates to heavy-light mesons are highly suppressed \cite{Buchheim:2014rpa}, which is useful for studying pure effects of the two-quark chiral condensate without suffering from contamination by little-known four-quark condensates.}

Recent lattice QCD simulations at finite temperature showed the thermal behaviors of $D$ and $D_s$ meson masses through the screening masses \cite{Bazavov:2014cta,Maezawa:2016pwo} and the peak positions on the spectral functions \cite{Kelly:2018hsi}.
One of the open questions is whether the pseudoscalar $D$ meson mass increases or decreases as the chiral symmetry is partially restored.
This is because the ``sign" of the mass shift is nontrivial even if the pseudoscalar $D$ and scalar $D_0^\ast$ are the chiral partners, and their masses become degenerate at the chiral-restored phase.
The results in Refs.~\cite{Bazavov:2014cta,Maezawa:2016pwo,Kelly:2018hsi} imply an increase of the pseudoscalar $D$ meson mass.

Furthermore, the relation between $D$ meson masses and the chiral condensate (or $\sigma$-mean field) at finite temperature/density has been discussed by using phenomenological approaches such as QCD sum rules \cite{Hayashigaki:2000es,Hilger:2008jg,Wang:2011mj,Hilger:2011cq,Wang:2011fv,Azizi:2014bba,Buchheim:2014rpa,Wang:2015uya,Suzuki:2015est,Buchheim:2018kss} and effective models \cite{Tsushima:1998ru,Sibirtsev:1999js,Mishra:2003se,Mishra:2008cd,Kumar:2010gb,Kumar:2011ff,Blaschke:2011yv,Suenaga:2014dia,Sasaki:2014asa,Suenaga:2014sga,Sasaki:2014wma,Suenaga:2015daa,Park:2016xrw,Harada:2016uca,Suenaga:2017deu,Suenaga:2018kta,Sugiura:2019ane} (see Refs.~\cite{Hosaka:2016ypm} for a recent review).\footnote{For $D$ meson mass shifts from phenomenological approaches {\it without} the modification of the chiral condensate (or $\sigma$-mean field), see also Refs.~\cite{Tolos:2004yg,Lutz:2005vx,Mizutani:2006vq,Tolos:2007vh,Molina:2008nh,Tolos:2009nn,JimenezTejero:2011fc,Yasui:2012rw} at finite baryon density and Refs.~\cite{Fuchs:2004fh,He:2011yi,Ghosh:2013xea,Cleven:2017fun} at finite temperature.
Such effects may be interpreted as an additional contribution different from the chiral symmetry restoration.}
Some of these theoretical studies are also motivated by future experiments of low-energy heavy-ion collisions which can create both the high-density baryonic matter and charmed hadrons, such as FAIR, NICA, and J-PARC.
In addition, $D$ mesons in a magnetic field can also be useful probes of chiral symmetry breaking in magnetic fields \cite{Machado:2013yaa,Gubler:2015qok,Yoshida:2016xgm,Reddy:2017pqp,Dhale:2018plh}.

\begin{figure}[t!]
    \begin{minipage}[t]{1.0\columnwidth}
        \begin{center}
            \includegraphics[clip, width=1.0\columnwidth]{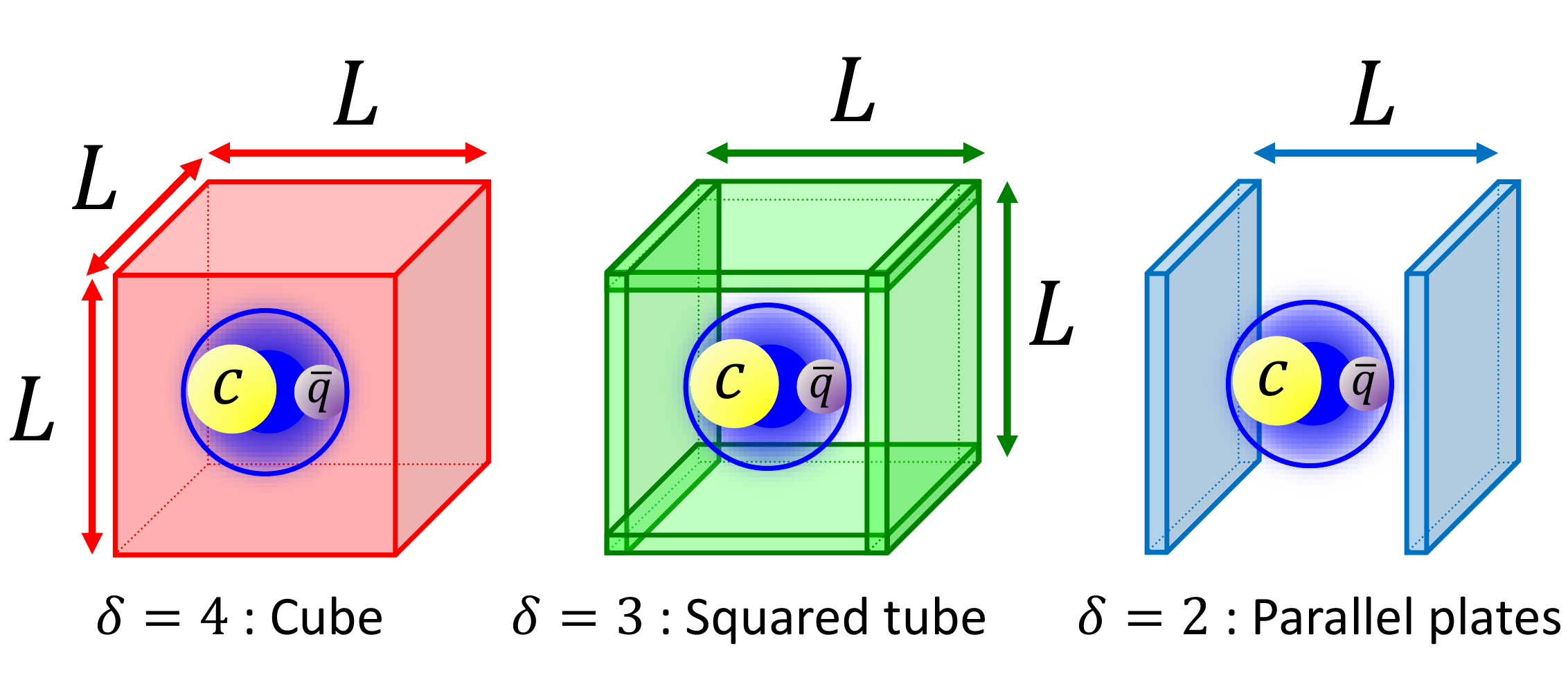}
        \end{center}
    \end{minipage}
    \caption{Sketches of setups studied in this paper: $D$ mesons in the $3+1$-dimensional space-time with compactified spatial length $L$.
$\delta$ is defined as the temporal and compactified spatial dimensions.
At $\delta=4$, $D$ mesons are affected by the finite-volume effect in the cubic geometry, which is a usual setup in lattice simulations.
At $\delta=3$, $D$ mesons are put in a squared tube (or a wave guide).
At $\delta=2$, $D$ mesons are affected by the finite-volume effect between two parallel plates, which is similar to the usual setup in the Casimir effect.}
    \label{D_in_box}
\vspace{-10pt}
\end{figure}

In this paper, we focus on a novel situation to study the $D$ meson chiral partners: finite-volume systems with a boundary condition (see Fig.~\ref{D_in_box}).
This is because chiral symmetry breaking is sensitive to the volume of the system.
Finite-volume systems with a ``box" geometry, in which all the spatial dimensions are compactified, are automatically realized in lattice simulations.
For advantages of lattice simulations, we can tune artificial parameters on the lattice, such as boundary conditions for quarks and the number of compactified dimensions (denoted as $\delta$ in this paper).
Examining responses of $D$ mesons to such tunable parameters will provide us with a deeper understanding of $D$ meson chiral partners.
In particular, the compactification of one spatial dimension is related to an energy shift of the nonperturbative QCD vacuum with two parallel plates, which is the so-called Casimir effect \cite{Casimir:1948dh}.
Such a situation will also be approximately realized in lattice QCD simulations ({\it e.g.}, see Refs.~\cite{Chernodub:2016owp,Chernodub:2017mhi,Chernodub:2017gwe,Chernodub:2018pmt,Chernodub:2018aix,Kitazawa:2019otp} as simulations for nonperturbative vacua such as Yang-Mills vacua).
Thus, finite-volume effects for $D$ mesons will be also useful as a hadronic observable of the Casimir effect for the chiral condensate.

In most of the previous works, the finite-volume effects for heavy-light mesons have been described by finite-volume corrections for dynamical pions coupled with the heavy-light mesons \cite{Goity:1990jb,Arndt:2004bg,Bernardoni:2009sx,Colangelo:2010ba,Briceno:2011rz}, using theoretical approaches such as the heavy-meson chiral perturbation theory (HM$\chi$PT) \cite{Arndt:2004bg,Bernardoni:2009sx,Briceno:2011rz} and the resummed L\"uscher formula \cite{Colangelo:2010ba}.
Here, we emphasize that in our model finite-volume effects are induced by that for a meson mean field (or chiral condensate), and it is essentially different from the understanding based on pion degrees of freedom.
Thus, our study will provide an alternative interpretation of finite-volume effects for $D$ mesons measured from lattice QCD simulations.

This paper is organized as follows.
In Sec.~\ref{sec:Model}, we introduce our model to describe $D$ mesons in a finite volume, which is based on the chiral-partner structure.
Numerical results are shown in Sec.~\ref{sec:Results}.
Section~\ref{sec:Concl} is devoted to our conclusion and outlook.

\section{Model}
\label{sec:Model}

\subsection{Linear sigma model}
\label{sec:LinearSigmaModel}
To investigate finite-temperature and/or finite-volume effects to $D$ mesons in the viewpoint of a chiral symmetry restoration, to begin with, we need to construct such environments. In the present work, the 2+1-flavor linear sigma model with quarks is employed for this purpose, which is given by
\begin{eqnarray}
{\cal L}_{\rm LS} = {\cal L}_{\rm q}  + {\cal L}_{\rm m}\ ,
\end{eqnarray}
with ${\cal L}_{\rm q}$ and ${\cal L}_{\rm m}$ being
\begin{eqnarray}
{\cal L}_{\rm q} &=&  \bar{q}_Li\Slash{\partial}q_L +\bar{q}_Ri\Slash{\partial}q_R  -g(\bar{q}_L\Sigma q_R+\bar{q}_R\Sigma^\dagger q_L) \ , \label{QuarkLagrangian}
\end{eqnarray}
and
\begin{eqnarray}
{\cal L}_{\rm m} &=&  {\rm tr}[\partial_\mu\Sigma^\dagger\partial^\mu\Sigma]-\bar{\mu}^2{\rm tr}[\Sigma^\dagger\Sigma]-\lambda_1\big({\rm tr}[\Sigma^\dagger\Sigma]\big)^2 \nonumber\\
&&-\lambda_2{\rm tr}[(\Sigma^\dagger\Sigma)^2] + c\big({\rm det}\Sigma+{\rm det}\Sigma^\dagger\big) \nonumber\\
&& +{\rm tr}[H(\Sigma+\Sigma^\dagger)]\ . \label{LSLagrangian}
\end{eqnarray}
In Eqs.~(\ref{QuarkLagrangian}) and~(\ref{LSLagrangian}), $q_{L(R)}$ denotes a left-handed (right-handed) quark column vector 
\begin{eqnarray}
q_{L(R)}= \left(
\begin{array}{c}
 u_{L(R)} \\
 d_{L(R)} \\
 s_{L(R)} \\
\end{array}
\right)\ ,
\end{eqnarray}
and $\Sigma$ denotes a scalar and pseudoscalar meson nonet matrix
\begin{eqnarray}
\Sigma = \sum_{a=0}^8 (\sigma^a+i\pi^a)\frac{\lambda^a}{2} \ . \label{SigmaDef}
\end{eqnarray}
In Eq.~(\ref{SigmaDef}), $\sigma^a$ and $\pi^a$ are scalar and pseudoscalar mesons, respectively, and $\lambda^a$ ($a=1, \dots, 8$) is the Gell-Mann matrix while $\lambda^0=\sqrt{\frac{2}{3}}{\bf 1}$.
The $3\times 3$ matrix $H$ is of the form
\begin{eqnarray}
H = \left(
\begin{array}{ccc}
\frac{h_q}{2} & 0 & 0 \\
0 & \frac{h_q}{2} & 0 \\
0 & 0 & \frac{h_s}{\sqrt{2}} \\
\end{array}
\right)\ .
\end{eqnarray}
Under the $U(3)_L\times U(3)_R$ chiral transformation, $q_L$ and $q_R$ are transformed as 
\begin{eqnarray}
q_{L} \to g_{L} q_{L} \ , \ \ q_R \to g_R q_R\ ,
\end{eqnarray} 
where $g_L$ $(g_R)$ is an element of the $U(3)_L$ ($U(3)_R$) chiral group, while the meson nonet $\Sigma$ is
\begin{eqnarray}
\Sigma \to g_L\Sigma g_R^\dagger\ .
\end{eqnarray}
Then, the Lagrangians~(\ref{QuarkLagrangian}) and~(\ref{LSLagrangian}) are invariant under the $U(3)_L\times U(3)_R$ chiral transformation except the last two terms in Eq.~(\ref{LSLagrangian}). The term proportional to $c$ is responsible for the $U(1)_A$ axial anomaly, while the one proportional to $H$ is included due to the explicit breaking of the chiral symmetry.  


One way to incorporate the spontaneous breakdown of the chiral symmetry is to replace the mesons matrix $\Sigma$ in Eq.~(\ref{SigmaDef}) into a mean field:
\begin{eqnarray}
\Sigma \to \left(
\begin{array}{ccc}
\frac{\sigma_q}{2} & 0 & 0 \\
0 & \frac{\sigma_q}{2} & 0 \\
0 & 0 & \frac{\sigma_s}{\sqrt{2}} \\
\end{array}
\right)\ . \label{SigmaMF}
\end{eqnarray}
Then, within this mean field approximation for mesons and the one-loop approximation for quarks,\footnote{We neglect the quantum and thermal fluctuations of meson fields, which are an approximation used in a conventional quark-meson model ({\it e.g.}, see Refs.~\cite{Scavenius:2000qd,Schaefer:2008hk}).
Studies including mesonic fluctuations are also interesting ({\it e.g.}, see Refs.~\cite{Mocsy:2004ab,Schaefer:2004en,Bowman:2008kc,Mitter:2013fxa} for studies of quark-meson model in the infinite volume).
} a thermodynamic potential per volume $\hat{\Omega}$ (the symbol $\hat{X}$ refers to the value of $X$ per volume: $\hat{X}=X/V$) reads
\begin{eqnarray}
\hat{\Omega} = \hat{\Omega}_{\rm q} + \hat{\Omega}_{\rm m}\ , \label{OmegaQandM}
\end{eqnarray}
with
\begin{eqnarray}
\hat{\Omega}_{\rm q} &=& - \frac{4N_c}{\beta}\sum_{l_0}\int\frac{d^3k}{(2\pi)^3}{\rm ln}(\omega_{l_0}^2+|\vec{k}|^2 +M_q^2 ) \nonumber\\
&& - \frac{2N_c}{\beta}\sum_{l_0}\int\frac{d^3k}{(2\pi)^3}{\rm ln}(\omega_{l_0}^2+|\vec{k}|^2 +M_s^2 )\ , \label{OmegaQ}
\end{eqnarray}
and
\begin{eqnarray}
\hat{\Omega}_{\rm m} &=& \frac{\bar{\mu}^2}{2}(\sigma_q^2+\sigma_s^2)+\frac{\lambda_1}{4}(\sigma_q^2+\sigma_s^2)^2+\frac{\lambda_2}{4}\left(\frac{\sigma_q^4}{2}+\sigma_s^4\right) \nonumber\\
&&-\frac{c}{2\sqrt{2}}\sigma_q^2\sigma_s-h_q\sigma_q-h_s\sigma_s \ . \label{OmegaM}
\end{eqnarray}
In Eqs.~(\ref{OmegaQ}) and~(\ref{OmegaM}), $\beta$ and $\omega_{l_0} $ are the inverse of the temperature $\beta=1/T$ and the Matsubara mode defined by $\omega_{l_0} = \frac{2\pi}{\beta} (l_0+\frac{1}{2})$ with $l_0=0,\pm1,\cdots$, respectively. $M_q=g\frac{\sigma_q}{2}$ and $M_s=g\frac{\sigma_s}{\sqrt{2}}$ are the constituent quark mass of the light quark and strange quark, respectively. The factor $4N_c$ and $2N_c$ denote the degrees of freedom of the systems for the light quarks and strange quark.

\begin{table}[t!]
\caption{Parameters of the linear sigma model.
These values are extracted from Ref.~\cite{Schaefer:2008hk} with $m_\sigma = 800$ MeV.}
  \begin{tabular}{lc} \hline\hline 
$c$ (MeV) & 4807.84 \\
$\bar{\mu}^2$ (MeV$^2$) & $-(306.26)^2$ \\
$\lambda_1$ & $13.49$ \\
$\lambda_2$ & $46.48$ \\
$h_q$ (MeV$^3$) & $(120.73)^3$ \\
$h_s$ (MeV$^3$) & $(336.41)^3$ \\
$g$ &  $6.5$ \\
\hline \hline
  \end{tabular}
  \label{tab:ParameterLSM}
\end{table}

For fixing the parameters in Eqs.~(\ref{QuarkLagrangian}) and~(\ref{LSLagrangian}), in this work, we use the same parameter set utilized in Ref~\cite{Schaefer:2008hk} with $m_\sigma=800$ MeV, which is listed in Table.~\ref{tab:ParameterLSM}. To get these values, the pion mass $m_\pi=138$ MeV, the kaon mass $m_K=496$ MeV, the averaged squared mass of $\eta$ and $\eta'$ mesons $m_{\eta}^2+m_{\eta'}^2=1.21\times 10^6$ MeV$^2$, the pion decay constant $f_\pi=92.4$ MeV, the kaon decay constant $f_K=113$ MeV, and the light-quark constituent quark mass $M_q=300$ MeV, together with the sigma meson mass $m_\sigma=800$ MeV are used as inputs.

By the minimum condition (or the gap equation) for the potential~(\ref{OmegaQandM}) at zero temperature, we can obtain the values of the mean fields: $\frac{\partial \hat{\Omega}}{\partial \sigma_q} = \frac{\partial \hat{\Omega}}{\partial \sigma_s} =0$.
As a result, $\sigma_q = 92.4 \, \mathrm{MeV}$ and $\sigma_s=94.5 \, \mathrm{MeV}$.

\subsection{Finite-volume effect}
\label{sec:finiteV}

$\hat{\Omega}_{\rm q}$ in Eq.~(\ref{OmegaQ}) includes UV divergences, which must be eliminated by a certain analytic continuation or regularization technique. Here, we make use of the regularization scheme with the Epstein-Hurwitz inhomogeneous zeta function.

Here, we demonstrate the case at $\delta=2$; {\it i.e.}, one spatial direction in addition to the temporal direction is compactified.
Then, for light and strange quarks, we impose the antiperiodic boundary condition [$\psi(\tau,x,y;z=0) = - \psi(\tau,x,y;z=L)$] or the periodic boundary condition [$\psi(\tau,x,y;z=0) = \psi(\tau,x,y;z=L)$], where only the $z$ component of the 3-momentum of quarks is discretized as
\begin{eqnarray}
k_z &&\to k_z^\mathrm{ap} = \frac{2\pi}{L}\left(l_1+\frac{1}{2}\right), \label{pz_anti} \\
k_z &&\to k_z^\mathrm{p} = \frac{2l_1 \pi}{L}, \label{pz_peri}
\end{eqnarray}
where $l_1=0, \pm1, \cdots$ is the label of discretized levels.
The discretized energy of the light (strange) quarks is given by $E_{q(s)}(k) = \sqrt{k_z^2 + k_\perp^2 + M_{q(s)}^2}$, where $k_\perp^2 =k_x^2+k_y^2$.
Then, the momentum integral with respect to $k_z$ in the thermodynamic potential (\ref{OmegaQ}) is replaced by the summation with respect to the discretized levels, and Eq.~(\ref{OmegaQ}) is rewritten as
\begin{eqnarray}
&& \hat{\Omega}_{\rm q}(\beta,L; \delta=2) = \nonumber\\
&& 2\hat{\Omega}_{\mathrm{Cas},q}^\mathrm{ap/p}(L) + \hat{\Omega}_{\mathrm{Cas},s}^\mathrm{ap/p}(L) - \frac{1}{L} \sum_{l_1=-\infty}^\infty \int \frac{d^2 k_\perp}{(2\pi)^2} \biggl
[ \Biggr. \nonumber\\
&& \left. \frac{1}{\beta} \left\{ 8 N_c \ln \left[ 1+e^{-\beta E_q(k)} \right] + 4N_c \ln \left[1+e^{-\beta E_s(k)} \right] \right\} \right],  \nonumber\\ \label{OmegaQ2}
\end{eqnarray}
where $\hat{\Omega}_{\mathrm{Cas},q(s)}(L)$ is the Casimir energy for light (strange) quarks per $V$, which is finite because the UV divergence is already subtracted.
Note that $\hat{\Omega}_{\mathrm{Cas},q(s)}(L)$ is independent of temperature, and all the temperature effects are included only in the third and fourth terms of Eq.~(\ref{OmegaQ2}).\footnote{In that sense, we can separate the Casimir effect at zero temperature from the combinatorial effect of finite temperature and volume.
The combinatorial effect may be also understood as a kind of ``thermal" Casimir effect.}

For the antiperiodic boundary condition, the Casimir energy for massive quarks is given by\footnote{The Casimir effect for massless fermions was first derived in the context of the MIT bag model by Johnson \cite{Johnson:1975zp}. For a derivation for massive fermions, see Refs.~\cite{Mamaev:1980jn,CougoPinto:1996bh,Elizalde:2002wg}.}
\begin{equation}
\hat{\Omega}_{\mathrm{Cas},q(s)}^\mathrm{ap} = 2N_c \sum_{n_1=1}^\infty (-1)^{n_1} \left( \frac{M_{q(s)}}{n_1 \pi L} \right)^2 K_2(n_1 M_{q(s)} L),
\label{Casimir_antiperiodic}
\end{equation}
where $K_2$ is the modified Bessel function.
For the periodic boundary condition, 
\begin{equation}
\hat{\Omega}_{\mathrm{Cas},q(s)}^\mathrm{p} = 2N_c \sum_{n_1=1}^\infty \left( \frac{M_{q(s)}}{n_1 \pi L} \right)^2 K_2(n_1 M_{q(s)} L).
\label{Casimir_periodic}
\end{equation}

For massive fermions with a mass $M$ and a degeneracy factor $\gamma$, a more general form including $\delta=1,2,3,4$ is given by (see Ref.~\cite{Ishikawa:2018yey} for a derivation)
\beqnn{
&& \hat{\Omega}_{\rm q}(\beta,L;\delta)
=
\hat{\Omega}_{0}\nonumber\\
&+&
\frac{2\gamma}{(2\pi)^2}\left[
2
\sum_{i=0} ^{\delta-1}
\sum_{n_i = 1} ^\infty
(-1)^{n_i\alpha_i} \left(\frac{M}{n_iL_{i}}\right)^2
K_2(n_i M L_{i}) \right.\nonumber\\
&+& \cdots \nonumber\\
&+&
2^\delta 
\sum_{n_0, \cdots, n_{\delta-1} = 1} ^\infty
\prod_{i=0} ^{\delta-1}
(-1)^{n_i\alpha_i} \nonumber\\
& \times& \left.
\left(
\frac{M}{\sqrt{\sum_{i = 0} ^{\delta-1} n_i ^2 L_{i} ^2}}
\right)^2
K_2 \left(
M
\sqrt{\sum_{i = 0} ^{\delta-1} n_i ^2 L_{i} ^2}
\right) \right], \nonumber\\
\label{Cas_ene_gene}
}
where $\hat{\Omega}_0$ is the UV divergence caused by the vacuum energy, which should be subtracted by hand.
The label $i=0,1,2,3$ denotes the direction of the compactified space-time.
$L_0=\beta$ is the temporal length (equivalent to the inverse temperature), and $L_i$ with $i=1,2,3$ is the spatial length for $i$ th direction.
The antiperiodic temporal boundary is $\alpha_0=1$.
For the antiperiodic (periodic) spatial boundary, we set $\alpha_i=1$ ($\alpha_i=0$).
When we focus on $\delta=2$, $\gamma =4N_c$ for light quarks, and $\gamma =2N_c$ for strange quarks, this form is equivalent to Eq.~(\ref{OmegaQ2}).

As seen in Eq.~(\ref{Cas_ene_gene}), the thermal effect and the Casimir effect with the antiperiodic boundary are similar to each other while the effect with the periodic boundary differs in a factor of $(-1)^{n_i\alpha_i}$.
The purpose of this paper is not only to investigate either the thermal or Casimir effect but also to examine the competition between the two effects.
Such a setup will be studied in future lattice simulations by using approaches similar to Refs.~\cite{Chernodub:2018aix,Kitazawa:2019otp}.

Practically, in our numerical calculation, we have to truncate the infinite series in Eq.~(\ref{Cas_ene_gene}) by introducing a cutoff for $n_i$.
The error caused by this truncation is discussed in Appendix~\ref{App_1}.


\subsection{$D$ meson Lagrangian}
\label{sec:DLagrangian}

Here, we introduce a Lagrangian for $D$ mesons based on the chiral-partner structure \cite{Nowak:1992um,Bardeen:1993ae}. We derive the Lagrangian assuming an exact heavy-quark spin symmetry (HQSS) to give a transparent argument in terms of masses of $D$ mesons although the HQSS is violated for the observed $D$. Such violations to the $D$ meson masses will be included in Sec.~\ref{sec:DMassFormula} in deriving mass formulas for $D$ mesons.

The fundamental fields in constructing the Lagrangian within the chiral-partner structure are heavy-light fields ${\cal H}_L$ ($\sim c\bar{q}_L$) and ${\cal H}_R$ ($\sim c\bar{q}_R)$, which transform under the $U(3)_L\times U(3)_R$ chiral transformation as
\begin{eqnarray}
{\cal H}_L \to {\cal H}_L g_L^\dagger\ , \ \ {\cal H}_R \to {\cal H}_R g_R^\dagger\ ,
\end{eqnarray}
and the $SU(2)_S$ heavy-quark spin transformation as
\begin{eqnarray}
{\cal H}_L \to {\cal S}{\cal H}_L \ , \ \ {\cal H}_R \to {\cal S}{\cal H}_R \ ,
\end{eqnarray}
[${\cal S}$ is an element of the $SU(2)_S$ heavy-quark spin group], respectively. Then, a Lagrangian invariant under the chiral transformation, heavy-quark spin transformation, and parity is given by \cite{Bardeen:2003kt,Nowak:2003ra,Harada:2003kt}
\begin{eqnarray}
{\cal L}_{D{\rm meson}} &=& -{\rm Tr}[{\cal H}_L iv\cdot\partial \bar{\cal H}_L+{\cal H}_Riv\cdot\partial \bar{\cal H}_R]  \nonumber\\
&& - \frac{g_\pi}{4}{\rm Tr}[{\cal H}_R\Sigma^\dagger\bar{\cal H}_L+{\cal H}_L\Sigma\bar{\cal H}_R] \nonumber\\
&& + i\frac{g_A}{2f_\pi}{\rm Tr}[{\cal H}_R\gamma_5\Slash{\partial}\Sigma^\dagger\bar{\cal H}_L-{\cal H}_L\gamma_5\Slash{\partial}\Sigma\bar{\cal H}_R]\ , \nonumber\\  \label{HMET}
\end{eqnarray}
up to one derivative with respect to $\Sigma$. In Eq.~(\ref{HMET}), $\bar{\cal H}_{L(R)}$ is defined by $\bar{\cal H}_{L(R)} \equiv \gamma_0{\cal H}_{L(R)}^\dagger\gamma_0$, and $v$ is a velocity of the heavy-light meson. $\Sigma$ is the meson nonet in Eq.~(\ref{SigmaDef}). Although the $g_A$ term is put for reproducing the $\Gamma_{D^*\to D\pi}$ decay width, in the following analysis, this term will be neglected in evaluating $D$ meson masses at finite-temperature and/or volume system within a mean field level.


\begin{table}[t!]
\caption{The masses of observed $D$ mesons. $J$ and $P$ are the total angular momentum and parity, respectively.}
  \begin{tabular}{llcc} \hline\hline
Meson & $J^P$ & Notation & Value (MeV) \\
\hline
$D$ & $ 0^-$ & $m_{P_q}$ & $1870$ \\
$D^\ast$ & $1^-$ & $m_{P_q^*}$  & $2010$ \\
$D_s$ & $0^-$ & $m_{P_s}$ & $1968$ \\
$D_s^\ast$ & $1^-$ & $m_{P_s^*}$ & $2112$ \\
$D_0^\ast$ & $0^+$ & $m_{D_q}$ & $2318$ \\
$D_1$ & $1^+$ & $m_{D_q^*}$ & $2427$ \\
$D_{s0}^\ast$ & $0^+$ & $m_{D_s}$ & $2317$ \\
$D_{s1}$ & $1^+$ & $m_{D_s^*}$ & $2460$ \\ \hline \hline
  \end{tabular}
  \label{tab:DMasses}
\end{table}

The Lagrangian~(\ref{HMET}) is convenient to get the chiral transformation laws since the heavy-light meson fields ${\cal H}_L$ and ${\cal H}_R$ are belonging to the fundamental representation of the $U(3)_L\times U(3)_R$ chiral group. However, ${\cal H}_L$ and ${\cal H}_R$ do not correspond to the physical $D$ meson states since these fields are not parity eigenstates. The parity-even state $G$ and the parity-odd state $H$ can be provided via
\begin{eqnarray}
{\cal H}_L = \frac{1}{\sqrt{2}}(G+iH\gamma_5)\ , \ \ {\cal H}_R = \frac{1}{\sqrt{2}}(G-iH\gamma_5)\ ,
\end{eqnarray}
and $G$ and $H$ are related to the observed $D$ meson fields as~\footnote{Throughout this paper, we use $P=(P_q,P_s)^T$, $P^*=(P_q^*,P_s^*)^T$, $D=(D_q,D_s)^T$, $D^*=(D_q^*,D_s^*)^T$ for referring to the pseudoscalar, vector, scalar, and axial vector $D$ mesons.}
\begin{eqnarray} 
H = \frac{1+\Slash{v}}{2}[\Slash{P}^*+iP\gamma_5] \ , \ \ G = \frac{1+\Slash{v}}{2}[-i\Slash{D}^*\gamma_5+D] \ . \nonumber\\
\end{eqnarray}
Due to the HQSS, $H$-doublet includes the pseudoscalar and vector $D$ mesons, while $G$ includes the scalar and axial-vector $D$ mesons: $H=(0^-,1^-)$ and $G=(0^+,1^+)$.

In Sec.~\ref{sec:DMassFormula}, mass formulas for $D$ mesons will be derived by the Lagrangian~(\ref{HMET}) incorporating the spontaneous breaking of the chiral symmetry.


\subsection{$D$ meson mass formula}
\label{sec:DMassFormula}

According to the heavy-light meson Lagrangian in Eq.~(\ref{HMET}), under the spontaneous breakdown of the chiral symmetry, {\it i.e.}, by replacing $\Sigma$ into its mean field as in Eq.~(\ref{SigmaMF}), mass formulas for $D$ mesons are provided by \cite{Bardeen:2003kt,Nowak:2003ra}
\begin{eqnarray}
m_{H_q} = m-\frac{g_\pi}{8}\sigma_q \ &,& \ m_{H_s} = m-\frac{g_\pi}{4\sqrt{2}}\sigma_s \ , \nonumber\\
m_{G_q} = m+\frac{g_\pi}{8}\sigma_q\ &,&\ m_{G_s} = m+\frac{g_\pi}{4\sqrt{2}}\sigma_s\ ,
 \label{ChiralPartner} 
\end{eqnarray}
with $H_q=(P_q,P_q^*)$, $H_s = (P_s,P_s^*)$, $G_q = (D_q, D_q^*)$, and $G_s=(D_s,D_s^*)$. As one can see in Eq.~(\ref{ChiralPartner}), when the chiral symmetry is restored, $\sigma_q=\sigma_s=0$, the masses of all $D$ mesons coincide, which clearly shows a peculiarity of the chiral partner-structure together with the HQSS.

The parameter $m$ can be determined by
\begin{eqnarray}
m =\frac{1}{4} \sum_i \bar{m}_i\ ,
\end{eqnarray}
in which $\bar{m}_i$ is a spin-averaged mass of the doublet $i$ with $i = H_q, H_s, G_q, G_s$:
\begin{eqnarray}
\bar{m}_{H_q} = \frac{3m_{P^*_q}+m_{P_q}}{4} \ &,&\ \bar{m}_{H_s}= \frac{3m_{P^*_s}+m_{P_s}}{4} \ , \nonumber\\
 \bar{m}_{G_q} = \frac{3m_{D^*_q}+m_{D_q}}{4}\  &,& \ \bar{m}_{G_s} = \frac{3m_{D^*_s}+m_{D_s}}{4} \ .
\end{eqnarray}
To fix $\bar{m}_i$, we put the physical values of $D$ meson masses summarized in Table.~\ref{tab:DMasses}.
On the other hand, $g_\pi$ can be determined by 
\begin{eqnarray}
g_\pi = \frac{1}{2}\sum_l g_\pi^l\ ,
\end{eqnarray}
with $l= q,s$, where $g_\pi^q$ and $g_\pi^s$ satisfy the extended Goldberger-Treiman relations
\begin{eqnarray}
4g_\pi^q\sigma_q &=& \bar{m}_{G_q}-\bar{m}_{H_q}, \nonumber\\
2\sqrt{2}g_\pi^s\sigma_s &=& \bar{m}_{G_s}-\bar{m}_{H_s},
\end{eqnarray}
with respect to the chiral-partner structure for each light flavor.


\begin{table}[t!]
\caption{Model parameters for $D$ meson mass formula.}
  \begin{tabular}{lc} \hline\hline
$m$ (MeV)              & $2220.44$ \\
$g_\pi$                & $14.5076$ \\      
$\Delta_{P_q}$ [for $D(0^-)$] (MeV)& $-182.869$ \\
$\Delta_{P_q^*}$ [for $D^\ast(1^-)$] (MeV)& $-42.8689$ \\
$\Delta_{P_s}$ [for $D_s(0^-)$] (MeV)& $-10.1526$ \\
$\Delta_{P_s^*}$ [for $D_s^\ast(1^-)$] (MeV)& $133.847$ \\
$\Delta_{D_q}$ [for $D_0^\ast(0^+)$] (MeV)& $-70 .0061$ \\
$\Delta_{D_q^*}$ [for $D_1(1^+)$] (MeV)& $38.9939$ \\
$\Delta_{D_s}$ [for $D_{s0}^\ast(0^+)$] (MeV) & $-145.722$ \\
$\Delta_{D_s^*}$ [for $D_{s1}(1^+)$] (MeV) & $-2.7224$ \\
\hline\hline
  \end{tabular}
  \label{tab:Delta_D}
\end{table}

The mass formulas in Eq.~(\ref{ChiralPartner}) determine the $D$ meson masses with the HQSS, while this symmetry is violated by a small mass difference between the doublets as indicated in the observed $D$ meson masses in Table.~\ref{tab:DMasses}. Such effects are included by adding small contributions to the mass formulas in Eq.~(\ref{ChiralPartner}), which leads to
\begin{eqnarray}
m_{P^{(*)}_q} = {m}_{H_q}+\Delta_{P^{(*)}_q}  \ &,&\ m_{P_s^{(*)}} = {m}_{H_s}+\Delta_{P^{(*)}_s} \ , \nonumber\\
m_{D^{(*)}_q} = {m}_{G_q}+\Delta_{D^{(*)}_q} \ &,&\ m_{D_s^{(*)}} = {m}_{G_s}+\Delta_{D^{(*)}_s} \ .
 \label{DMassWithDelta} 
\end{eqnarray}
The values of $m$, $g_\pi$, and $\Delta_{D}$ are summarized in Table.~\ref{tab:Delta_D}.
In Sec~\ref{sec:Results}, numerical computations of $D$ mesons mass shifts in finite temperature and/or volume will be performed with the mass formulas in Eq.~(\ref{DMassWithDelta}).

\section{Numerical results}
\label{sec:Results}
In this section, we show the numerical results of $D$ meson masses after discussing the volume dependence of $\sigma$ mean fields.

\subsection{Dependence of $\sigma$ mean fields}
\label{subsec:Sigma}

\begin{figure}[t!]
    \begin{minipage}[t]{1.0\columnwidth}
        \begin{center}
            \includegraphics[clip, width=1.0\columnwidth]{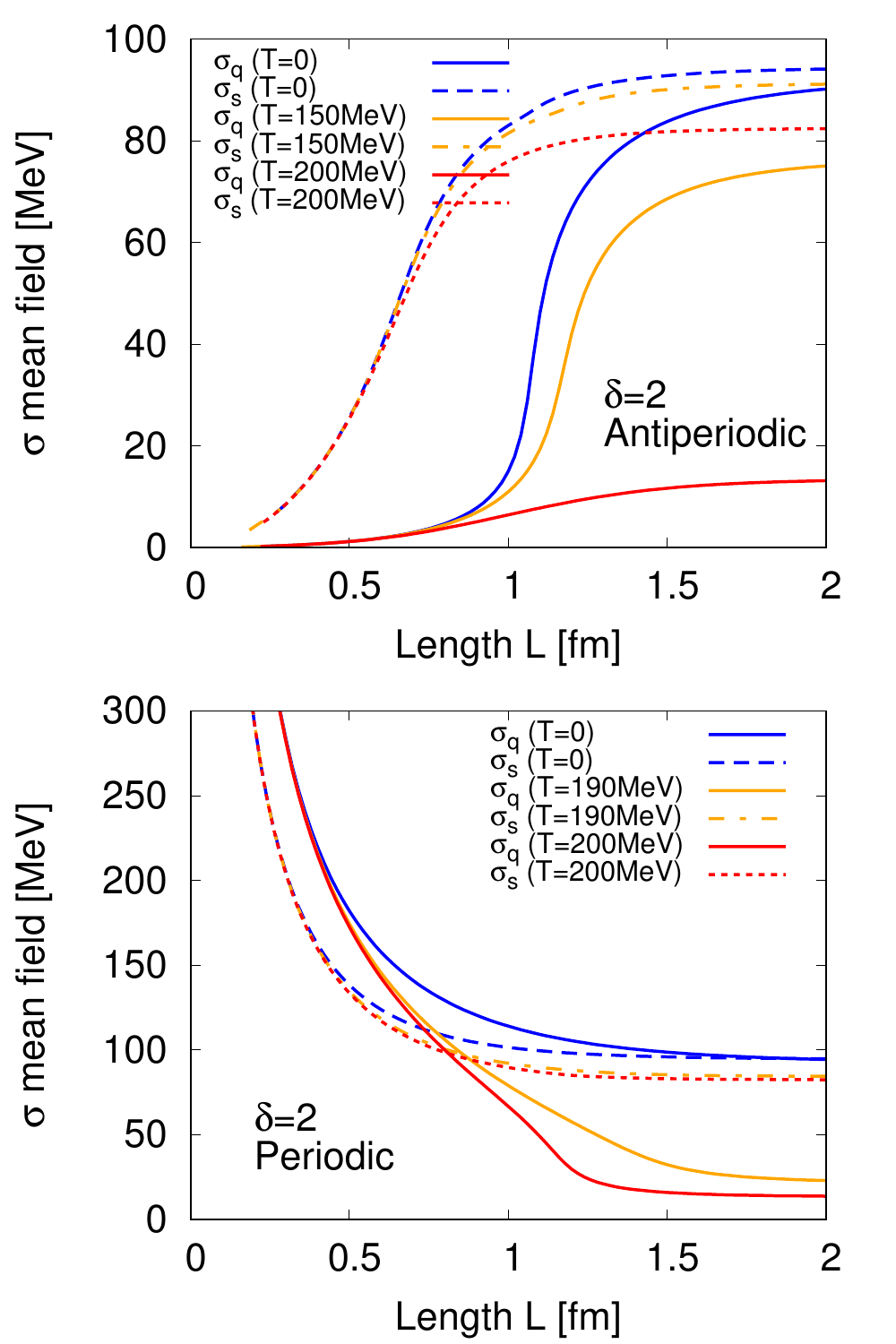}
        \end{center}
    \end{minipage}
    \caption{Volume dependence of $\sigma$ mean fields in a finite box at $\delta=2$.
Upper: antiperiodic boundary. Lower: periodic boundary.}
    \label{sigma_LM_delta2}
    \vspace{-10pt}
\end{figure}

\begin{figure}[t!]
    \begin{minipage}[t]{0.75\columnwidth}
        \begin{center}
            \includegraphics[clip, width=1.0\columnwidth]{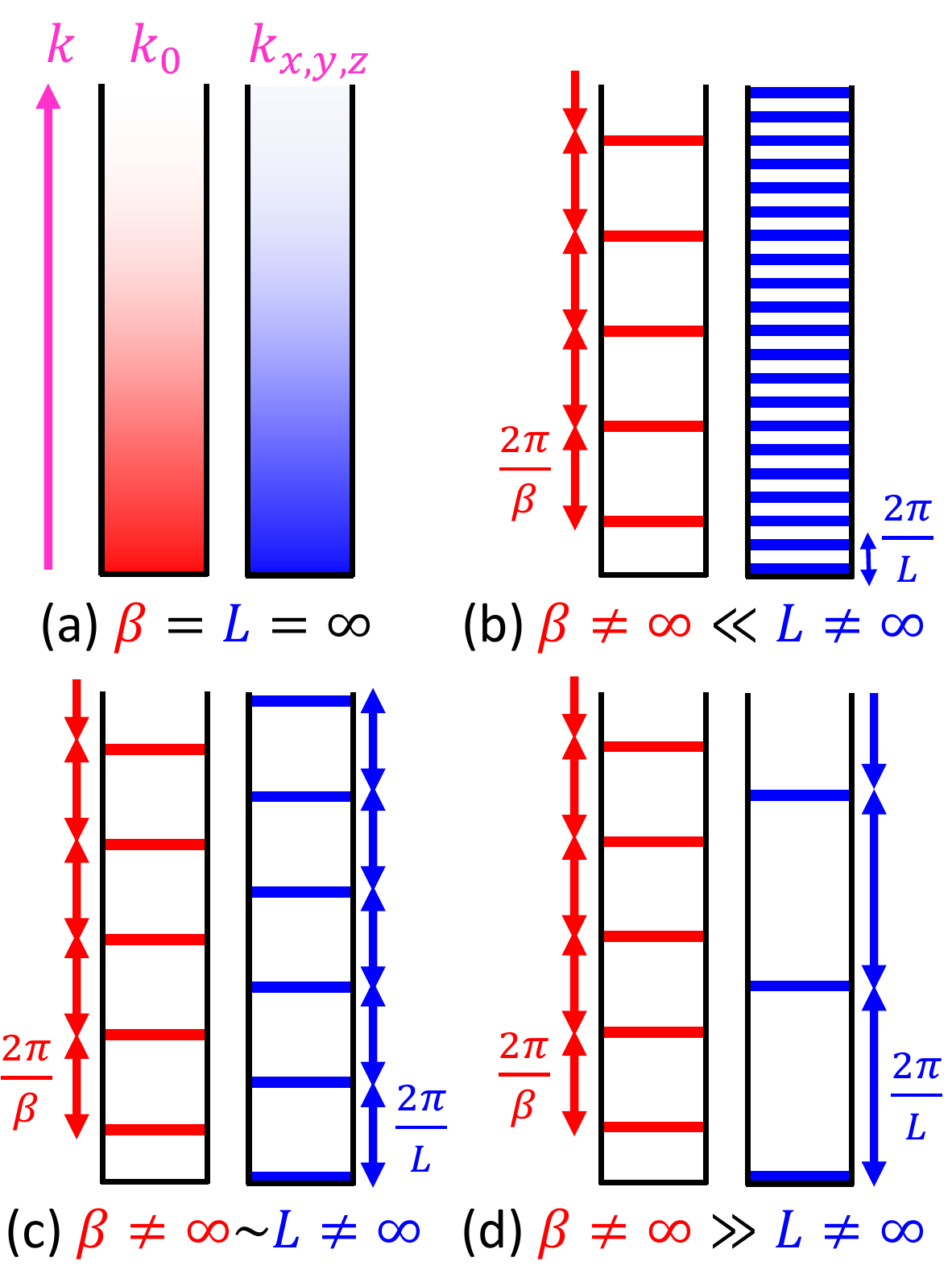}
        \end{center}
    \end{minipage}
    \caption{Sketches of discretization of momentum at finite temperature and volume {\it with the periodic boundary condition}.
(a) At zero temperature and infinite volume ($\beta=L=\infty$), all the components of the 4-momentum $k$ of a quark are continuous.
(b) At larger volume and high temperature ($L \gg \beta$), the lowest mode of temporal component $k_0$ is dominant in infrared dynamics of quarks.
(c) A volume comparable to temperature scale ($L \sim \beta$).
(d) At smaller volume and high temperature ($L \ll \beta$), the zero mode of spatial component $k_{x,y,z}$ becomes dominant.
}
    \label{anomalous}
\vspace{-10pt}
\end{figure}

In Fig.~\ref{sigma_LM_delta2}, we show the $L$ dependences of the $\sigma$ mean fields at $\delta=2$.
From these figures, our findings are as follows:
\begin{enumerate}
\item {\it Chiral symmetry restoration with antiperiodic boundary.}---For the antiperiodic boundary, a smaller length leads to the chiral symmetry restoration, and the values of the $\sigma$ mean fields decrease.
At the small volume limit, the mean fields go to zero, and the chiral symmetry is completely restored.
The transition length of $\sigma_q$ at $T=0$ is about $L \sim 1 \mathrm{fm}$.
\item {\it Chiral symmetry breaking enhancement with periodic boundary.}---For the periodic boundary, a small length catalyzes the chiral symmetry breaking, and the values of the $\sigma$ mean fields increase.
This effect is caused by the appearance of the ``zero mode" in discretized momenta [see Figs.~\ref{anomalous}(b)--\ref{anomalous}(d)], which is a different situation from the antiperiodic boundary without momentum zero modes. 
The enhancement of chiral symmetry breaking with the periodic boundary condition for fermions is suggested also by other effective models such as four-Fermi models \cite{Kim:1987db,Song:1990dm,Song:1993da,Kim:1994es,Vshivtsev:1995xx,Vdovichenko:1998ev,Vshivtsev:1998fg,Hayashi:2010ru,Ebert:2010eq,Wang:2018ovx,Inagaki:2019kbc,Xu:2019gia}, the linear sigma (or quark-meson) model \cite{Braun:2005gy,Braun:2005fj,Palhares:2009tf,Braun:2010vd,Tripolt:2013zfa,Phat:2014asa,Almasi:2016zqf}, the Walecka model \cite{Ishikawa:2018yey}, and the parity-doublet model for nucleons \cite{Ishikawa:2018yey}.
\item {\it Comparison between $\sigma_q$ and $\sigma_s$.}---The $\sigma_s$ mean field is less sensitive to the length $L$ than $\sigma_q$.
This is a situation similar to finite temperature.
For the antiperiodic boundary, the transition length of $\sigma_s$ is lower than that of $\sigma_q$.
At a small volume with the periodic boundary, $\sigma_q$ is enhanced by the chiral symmetry breaking more rapidly than $\sigma_s$, which leads to $\sigma_q > \sigma_s$.
Thus the reverse of the magnitudes of $\sigma_q$ and $\sigma_s$ would be a useful signal for examining the boundary dependence of chiral condensates from lattice simulations.
For instance, at zero temperature, we find $\sigma_q > \sigma_s$ even at $L<1.9 \, \mathrm{fm}$, while in a large volume we observe $\sigma_q < \sigma_s$ since $\sigma_q = 92.4 \, \mathrm{MeV}$ and $\sigma_s=94.5 \, \mathrm{MeV}$ at $T=0$ and $L \to \infty$.
\item {\it Anomalous behavior at high temperature.}---We find a difference between $T=190 \ \mathrm{MeV}$ and $T=200 \ \mathrm{MeV}$ with the periodic boundary, where the $L$ dependence of $\sigma_q$ at $T=200 \ \mathrm{MeV}$ and around $1<L<1.5  \ \mathrm{fm} $ shows an anomalous behavior.
This is induced by competition between finite length and temperature effects.
In the large-volume region ($L> 1.2 \, \mathrm{fm}$), the infrared dynamics of constituent quarks is dominated by the low Matsubara modes [$\omega_{l_0} = \frac{2\pi}{\beta} (l_0+\frac{1}{2})$], particularly the lowest mode ($\omega_0 = \frac{\pi}{\beta} $), as shown in (b) of Fig.~\ref{anomalous}.
Then, the $\sigma_q$ mean field is sensitive to $T$ but insensitive to $L$.
Next, in the intermediate-volume region ($0.6<L < 1.2 \, \mathrm{fm}$), the quark dynamics is determined by not only the lowest Matsubara mode but also discretized modes of spatial momentum ($k_{zl_1} = \frac{2\pi}{L} l_1$) [see Fig.~\ref{anomalous}(c)].
In the small-volume region ($L < 0.6 \, \mathrm{fm}$), the dynamics is dominated by the (gapless) zero mode induced by the periodic boundary [see Fig.~\ref{anomalous}(d)].
As a result, $T$ dependence from the lowest Matubara mode becomes invisible.
In other words, the enhancement of chiral symmetry breaking by the spatial zero mode overcomes the suppression of the chiral condensate by the lowest Matsubara mode.
Thus, the $L$ dependence of the mean field shows an anomalous shift at a ``boundary" region in which dominated modes interchange.
We will discuss that such behavior of mean fields can be observed even in $D$ meson masses as an anomalous mass shift.
Note that this behavior was also observed from other models with $\sigma$ mean field \cite{Ishikawa:2018yey}.
\end{enumerate}

\begin{figure}[t!]
    \begin{minipage}[t]{1.0\columnwidth}
        \begin{center}
            \includegraphics[clip, width=1.0\columnwidth]{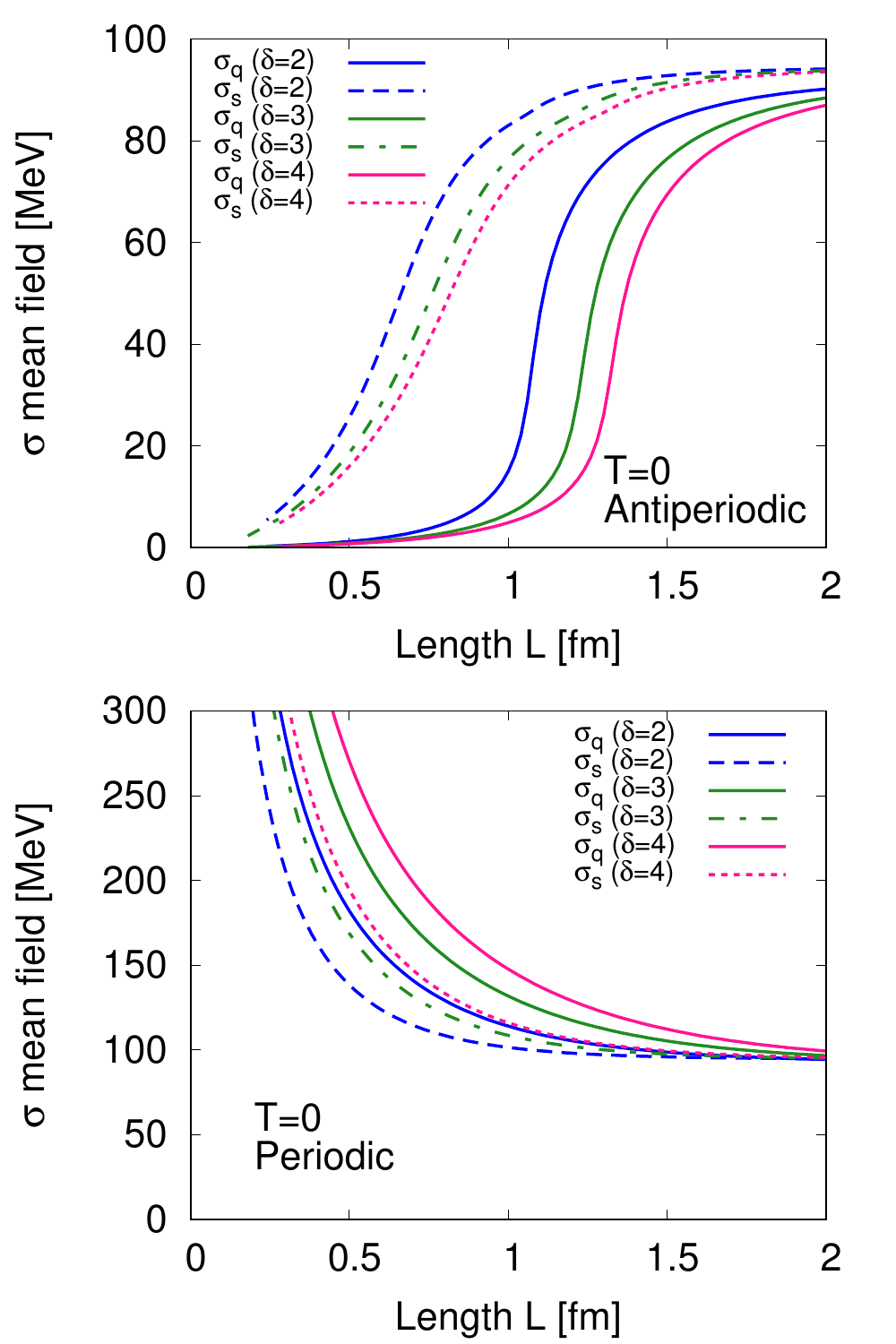}
        \end{center}
    \end{minipage}
    \caption{$\delta$ dependence of length transition for $\sigma$ mean fields in a finite box at $T=0$.
Upper: antiperiodic boundary. Lower: periodic boundary.}
    \label{sigma_LM_delta}
    \vspace{-10pt}
\end{figure}

\begin{figure}[b!]
    \begin{minipage}[t]{1.0\columnwidth}
        \begin{center}
            \includegraphics[clip, width=1.0\columnwidth]{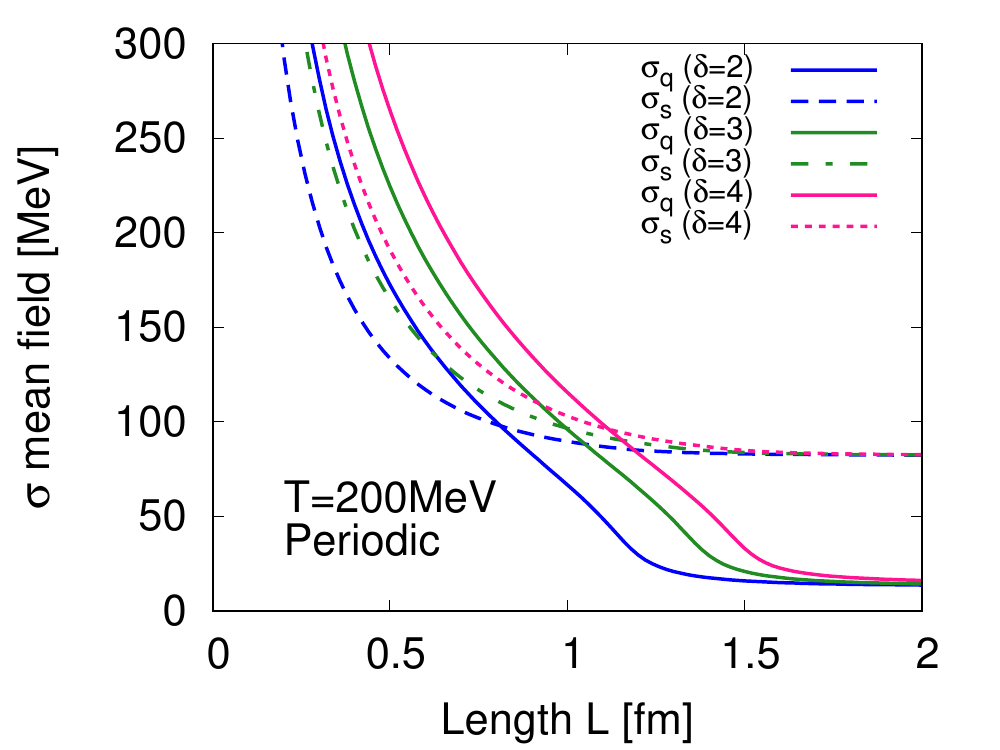}
        \end{center}
    \end{minipage}
    \caption{$\delta$ dependence of length transition for $\sigma$ mean fields in a finite box with periodic boundary at $T=200 \, \mathrm{MeV}$.}
    \label{sigma_LM_delta_T=200}
    \vspace{-10pt}
\end{figure}

\begin{figure}[t!]
    \begin{minipage}[t]{1.0\columnwidth}
        \begin{center}
            \includegraphics[clip, width=1.0\columnwidth]{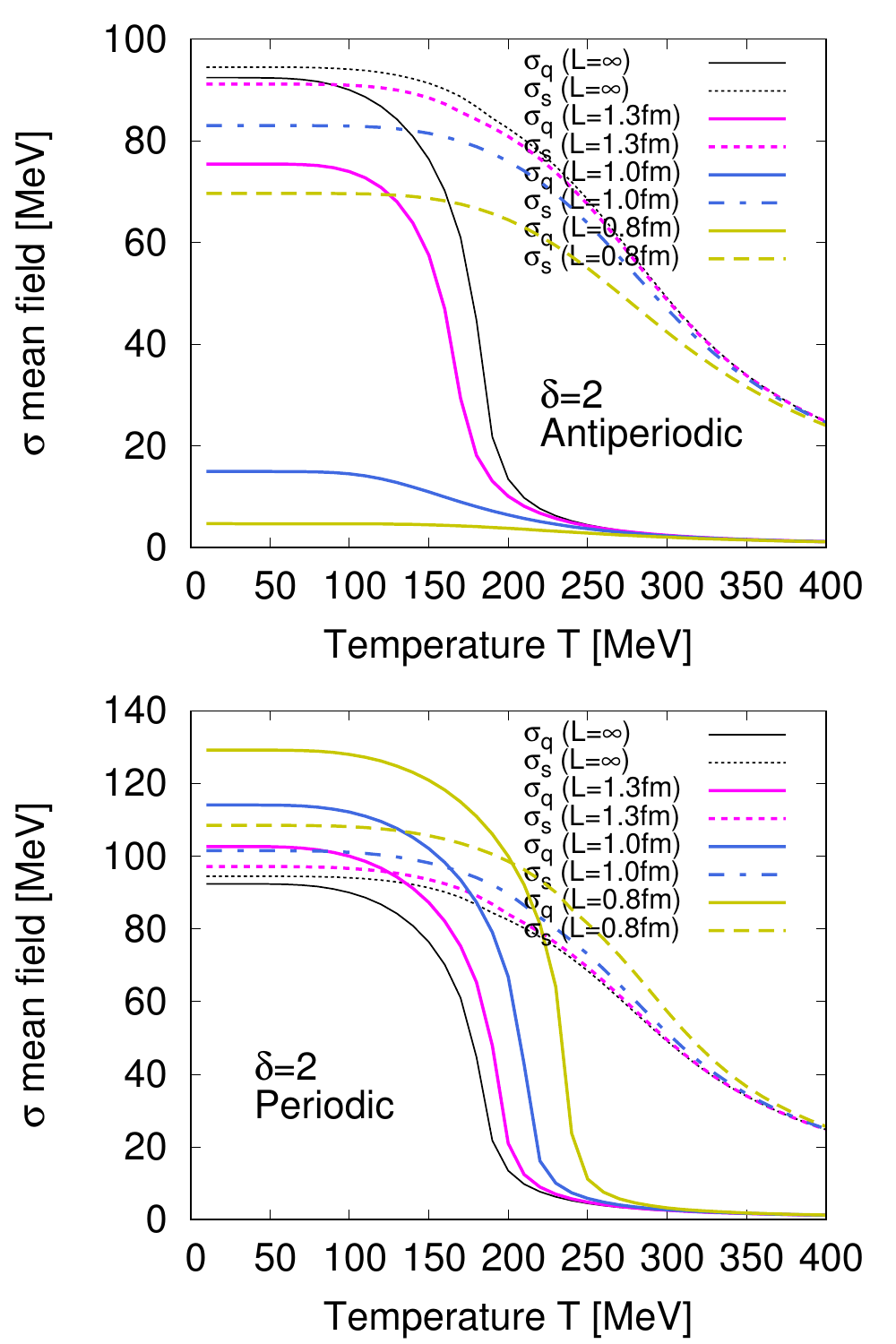}
        \end{center}
    \end{minipage}
    \caption{Temperature dependence of $\sigma$ mean fields in a finite box at $\delta=2$.
Upper: antiperiodic boundary. Lower: periodic boundary.}
    \label{sigma_TM_delta2}
    \vspace{-10pt}
\end{figure}

\begin{figure}[t!]
    \begin{minipage}[t]{1.0\columnwidth}
        \begin{center}
            \includegraphics[clip, width=1.0\columnwidth]{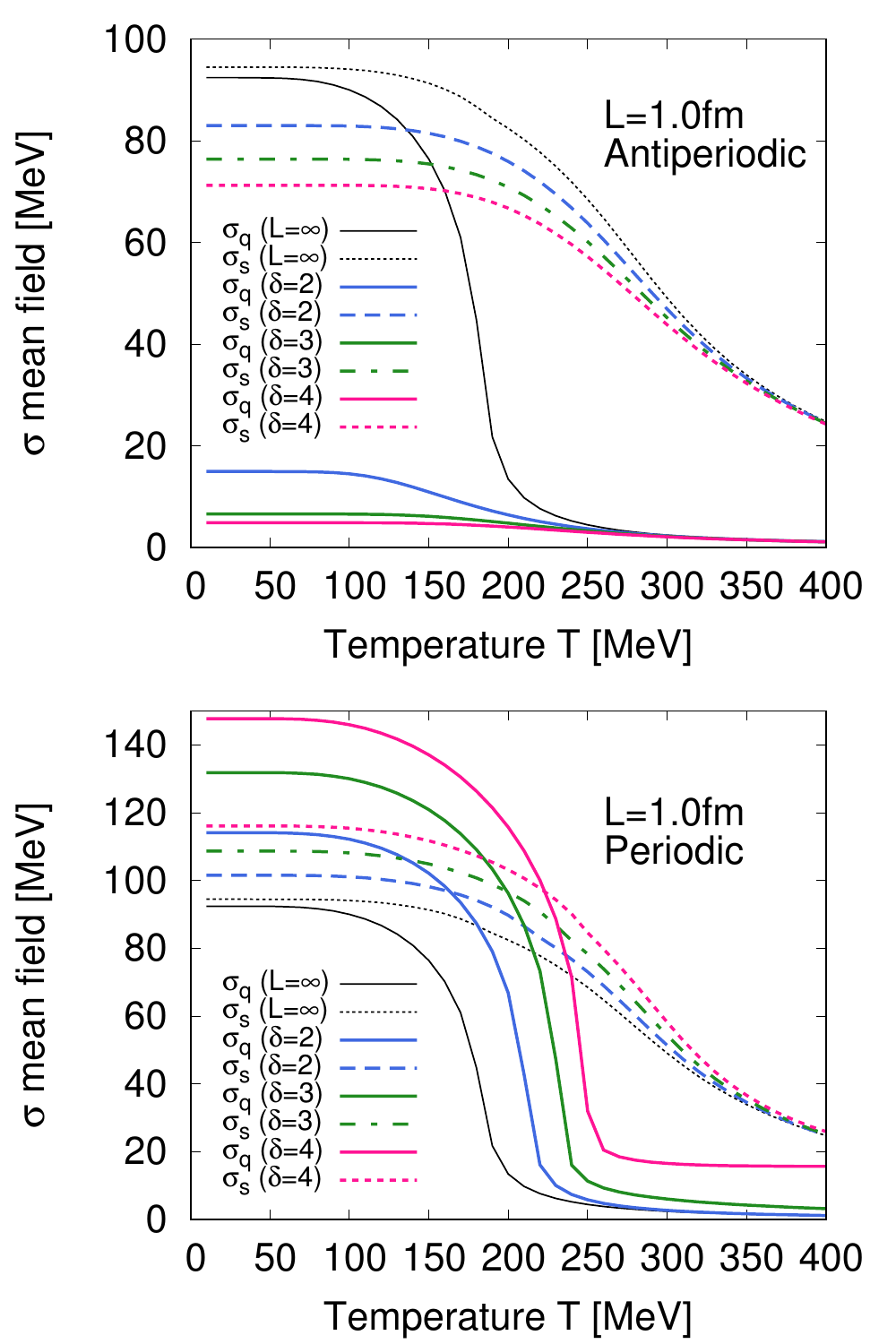}
        \end{center}
    \end{minipage}
    \caption{$\delta$ dependence of thermal transition for $\sigma$ mean fields in a finite box at $L=1.0 \, \mathrm{fm}$.
Upper: antiperiodic boundary. Lower: periodic boundary.}
    \label{sigma_TM_delta}
    \vspace{-10pt}
\end{figure}

\begin{figure}[t!]
    \begin{minipage}[t]{1.0\columnwidth}
        \begin{center}
            \includegraphics[clip, width=1.0\columnwidth]{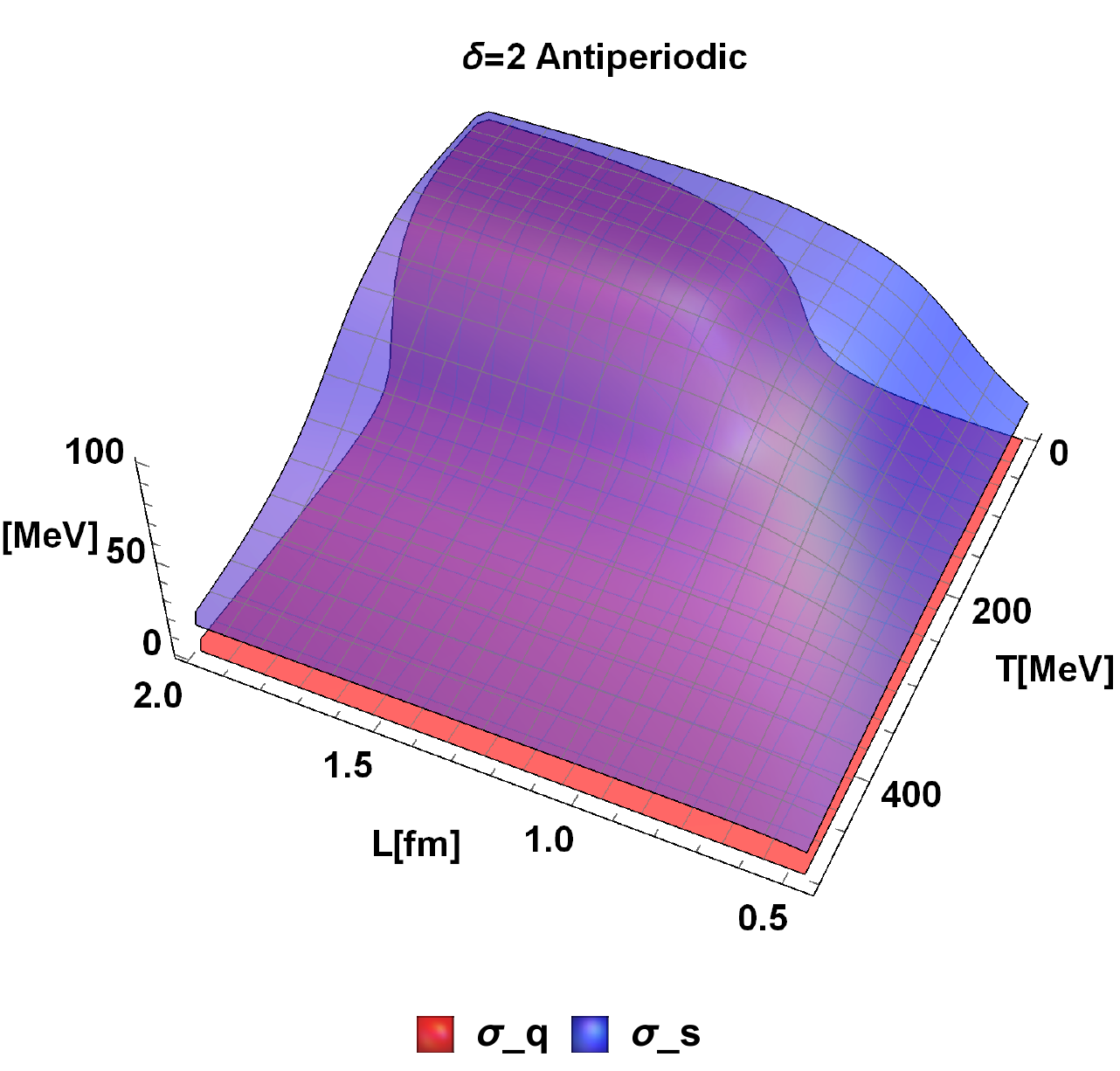}
            \includegraphics[clip, width=1.0\columnwidth]{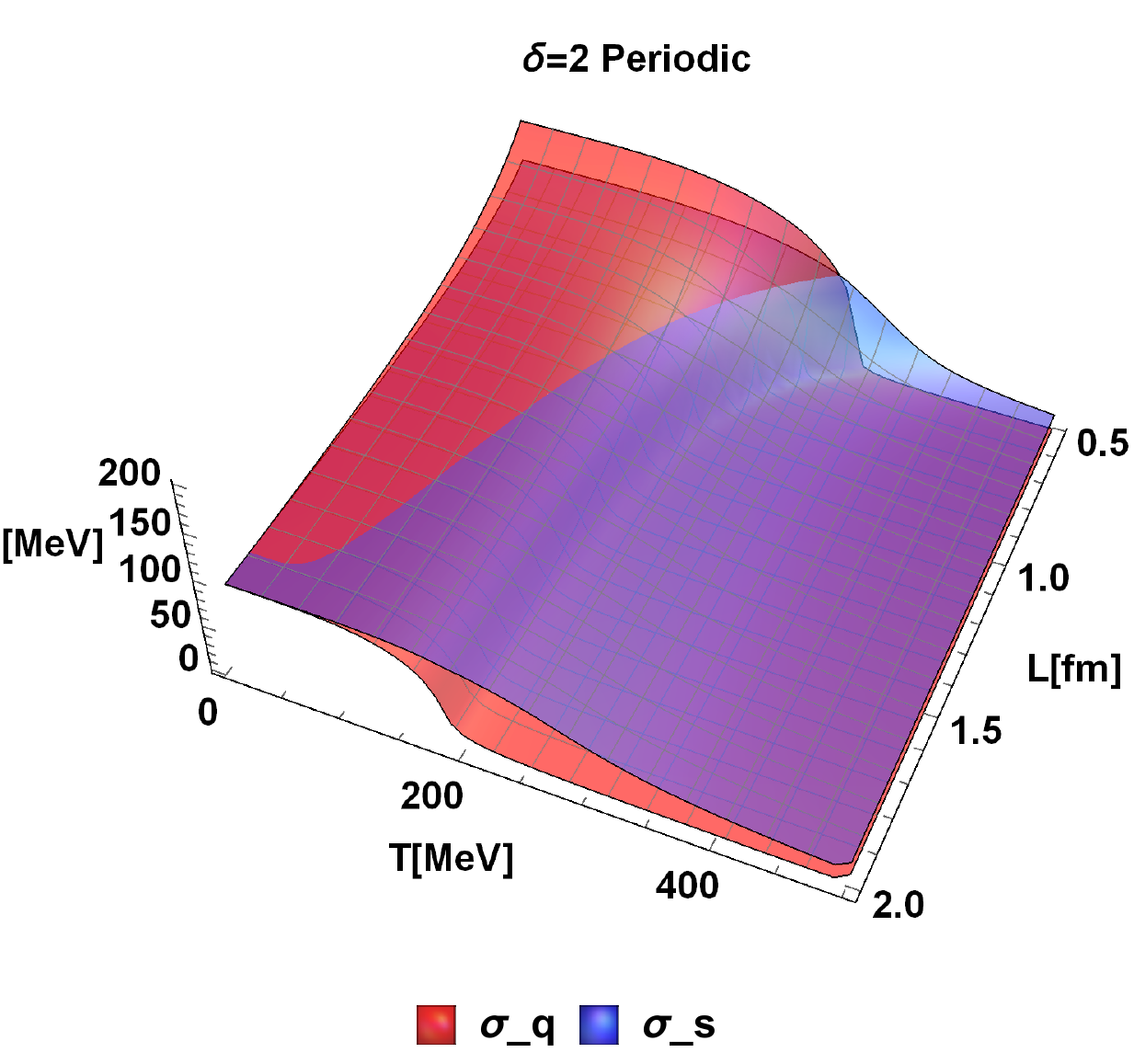}
        \end{center}
    \end{minipage}
    \caption{$\sigma_q$ (red) and $\sigma_s$ (blue) mean fields on the $L$-$T$ plane at $\delta=2$.
Upper: antiperiodic boundary. Lower: periodic boundary.}
    \label{3d}
    \vspace{-10pt}
\end{figure}

Next, we examine the dependence on the number of the compactified spatial dimensions; we compare $\delta=2,3,4$.
In Fig.~\ref{sigma_LM_delta}, we focus on the results at $T=0$.
For the antiperiodic boundary, larger $\delta$ leads to stronger restoration of the chiral symmetry.
As a result, the chiral transition length for $\sigma$ mean fields gets larger as $\delta$ increases.
For the periodic boundary, larger $\delta$ enhances the chiral symmetry breaking, and it leads to larger $\sigma$ mean fields.  

In Fig.~\ref{sigma_LM_delta_T=200}, we focus on the results at $T=200 \, \mathrm{MeV}$.
We find that the anomalous shift of $\sigma_q$ appears at $\delta=3,4$ as well as $\delta=2$, which means that even the usual cubic geometry ($\delta=4$) in lattice QCD simulations can lead to the anomalous shift of $\sigma_q$ by carefully examining the region around $L\sim 1.5 \, \mathrm{fm}$.

\begin{figure*}[t!]
    \begin{minipage}[t]{2.0\columnwidth}
        \begin{center}
            \includegraphics[clip, width=1.0\columnwidth]{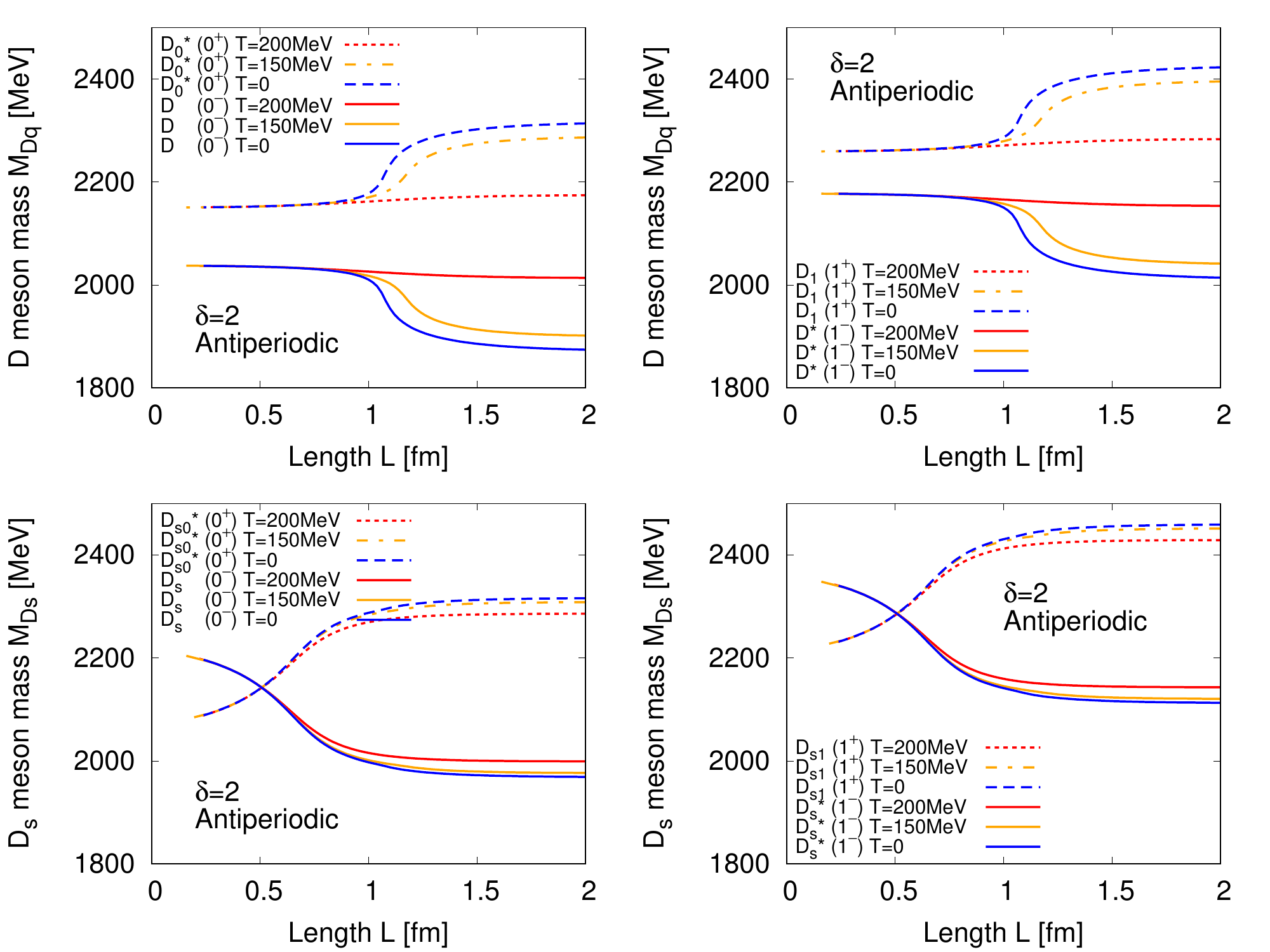}
        \end{center}
    \end{minipage}
    \caption{Volume dependences of $D_q$ and $D_s$ meson masses in a finite box at $\delta=2$ with antiperiodic boundary condition.}
    \label{D_LM_ap_delta2}
    \vspace{-10pt}
\end{figure*}

\begin{figure*}[t!]
    \begin{minipage}[t]{2.0\columnwidth}
        \begin{center}
            \includegraphics[clip, width=1.0\columnwidth]{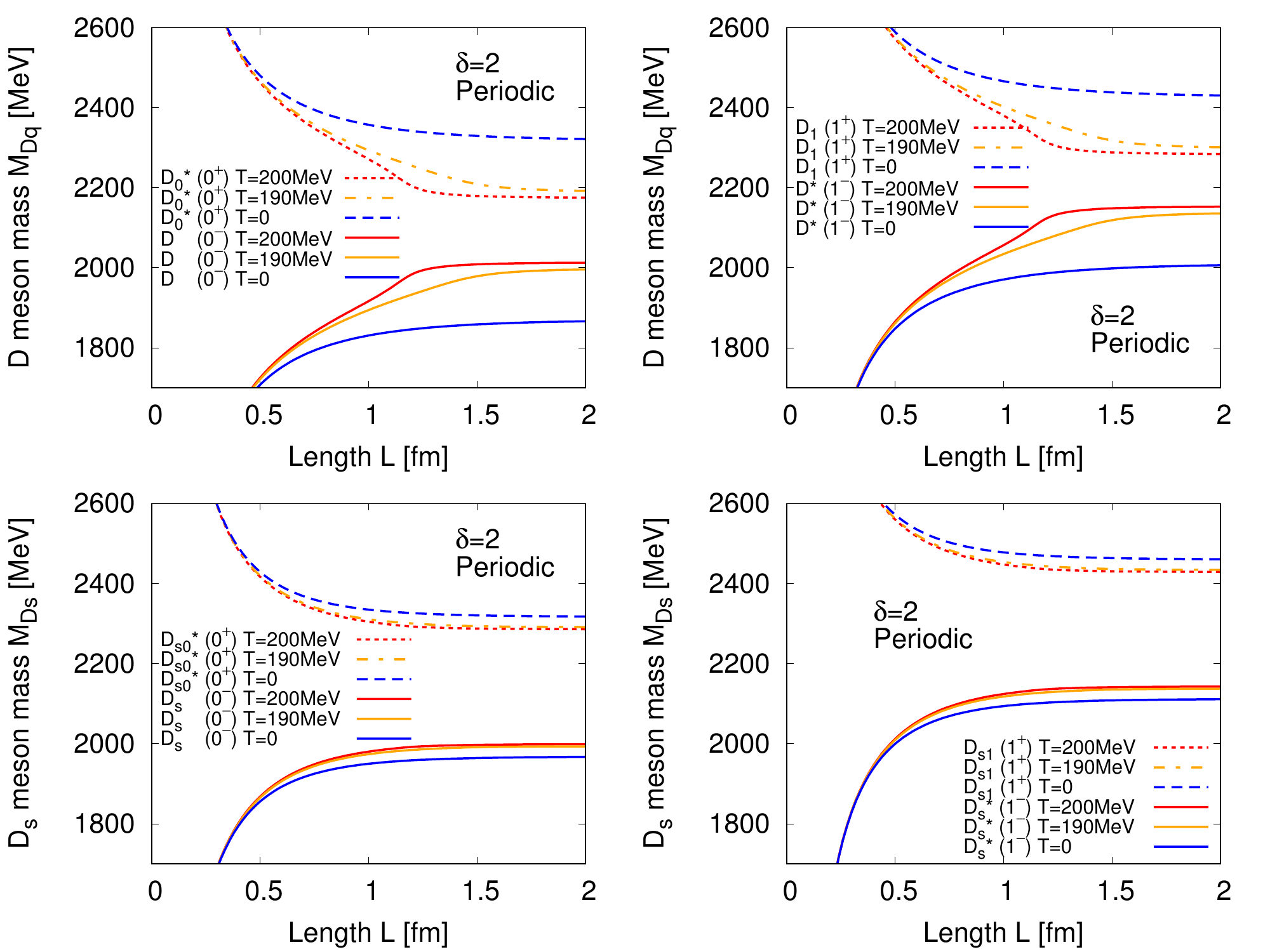}
        \end{center}
    \end{minipage}
    \caption{Volume dependences of $D_q$ and $D_s$ meson masses in a finite box at $\delta=2$ with periodic boundary condition.}
    \label{D_LM_p_delta2}
    \vspace{-10pt}
\end{figure*}

In Fig.~\ref{sigma_TM_delta2}, we show the $T$ dependence (or thermal phase transition) of the $\sigma$ mean fields at some fixed $L$ and $\delta=2$.
In infinite volume ($L \to \infty$), the $\sigma_q$ mean field is suddenly suppressed around $T=200\, \mathrm{MeV}$ that is the chiral phase transition at finite temperature.
The transition temperature for $\sigma_s$ is higher than that for $\sigma_q$.
In a finite volume with the antiperiodic boundary, since the chiral condensates decrease by the finite-volume effect, the transition temperature becomes lower as $L$ decreases.
For the periodic boundary, since the chiral condensates increase, the transition temperatures also increase.

We note that, for the antiperiodic boundary, the finite-volume effect for the $T$ dependence of $\sigma$ and its transition temperature is similar to the thermal effect for the $L$ dependence of $\sigma$ and its transition length.
In this sense, the upper panel of Fig.~\ref{sigma_TM_delta2} is similar to the upper panel of Fig.~\ref{sigma_LM_delta2}.
Such a similarity between $L$ and $\beta=1/T$ can be also understood in the form of Eq.~(\ref{Cas_ene_gene}).

In Fig.~\ref{sigma_TM_delta}, we focus on the $\delta$ dependence of the thermal phase transition at $L=1.0 \, \mathrm{fm}$.
As with the finite-volume transitions, for the antiperiodic boundary, larger $\delta$ leads to the lower transition temperature.
For the periodic boundary, larger $\delta$ enhances the chiral symmetry breaking, and the transition temperature becomes higher.

Finally, in Fig.~\ref{3d}, we summarize $\sigma_q$ and $\sigma_s$ on the $L$-$T$ plane at $\delta=2$.
For the antiperiodic boundary, $\sigma_q < \sigma_s$ is realized in any $L$ and $T$.
On the other hand, for the periodic boundary, we can observe the domain of $\sigma_q > \sigma_s$ in smaller volumes.

\subsection{Finite-$L$ transition with antiperiodic boundary}
In Fig.~\ref{D_LM_ap_delta2}, we show the $L$ dependence of the $D$ mesons for the antiperiodic boundary.
The antiperiodic boundary condition leads to the chiral symmetry restoration (or the reduction of the $\sigma$ mean fields), so it affects $D$ meson masses.
The masses of $D$ meson chiral partners (the scalar partners, $0^-$ and $0^+$, and the vector partners $1^-$ and $1^+$) become degenerate as $L$ decreases.
Thus, the degeneracy of $D$ meson masses induced by a small-volume system will be a useful signal for elucidating the chiral-partner structures of $D$ mesons from lattice QCD simulations.

In the small-$L$ region, $L< 0.5 \, \mathrm{fm}$, the $D_q$ meson mass converges to an $L$-independent constant, and the mass splitting between chiral partners survives.
The $D_s$ meson mass also converges to constants after a level crossing between the chiral partners.
Within our model, these masses in the small-$L$ limit are determined by the averaged mass $m$ and the violation parameter $\Delta_D$ of the HQSS [as defined in Eq.~(\ref{DMassWithDelta})].
However, these behaviors should be regarded as an artifact of our model because $m$ and $\Delta_D$ are fixed to reproduce the experimental values of the $D$ meson masses in infinite volume at zero temperature.
Therefore, the splitting between the chiral partners in the small-$L$ limit and the level crossing for $D_s$ mesons, shown in Fig.~\ref{D_LM_ap_delta2}, should be not physical but artificial.
In principle, the $L$ dependences of these parameters (particularly, $\Delta_D$) would be interesting, but they are beyond the scope of this paper.

\subsection{Finite-$L$ transition with periodic boundary}
In Fig.~\ref{D_LM_p_delta2}, we show the $L$ dependence of the $D$ mesons for the periodic boundary.
Since the periodic boundary condition leads to the enhancement of chiral symmetry breaking, the masses of $D$ meson chiral partners split with decreasing $L$.
In the region of $L< 0.5 \, \mathrm{fm}$, the temperature dependence is lost because the finite-volume effect overcomes the thermal effects.

For the periodic boundary, the $L$ dependences of $D_q$ and $D_s$ mesons are qualitatively similar to each other at low temperature, but the $D_q$ mesons are more sensitive to the finite-volume effects than $D_s$ mesons.
Therefore, $D_q$ mesons could be better as a probe of finite-volume effects for chiral partner structures.

We comment on ``anomalous" mass shifts at high temperature.
For all the $D_q$ mesons at $T=200 \, \mathrm{MeV}$, we find an anomalous mass shift, which is absent at lower temperature, $T=190 \, \mathrm{MeV}$.
This mass shift is induced by the anomalous $L$ dependence of $\sigma_q$, as already mentioned in Figs.~\ref{sigma_LM_delta2} and \ref{sigma_LM_delta_T=200}.
Such anomalous behaviors would be a qualitative signal for studying the finite-volume effect for $D$ mesons by using lattice QCD simulations.
Note that $T=200 \, \mathrm{MeV}$ is near the pseudocritical temperature of the chiral phase transition in our model, so the $D$ mesons may be dissolved, and we cannot measure the pole masses from the temporal correlators.
Even if that is the case, an anomalous mass shift could be observed in $D$ meson screening masses from the spatial correlators.

\begin{figure*}[t!]
    \begin{minipage}[t]{2.0\columnwidth}
        \begin{center}
            \includegraphics[clip, width=1.0\columnwidth]{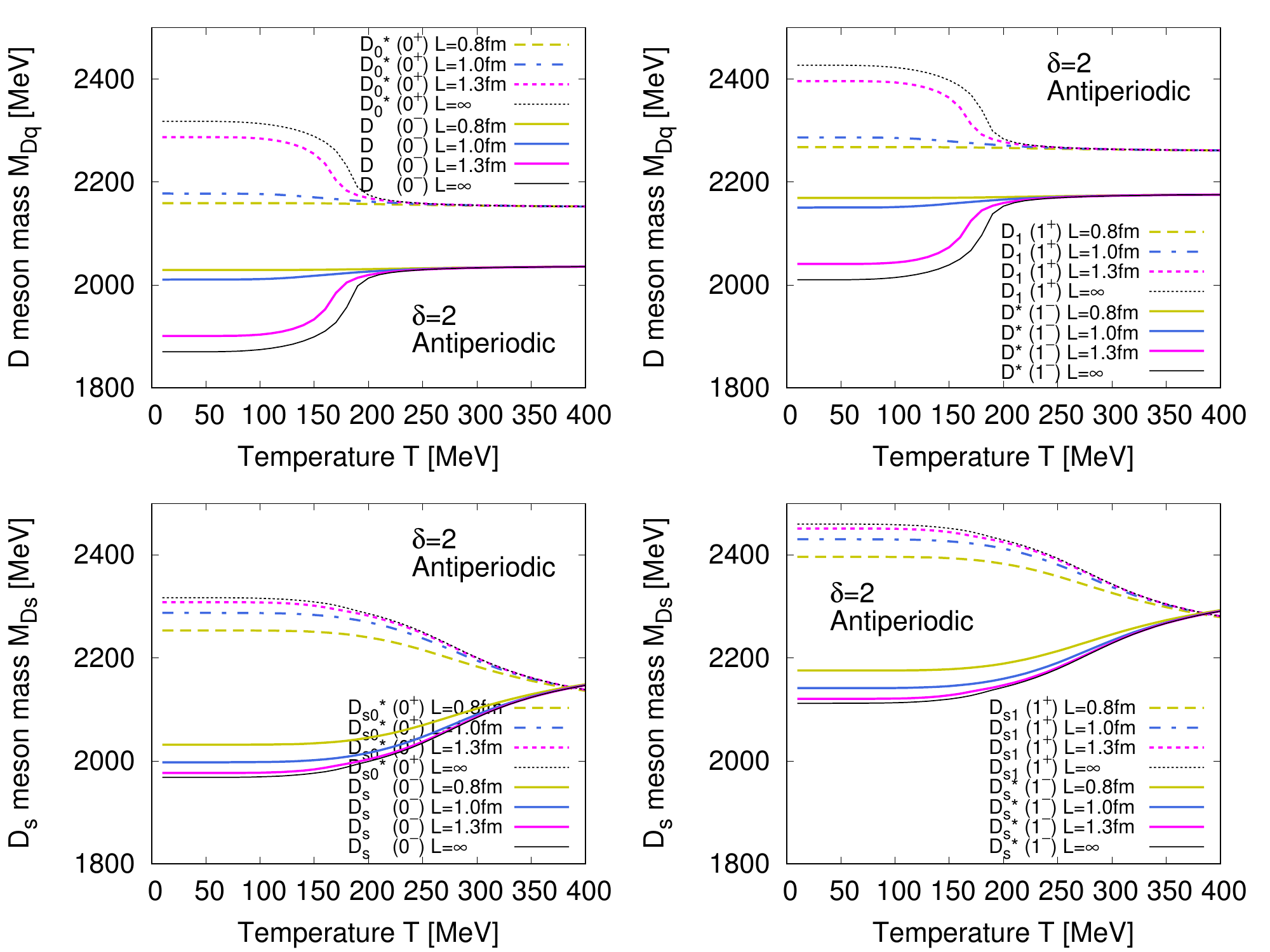}
        \end{center}
    \end{minipage}
    \caption{Temperature dependences of $D_q$ and $D_s$ meson masses in a finite box at $\delta=2$ with antiperiodic boundary condition.}
    \label{D_TM_ap_delta2}
    \vspace{-10pt}
\end{figure*}

\subsection{Finite-$T$ transition with antiperiodic boundary}
In Fig.~\ref{D_TM_ap_delta2}, we show the $T$ dependence of the $D$ mesons for the antiperiodic boundary.
Since the antiperiodic boundary condition reduces the chiral symmetry breaking, and the transition temperatures of $\sigma$ mean fields decrease, the degeneracy temperature of $D$ meson chiral partners becomes lower as $L$ decreases.

In the high-temperature phase, the values of the $\sigma$ mean fields are almost zero by the thermal effects.
Here, the infrared dynamics of quarks is dominated only by the thermal effects, and the finite-volume effects are invisible. In this phase, $D$ meson masses are determined only by the averaged mass $m$ and the violation parameters $\Delta_D$ of the HQSS.

\begin{figure*}[t!]
    \begin{minipage}[t]{2.0\columnwidth}
        \begin{center}
            \includegraphics[clip, width=1.0\columnwidth]{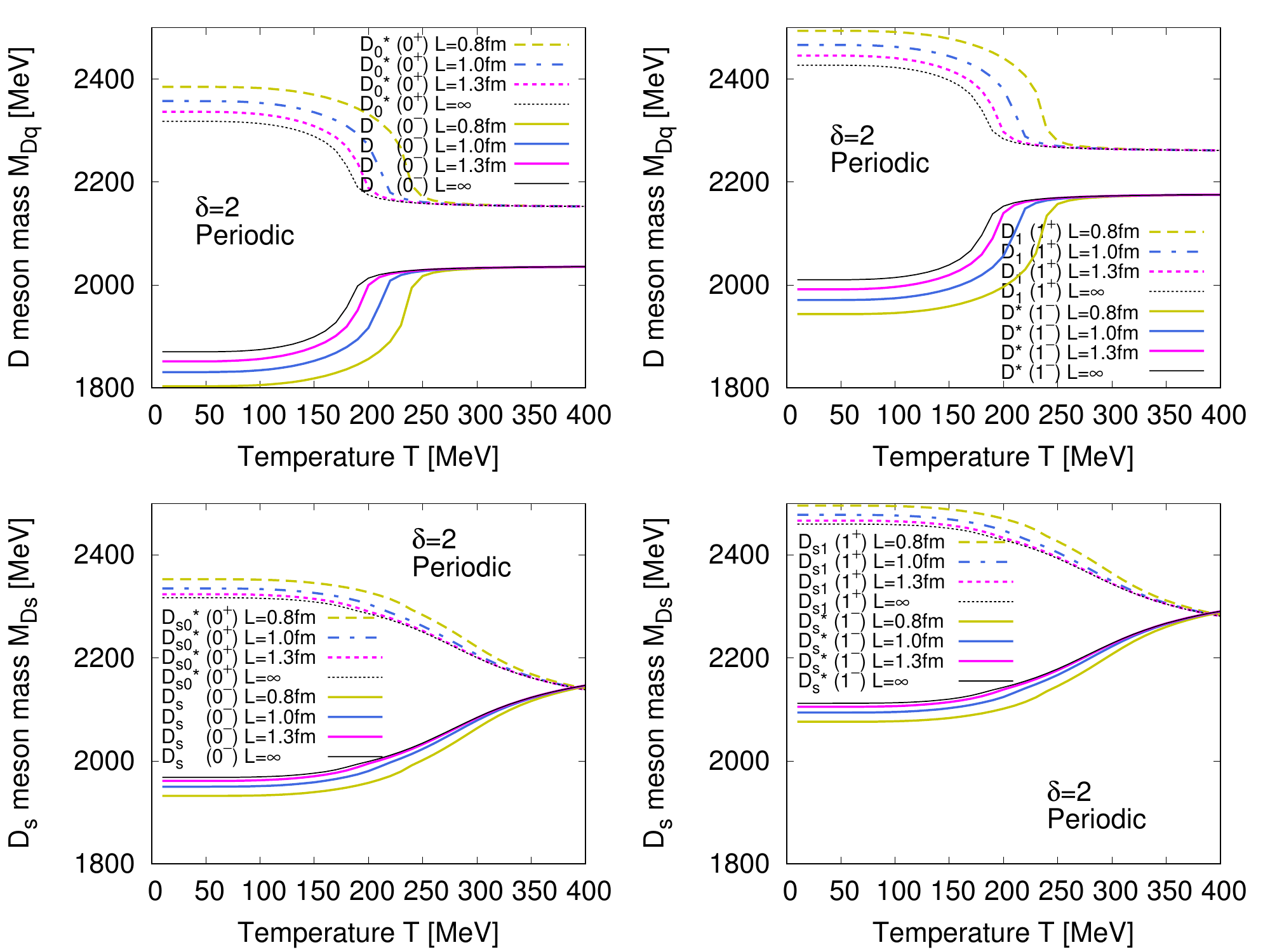}
        \end{center}
    \end{minipage}
    \caption{Temperature dependences of $D_q$ and $D_s$ meson masses in a finite box at $\delta=2$ with periodic boundary condition.}
    \label{D_TM_p_delta2}
    \vspace{-10pt}
\end{figure*}

\subsection{Finite-$T$ transition with periodic boundary}
In Fig.~\ref{D_TM_p_delta2}, we show the $T$ dependence of the $D$ mesons for the periodic boundary.
Since the periodic boundary condition catalyzes the chiral symmetry breaking, and the transition temperature of $\sigma$ mean fields increases, the degeneracy temperature of $D$ meson chiral partners gets higher as $L$ decreases.
Thus, the difference between the degeneracy temperatures with the periodic and antiperiodic boundaries will be useful in studies with lattice simulations.

$D$ meson masses in the high-temperature phase are dominated by the averaged mass $m$ and the violation parameters $\Delta_D$ of the HQSS.
This is the same as the situation with the antiperiodic boundary, as discussed in the previous subsection.

\section{Conclusions}
\label{sec:Concl}
In this work, we constructed a formalism for investigating possible finite-volume effects for $D$ meson systems.
We analyzed the finite-volume and temperature dependences of the $\sigma_q$ and $\sigma_s$ mean fields based on the linear sigma model with $2+1$-flavor quarks.
Here, the finite-volume (or Casimir) effects for constituent light quarks were introduced by using the regularization scheme with the Epstein-Hurwitz inhomogeneous zeta function.
In other words, such an effect means the Casimir effects for the chiral condensates (or $\sigma$ mean fields).
We have shown the phase diagram of $\sigma_q$ and $\sigma_s$ on the volume-temperature plane at $\delta=2$, as drawn in Fig.~\ref{3d}.
Here, for the periodic boundary, we found the region of $\sigma_q > \sigma_s$ while $\sigma_q < \sigma_s$ at any $T$ in infinite volume.

The Lagrangian for $D$ mesons was formulated based on the chiral-partner structure and heavy-quark spin symmetry for $D$ mesons.
As a result, we found the mass shifts of $D$ mesons induced by the finite-volume effects with the periodic or antiperiodic boundary.
The antiperiodic boundary leads to the chiral symmetry restoration in a small volume, so the masses of the $D$ meson chiral partners degenerate.
The periodic boundary enhances the chiral symmetry breaking, and the masses of the chiral partners split.
Furthermore, we pointed out that the anomalous mass shifts of $D$ mesons with the periodic boundary at high temperature would be a useful signal for examining finite-volume effects in lattice QCD simulations for a small volume. 

We emphasize again that $D$ mesons could be one of the clearest probes of the chiral condensates, which will be confirmed by future lattice QCD simulations.
The pole masses of $D$ mesons in small volume at low temperature can be extracted from the temporal correlators of the $D$ meson currents.
At higher temperature, the extraction of the pole masses would be difficult because of the short distance of the temporal correlators, but these can be related to other observables such as the screening masses from the spatial correlators \cite{Bazavov:2014cta,Maezawa:2016pwo} and spectral functions \cite{Kelly:2018hsi}.
To investigate screening masses and spectral functions within mean-field models will also be interesting.

\section*{Acknowledgments}
This work was supported by JSPS KAKENHI (Grant No. JP17K14277).
K.N. is supported partly by the Grant-in-Aid for Japan Society for the Promotion of Science Research Fellow (Grant No. 18J11457).
D.S. is supported by NSFC grant 20201191997.

\appendix
\section{Truncation error for Casimir energy} \label{App_1}

\begin{figure}[t!]
    \begin{minipage}[t]{1.0\columnwidth}
        \begin{center}
            \includegraphics[clip, width=1.0\columnwidth]{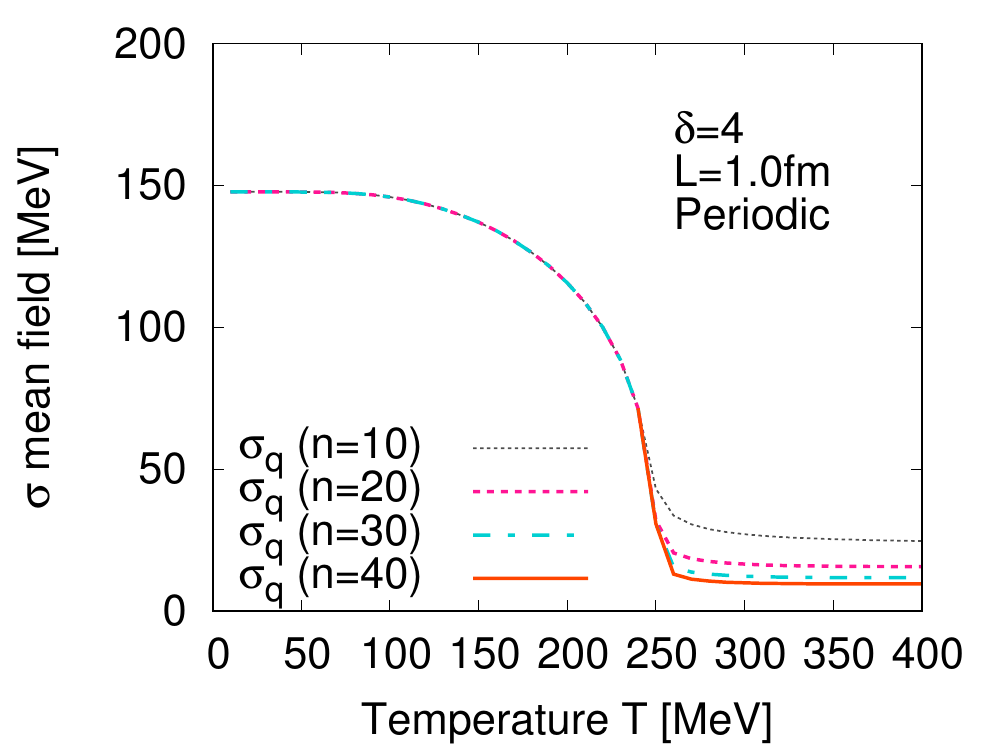}
        \end{center}
    \end{minipage}
    \caption{Truncation dependence of thermal transition for $\sigma_q$ mean field in a finite box at $L = 1.0$ fm and $\delta=4$ with periodic boundary.
The line at $n=20$ is the same as that in the lower panel of Fig.~\ref{sigma_TM_delta}.}
    \label{truncation}
    \vspace{-10pt}
\end{figure}

In this Appendix, we discuss a truncation error in our calculation.
Equation (\ref{Cas_ene_gene}), which represents the Casimir energy, has the infinite series with indices $n_i$.
In the numerical calculation, we practically have to truncate the series by introducing cutoff parameters for $n_i$.
Since this cutoff could be an origin of systematic uncertainty, we examine the error from this truncation.

In the figures shown in this paper, we plotted the results using a cutoff of $n_i \leq 20$.
To quantitatively estimate uncertainty from the truncation, we compare the difference between $n_i \leq 20$ and $n_i \leq 10$.
By using this estimate, we find that the numerical results of $\sigma_q$ and $\sigma_s$ lead to the error less than $1 \, \mathrm{MeV}$, except for $\sigma_q$ with the periodic boundary condition in Figs.~\ref{sigma_LM_delta_T=200} and \ref{sigma_TM_delta}.
In the following, we focus on the errors in Figs.~\ref{sigma_LM_delta_T=200} and \ref{sigma_TM_delta}.

First, we discuss $\sigma_q$ in Fig.~\ref{sigma_LM_delta_T=200}.
The results at $\delta = 2$ still have the precision with the error less than $1$ MeV.
The uncertainty at $\delta = 3$ is approximately $3$ MeV at $L > 1.4$ fm, and that at $\delta = 4$ is approximately $7$ MeV at $1.5$ fm.
Both the results suffer from the uncertainty in a volume larger than the transition length, and its uncertainty becomes more severe in larger volumes.

For the periodic boundary in Fig.~\ref{sigma_TM_delta}, $\sigma_q$ at $\delta = 2$ also still has the precision with the error less than $1$ MeV.
The error at $\delta = 3$ is less than $5$ MeV near the transition temperature.
At $\delta = 4$ and temperature higher than the transition, $\sigma_q$ shows a most serious error, approximately $10$ MeV.
To discuss the error convergence at $\delta = 4$, in Fig.~\ref{truncation}, we show the cutoff dependence ($n \leq 10$, $20$, $30$, and $40$).
In the high-temperature region, there is a sizable error from the truncation while the low-temperature region does not suffer from the error.
When we estimate the error from the difference between $n\leq 10$ and $n\leq 20$, the uncertainty is approximately $10$ MeV, and the difference between $n\leq 20$ and $n\leq 40$ is approximately $5$ MeV.
According to these dependences, we expect that the uncertainty is approximately $10$ MeV.

Finally, we comment on the convergence of the infinite series.
The truncation uncertainty is expected to be large at smaller $\sigma_q$ and larger $T$.
This is because the series in Eq.~(\ref{Cas_ene_gene}) converges by the modified Bessel function $K_2(\frac{n_0 M}{T})$, and this function is exponentially suppressed as $K_2(\frac{n_0 M}{T}) \sim \sqrt{\frac{T \pi}{2n_0 M}}\mathrm{exp}^{-\frac{n_0 M}{T}}$ at large $\frac{n_0 M}{T}$, where $n_0$ is large enough.
Therefore, we should pay attention to the uncertainty in that case.
Such uncertainty is serious in Figs.~\ref{sigma_LM_delta_T=200} and \ref{sigma_TM_delta}.

\bibliography{Dmeson_casimir_ref}

\begin{thebibliography}{102}%
\makeatletter
\providecommand \@ifxundefined [1]{%
 \@ifx{#1\undefined}
}%
\providecommand \@ifnum [1]{%
 \ifnum #1\expandafter \@firstoftwo
 \else \expandafter \@secondoftwo
 \fi
}%
\providecommand \@ifx [1]{%
 \ifx #1\expandafter \@firstoftwo
 \else \expandafter \@secondoftwo
 \fi
}%
\providecommand \natexlab [1]{#1}%
\providecommand \enquote  [1]{``#1''}%
\providecommand \bibnamefont  [1]{#1}%
\providecommand \bibfnamefont [1]{#1}%
\providecommand \citenamefont [1]{#1}%
\providecommand \href@noop [0]{\@secondoftwo}%
\providecommand \href [0]{\begingroup \@sanitize@url \@href}%
\providecommand \@href[1]{\@@startlink{#1}\@@href}%
\providecommand \@@href[1]{\endgroup#1\@@endlink}%
\providecommand \@sanitize@url [0]{\catcode `\\12\catcode `\$12\catcode
  `\&12\catcode `\#12\catcode `\^12\catcode `\_12\catcode `\%12\relax}%
\providecommand \@@startlink[1]{}%
\providecommand \@@endlink[0]{}%
\providecommand \url  [0]{\begingroup\@sanitize@url \@url }%
\providecommand \@url [1]{\endgroup\@href {#1}{\urlprefix }}%
\providecommand \urlprefix  [0]{URL }%
\providecommand \Eprint [0]{\href }%
\providecommand \doibase [0]{http://dx.doi.org/}%
\providecommand \selectlanguage [0]{\@gobble}%
\providecommand \bibinfo  [0]{\@secondoftwo}%
\providecommand \bibfield  [0]{\@secondoftwo}%
\providecommand \translation [1]{[#1]}%
\providecommand \BibitemOpen [0]{}%
\providecommand \bibitemStop [0]{}%
\providecommand \bibitemNoStop [0]{.\EOS\space}%
\providecommand \EOS [0]{\spacefactor3000\relax}%
\providecommand \BibitemShut  [1]{\csname bibitem#1\endcsname}%
\let\auto@bib@innerbib\@empty
\bibitem [{\citenamefont {Nowak}\ \emph {et~al.}(1993)\citenamefont {Nowak},
  \citenamefont {Rho},\ and\ \citenamefont {Zahed}}]{Nowak:1992um}%
  \BibitemOpen
  \bibfield  {author} {\bibinfo {author} {\bibfnamefont {M.~A.}\ \bibnamefont
  {Nowak}}, \bibinfo {author} {\bibfnamefont {M.}~\bibnamefont {Rho}}, \ and\
  \bibinfo {author} {\bibfnamefont {I.}~\bibnamefont {Zahed}},\ }\href
  {\doibase 10.1103/PhysRevD.48.4370} {\bibfield  {journal} {\bibinfo
  {journal} {Phys. Rev.}\ }\textbf {\bibinfo {volume} {D48}},\ \bibinfo {pages}
  {4370} (\bibinfo {year} {1993})},\ \Eprint
  {http://arxiv.org/abs/hep-ph/9209272} {arXiv:hep-ph/9209272 [hep-ph]}
  \BibitemShut {NoStop}%
\bibitem [{\citenamefont {Bardeen}\ and\ \citenamefont
  {Hill}(1994)}]{Bardeen:1993ae}%
  \BibitemOpen
  \bibfield  {author} {\bibinfo {author} {\bibfnamefont {W.~A.}\ \bibnamefont
  {Bardeen}}\ and\ \bibinfo {author} {\bibfnamefont {C.~T.}\ \bibnamefont
  {Hill}},\ }\href {\doibase 10.1103/PhysRevD.49.409} {\bibfield  {journal}
  {\bibinfo  {journal} {Phys. Rev.}\ }\textbf {\bibinfo {volume} {D49}},\
  \bibinfo {pages} {409} (\bibinfo {year} {1994})},\ \Eprint
  {http://arxiv.org/abs/hep-ph/9304265} {arXiv:hep-ph/9304265 [hep-ph]}
  \BibitemShut {NoStop}%
\bibitem [{\citenamefont {Bardeen}\ \emph {et~al.}(2003)\citenamefont
  {Bardeen}, \citenamefont {Eichten},\ and\ \citenamefont
  {Hill}}]{Bardeen:2003kt}%
  \BibitemOpen
  \bibfield  {author} {\bibinfo {author} {\bibfnamefont {W.~A.}\ \bibnamefont
  {Bardeen}}, \bibinfo {author} {\bibfnamefont {E.~J.}\ \bibnamefont
  {Eichten}}, \ and\ \bibinfo {author} {\bibfnamefont {C.~T.}\ \bibnamefont
  {Hill}},\ }\href {\doibase 10.1103/PhysRevD.68.054024} {\bibfield  {journal}
  {\bibinfo  {journal} {Phys. Rev.}\ }\textbf {\bibinfo {volume} {D68}},\
  \bibinfo {pages} {054024} (\bibinfo {year} {2003})},\ \Eprint
  {http://arxiv.org/abs/hep-ph/0305049} {arXiv:hep-ph/0305049 [hep-ph]}
  \BibitemShut {NoStop}%
\bibitem [{\citenamefont {Nowak}\ \emph {et~al.}(2004)\citenamefont {Nowak},
  \citenamefont {Rho},\ and\ \citenamefont {Zahed}}]{Nowak:2003ra}%
  \BibitemOpen
  \bibfield  {author} {\bibinfo {author} {\bibfnamefont {M.~A.}\ \bibnamefont
  {Nowak}}, \bibinfo {author} {\bibfnamefont {M.}~\bibnamefont {Rho}}, \ and\
  \bibinfo {author} {\bibfnamefont {I.}~\bibnamefont {Zahed}},\ }\href@noop {}
  {\bibfield  {journal} {\bibinfo  {journal} {Acta Phys. Polon.}\ }\textbf
  {\bibinfo {volume} {B35}},\ \bibinfo {pages} {2377} (\bibinfo {year}
  {2004})},\ \Eprint {http://arxiv.org/abs/hep-ph/0307102}
  {arXiv:hep-ph/0307102 [hep-ph]} \BibitemShut {NoStop}%
\bibitem [{\citenamefont {Harada}\ \emph {et~al.}(2004)\citenamefont {Harada},
  \citenamefont {Rho},\ and\ \citenamefont {Sasaki}}]{Harada:2003kt}%
  \BibitemOpen
  \bibfield  {author} {\bibinfo {author} {\bibfnamefont {M.}~\bibnamefont
  {Harada}}, \bibinfo {author} {\bibfnamefont {M.}~\bibnamefont {Rho}}, \ and\
  \bibinfo {author} {\bibfnamefont {C.}~\bibnamefont {Sasaki}},\ }\href
  {\doibase 10.1103/PhysRevD.70.074002} {\bibfield  {journal} {\bibinfo
  {journal} {Phys. Rev.}\ }\textbf {\bibinfo {volume} {D70}},\ \bibinfo {pages}
  {074002} (\bibinfo {year} {2004})},\ \Eprint
  {http://arxiv.org/abs/hep-ph/0312182} {arXiv:hep-ph/0312182 [hep-ph]}
  \BibitemShut {NoStop}%
\bibitem [{\citenamefont {Wise}(1992)}]{Wise:1992hn}%
  \BibitemOpen
  \bibfield  {author} {\bibinfo {author} {\bibfnamefont {M.~B.}\ \bibnamefont
  {Wise}},\ }\href {\doibase 10.1103/PhysRevD.45.R2188} {\bibfield  {journal}
  {\bibinfo  {journal} {Phys. Rev.}\ }\textbf {\bibinfo {volume} {D45}},\
  \bibinfo {pages} {R2188} (\bibinfo {year} {1992})}\BibitemShut {NoStop}%
\bibitem [{\citenamefont {Burdman}\ and\ \citenamefont
  {Donoghue}(1992)}]{Burdman:1992gh}%
  \BibitemOpen
  \bibfield  {author} {\bibinfo {author} {\bibfnamefont {G.}~\bibnamefont
  {Burdman}}\ and\ \bibinfo {author} {\bibfnamefont {J.~F.}\ \bibnamefont
  {Donoghue}},\ }\href {\doibase 10.1016/0370-2693(92)90068-F} {\bibfield
  {journal} {\bibinfo  {journal} {Phys. Lett.}\ }\textbf {\bibinfo {volume}
  {B280}},\ \bibinfo {pages} {287} (\bibinfo {year} {1992})}\BibitemShut
  {NoStop}%
\bibitem [{\citenamefont {Yan}\ \emph {et~al.}(1992)\citenamefont {Yan},
  \citenamefont {Cheng}, \citenamefont {Cheung}, \citenamefont {Lin},
  \citenamefont {Lin},\ and\ \citenamefont {Yu}}]{Yan:1992gz}%
  \BibitemOpen
  \bibfield  {author} {\bibinfo {author} {\bibfnamefont {T.-M.}\ \bibnamefont
  {Yan}}, \bibinfo {author} {\bibfnamefont {H.-Y.}\ \bibnamefont {Cheng}},
  \bibinfo {author} {\bibfnamefont {C.-Y.}\ \bibnamefont {Cheung}}, \bibinfo
  {author} {\bibfnamefont {G.-L.}\ \bibnamefont {Lin}}, \bibinfo {author}
  {\bibfnamefont {Y.~C.}\ \bibnamefont {Lin}}, \ and\ \bibinfo {author}
  {\bibfnamefont {H.-L.}\ \bibnamefont {Yu}},\ }\href {\doibase
  10.1103/PhysRevD.46.1148, 10.1103/PhysRevD.55.5851} {\bibfield  {journal}
  {\bibinfo  {journal} {Phys. Rev.}\ }\textbf {\bibinfo {volume} {D46}},\
  \bibinfo {pages} {1148} (\bibinfo {year} {1992})},\ \bibinfo {note}
  {[Erratum: Phys. Rev.D55,5851(1997)]}\BibitemShut {NoStop}%
\bibitem [{\citenamefont {Eichten}\ and\ \citenamefont
  {Hill}(1990)}]{Eichten:1989zv}%
  \BibitemOpen
  \bibfield  {author} {\bibinfo {author} {\bibfnamefont {E.}~\bibnamefont
  {Eichten}}\ and\ \bibinfo {author} {\bibfnamefont {B.~R.}\ \bibnamefont
  {Hill}},\ }\href {\doibase 10.1016/0370-2693(90)92049-O} {\bibfield
  {journal} {\bibinfo  {journal} {Phys. Lett.}\ }\textbf {\bibinfo {volume}
  {B234}},\ \bibinfo {pages} {511} (\bibinfo {year} {1990})}\BibitemShut
  {NoStop}%
\bibitem [{\citenamefont {Georgi}(1990)}]{Georgi:1990um}%
  \BibitemOpen
  \bibfield  {author} {\bibinfo {author} {\bibfnamefont {H.}~\bibnamefont
  {Georgi}},\ }\href {\doibase 10.1016/0370-2693(90)91128-X} {\bibfield
  {journal} {\bibinfo  {journal} {Phys. Lett.}\ }\textbf {\bibinfo {volume}
  {B240}},\ \bibinfo {pages} {447} (\bibinfo {year} {1990})}\BibitemShut
  {NoStop}%
\bibitem [{\citenamefont {Isgur}\ and\ \citenamefont
  {Wise}(1989)}]{Isgur:1989vq}%
  \BibitemOpen
  \bibfield  {author} {\bibinfo {author} {\bibfnamefont {N.}~\bibnamefont
  {Isgur}}\ and\ \bibinfo {author} {\bibfnamefont {M.~B.}\ \bibnamefont
  {Wise}},\ }\href {\doibase 10.1016/0370-2693(89)90566-2} {\bibfield
  {journal} {\bibinfo  {journal} {Phys. Lett.}\ }\textbf {\bibinfo {volume}
  {B232}},\ \bibinfo {pages} {113} (\bibinfo {year} {1989})}\BibitemShut
  {NoStop}%
\bibitem [{\citenamefont {Isgur}\ and\ \citenamefont
  {Wise}(1990)}]{Isgur:1989ed}%
  \BibitemOpen
  \bibfield  {author} {\bibinfo {author} {\bibfnamefont {N.}~\bibnamefont
  {Isgur}}\ and\ \bibinfo {author} {\bibfnamefont {M.~B.}\ \bibnamefont
  {Wise}},\ }\href {\doibase 10.1016/0370-2693(90)91219-2} {\bibfield
  {journal} {\bibinfo  {journal} {Phys. Lett.}\ }\textbf {\bibinfo {volume}
  {B237}},\ \bibinfo {pages} {527} (\bibinfo {year} {1990})}\BibitemShut
  {NoStop}%
\bibitem [{\citenamefont {Buchheim}\ \emph {et~al.}(2015)\citenamefont
  {Buchheim}, \citenamefont {Hilger},\ and\ \citenamefont {K{\"
  a}mpfer}}]{Buchheim:2014rpa}%
  \BibitemOpen
  \bibfield  {author} {\bibinfo {author} {\bibfnamefont {T.}~\bibnamefont
  {Buchheim}}, \bibinfo {author} {\bibfnamefont {T.}~\bibnamefont {Hilger}}, \
  and\ \bibinfo {author} {\bibfnamefont {B.}~\bibnamefont {K{\" a}mpfer}},\
  }\href {\doibase 10.1103/PhysRevC.91.015205} {\bibfield  {journal} {\bibinfo
  {journal} {Phys. Rev.}\ }\textbf {\bibinfo {volume} {C91}},\ \bibinfo {pages}
  {015205} (\bibinfo {year} {2015})},\ \Eprint {http://arxiv.org/abs/1411.7863}
  {arXiv:1411.7863 [nucl-th]} \BibitemShut {NoStop}%
\bibitem [{\citenamefont {Bazavov}\ \emph {et~al.}(2015)\citenamefont
  {Bazavov}, \citenamefont {Karsch}, \citenamefont {Maezawa}, \citenamefont
  {Mukherjee},\ and\ \citenamefont {Petreczky}}]{Bazavov:2014cta}%
  \BibitemOpen
  \bibfield  {author} {\bibinfo {author} {\bibfnamefont {A.}~\bibnamefont
  {Bazavov}}, \bibinfo {author} {\bibfnamefont {F.}~\bibnamefont {Karsch}},
  \bibinfo {author} {\bibfnamefont {Y.}~\bibnamefont {Maezawa}}, \bibinfo
  {author} {\bibfnamefont {S.}~\bibnamefont {Mukherjee}}, \ and\ \bibinfo
  {author} {\bibfnamefont {P.}~\bibnamefont {Petreczky}},\ }\href {\doibase
  10.1103/PhysRevD.91.054503} {\bibfield  {journal} {\bibinfo  {journal} {Phys.
  Rev.}\ }\textbf {\bibinfo {volume} {D91}},\ \bibinfo {pages} {054503}
  (\bibinfo {year} {2015})},\ \Eprint {http://arxiv.org/abs/1411.3018}
  {arXiv:1411.3018 [hep-lat]} \BibitemShut {NoStop}%
\bibitem [{\citenamefont {Maezawa}\ \emph {et~al.}(2016)\citenamefont
  {Maezawa}, \citenamefont {Karsch}, \citenamefont {Mukherjee},\ and\
  \citenamefont {Petreczky}}]{Maezawa:2016pwo}%
  \BibitemOpen
  \bibfield  {author} {\bibinfo {author} {\bibfnamefont {Y.}~\bibnamefont
  {Maezawa}}, \bibinfo {author} {\bibfnamefont {F.}~\bibnamefont {Karsch}},
  \bibinfo {author} {\bibfnamefont {S.}~\bibnamefont {Mukherjee}}, \ and\
  \bibinfo {author} {\bibfnamefont {P.}~\bibnamefont {Petreczky}},\ }\href
  {\doibase 10.22323/1.251.0199} {\bibfield  {journal} {\bibinfo  {journal}
  {PoS}\ }\textbf {\bibinfo {volume} {LATTICE2015}},\ \bibinfo {pages} {199}
  (\bibinfo {year} {2016})}\BibitemShut {NoStop}%
\bibitem [{\citenamefont {Kelly}\ \emph {et~al.}(2018)\citenamefont {Kelly},
  \citenamefont {Rothkopf},\ and\ \citenamefont {Skullerud}}]{Kelly:2018hsi}%
  \BibitemOpen
  \bibfield  {author} {\bibinfo {author} {\bibfnamefont {A.}~\bibnamefont
  {Kelly}}, \bibinfo {author} {\bibfnamefont {A.}~\bibnamefont {Rothkopf}}, \
  and\ \bibinfo {author} {\bibfnamefont {J.-I.}\ \bibnamefont {Skullerud}},\
  }\href {\doibase 10.1103/PhysRevD.97.114509} {\bibfield  {journal} {\bibinfo
  {journal} {Phys. Rev.}\ }\textbf {\bibinfo {volume} {D97}},\ \bibinfo {pages}
  {114509} (\bibinfo {year} {2018})},\ \Eprint
  {http://arxiv.org/abs/1802.00667} {arXiv:1802.00667 [hep-lat]} \BibitemShut
  {NoStop}%
\bibitem [{\citenamefont {Hayashigaki}(2000)}]{Hayashigaki:2000es}%
  \BibitemOpen
  \bibfield  {author} {\bibinfo {author} {\bibfnamefont {A.}~\bibnamefont
  {Hayashigaki}},\ }\href {\doibase 10.1016/S0370-2693(00)00760-7} {\bibfield
  {journal} {\bibinfo  {journal} {Phys. Lett.}\ }\textbf {\bibinfo {volume}
  {B487}},\ \bibinfo {pages} {96} (\bibinfo {year} {2000})},\ \Eprint
  {http://arxiv.org/abs/nucl-th/0001051} {arXiv:nucl-th/0001051 [nucl-th]}
  \BibitemShut {NoStop}%
\bibitem [{\citenamefont {Hilger}\ \emph {et~al.}(2009)\citenamefont {Hilger},
  \citenamefont {Thomas},\ and\ \citenamefont {K{\" a}mpfer}}]{Hilger:2008jg}%
  \BibitemOpen
  \bibfield  {author} {\bibinfo {author} {\bibfnamefont {T.}~\bibnamefont
  {Hilger}}, \bibinfo {author} {\bibfnamefont {R.}~\bibnamefont {Thomas}}, \
  and\ \bibinfo {author} {\bibfnamefont {B.}~\bibnamefont {K{\" a}mpfer}},\
  }\href {\doibase 10.1103/PhysRevC.79.025202} {\bibfield  {journal} {\bibinfo
  {journal} {Phys. Rev.}\ }\textbf {\bibinfo {volume} {C79}},\ \bibinfo {pages}
  {025202} (\bibinfo {year} {2009})},\ \Eprint {http://arxiv.org/abs/0809.4996}
  {arXiv:0809.4996 [nucl-th]} \BibitemShut {NoStop}%
\bibitem [{\citenamefont {Wang}\ and\ \citenamefont
  {Huang}(2011)}]{Wang:2011mj}%
  \BibitemOpen
  \bibfield  {author} {\bibinfo {author} {\bibfnamefont {Z.-G.}\ \bibnamefont
  {Wang}}\ and\ \bibinfo {author} {\bibfnamefont {T.}~\bibnamefont {Huang}},\
  }\href {\doibase 10.1103/PhysRevC.84.048201} {\bibfield  {journal} {\bibinfo
  {journal} {Phys. Rev.}\ }\textbf {\bibinfo {volume} {C84}},\ \bibinfo {pages}
  {048201} (\bibinfo {year} {2011})},\ \Eprint {http://arxiv.org/abs/1107.5889}
  {arXiv:1107.5889 [hep-ph]} \BibitemShut {NoStop}%
\bibitem [{\citenamefont {Hilger}\ \emph {et~al.}(2011)\citenamefont {Hilger},
  \citenamefont {K{\" a}mpfer},\ and\ \citenamefont {Leupold}}]{Hilger:2011cq}%
  \BibitemOpen
  \bibfield  {author} {\bibinfo {author} {\bibfnamefont {T.}~\bibnamefont
  {Hilger}}, \bibinfo {author} {\bibfnamefont {B.}~\bibnamefont {K{\"
  a}mpfer}}, \ and\ \bibinfo {author} {\bibfnamefont {S.}~\bibnamefont
  {Leupold}},\ }\href {\doibase 10.1103/PhysRevC.84.045202} {\bibfield
  {journal} {\bibinfo  {journal} {Phys. Rev.}\ }\textbf {\bibinfo {volume}
  {C84}},\ \bibinfo {pages} {045202} (\bibinfo {year} {2011})},\ \Eprint
  {http://arxiv.org/abs/1109.1229} {arXiv:1109.1229 [nucl-th]} \BibitemShut
  {NoStop}%
\bibitem [{\citenamefont {Wang}(2013)}]{Wang:2011fv}%
  \BibitemOpen
  \bibfield  {author} {\bibinfo {author} {\bibfnamefont {Z.-G.}\ \bibnamefont
  {Wang}},\ }\href {\doibase 10.1142/S0217751X13500498} {\bibfield  {journal}
  {\bibinfo  {journal} {Int. J. Mod. Phys.}\ }\textbf {\bibinfo {volume}
  {A28}},\ \bibinfo {pages} {1350049} (\bibinfo {year} {2013})},\ \Eprint
  {http://arxiv.org/abs/1109.5219} {arXiv:1109.5219 [hep-ph]} \BibitemShut
  {NoStop}%
\bibitem [{\citenamefont {Azizi}\ \emph {et~al.}(2014)\citenamefont {Azizi},
  \citenamefont {Er},\ and\ \citenamefont {Sundu}}]{Azizi:2014bba}%
  \BibitemOpen
  \bibfield  {author} {\bibinfo {author} {\bibfnamefont {K.}~\bibnamefont
  {Azizi}}, \bibinfo {author} {\bibfnamefont {N.}~\bibnamefont {Er}}, \ and\
  \bibinfo {author} {\bibfnamefont {H.}~\bibnamefont {Sundu}},\ }\href
  {\doibase 10.1140/epjc/s10052-014-3021-1} {\bibfield  {journal} {\bibinfo
  {journal} {Eur. Phys. J.}\ }\textbf {\bibinfo {volume} {C74}},\ \bibinfo
  {pages} {3021} (\bibinfo {year} {2014})},\ \Eprint
  {http://arxiv.org/abs/1405.3058} {arXiv:1405.3058 [hep-ph]} \BibitemShut
  {NoStop}%
\bibitem [{\citenamefont {Wang}(2015)}]{Wang:2015uya}%
  \BibitemOpen
  \bibfield  {author} {\bibinfo {author} {\bibfnamefont {Z.-G.}\ \bibnamefont
  {Wang}},\ }\href {\doibase 10.1103/PhysRevC.92.065205} {\bibfield  {journal}
  {\bibinfo  {journal} {Phys. Rev.}\ }\textbf {\bibinfo {volume} {C92}},\
  \bibinfo {pages} {065205} (\bibinfo {year} {2015})},\ \Eprint
  {http://arxiv.org/abs/1501.05093} {arXiv:1501.05093 [hep-ph]} \BibitemShut
  {NoStop}%
\bibitem [{\citenamefont {Suzuki}\ \emph {et~al.}(2016)\citenamefont {Suzuki},
  \citenamefont {Gubler},\ and\ \citenamefont {Oka}}]{Suzuki:2015est}%
  \BibitemOpen
  \bibfield  {author} {\bibinfo {author} {\bibfnamefont {K.}~\bibnamefont
  {Suzuki}}, \bibinfo {author} {\bibfnamefont {P.}~\bibnamefont {Gubler}}, \
  and\ \bibinfo {author} {\bibfnamefont {M.}~\bibnamefont {Oka}},\ }\href
  {\doibase 10.1103/PhysRevC.93.045209} {\bibfield  {journal} {\bibinfo
  {journal} {Phys. Rev.}\ }\textbf {\bibinfo {volume} {C93}},\ \bibinfo {pages}
  {045209} (\bibinfo {year} {2016})},\ \Eprint
  {http://arxiv.org/abs/1511.04513} {arXiv:1511.04513 [hep-ph]} \BibitemShut
  {NoStop}%
\bibitem [{\citenamefont {Buchheim}\ \emph {et~al.}(2018)\citenamefont
  {Buchheim}, \citenamefont {Hilger}, \citenamefont {K{\" a}mpfer},\ and\
  \citenamefont {Leupold}}]{Buchheim:2018kss}%
  \BibitemOpen
  \bibfield  {author} {\bibinfo {author} {\bibfnamefont {T.}~\bibnamefont
  {Buchheim}}, \bibinfo {author} {\bibfnamefont {T.}~\bibnamefont {Hilger}},
  \bibinfo {author} {\bibfnamefont {B.}~\bibnamefont {K{\" a}mpfer}}, \ and\
  \bibinfo {author} {\bibfnamefont {S.}~\bibnamefont {Leupold}},\ }\href
  {\doibase 10.1088/1361-6471/aab44e} {\bibfield  {journal} {\bibinfo
  {journal} {J. Phys.}\ }\textbf {\bibinfo {volume} {G45}},\ \bibinfo {pages}
  {085104} (\bibinfo {year} {2018})},\ \Eprint
  {http://arxiv.org/abs/1801.01472} {arXiv:1801.01472 [nucl-th]} \BibitemShut
  {NoStop}%
\bibitem [{\citenamefont {Tsushima}\ \emph {et~al.}(1999)\citenamefont
  {Tsushima}, \citenamefont {Lu}, \citenamefont {Thomas}, \citenamefont
  {Saito},\ and\ \citenamefont {Landau}}]{Tsushima:1998ru}%
  \BibitemOpen
  \bibfield  {author} {\bibinfo {author} {\bibfnamefont {K.}~\bibnamefont
  {Tsushima}}, \bibinfo {author} {\bibfnamefont {D.-H.}\ \bibnamefont {Lu}},
  \bibinfo {author} {\bibfnamefont {A.~W.}\ \bibnamefont {Thomas}}, \bibinfo
  {author} {\bibfnamefont {K.}~\bibnamefont {Saito}}, \ and\ \bibinfo {author}
  {\bibfnamefont {R.~H.}\ \bibnamefont {Landau}},\ }\href {\doibase
  10.1103/PhysRevC.59.2824} {\bibfield  {journal} {\bibinfo  {journal} {Phys.
  Rev.}\ }\textbf {\bibinfo {volume} {C59}},\ \bibinfo {pages} {2824} (\bibinfo
  {year} {1999})},\ \Eprint {http://arxiv.org/abs/nucl-th/9810016}
  {arXiv:nucl-th/9810016 [nucl-th]} \BibitemShut {NoStop}%
\bibitem [{\citenamefont {Sibirtsev}\ \emph {et~al.}(1999)\citenamefont
  {Sibirtsev}, \citenamefont {Tsushima},\ and\ \citenamefont
  {Thomas}}]{Sibirtsev:1999js}%
  \BibitemOpen
  \bibfield  {author} {\bibinfo {author} {\bibfnamefont {A.}~\bibnamefont
  {Sibirtsev}}, \bibinfo {author} {\bibfnamefont {K.}~\bibnamefont {Tsushima}},
  \ and\ \bibinfo {author} {\bibfnamefont {A.~W.}\ \bibnamefont {Thomas}},\
  }\href {\doibase 10.1007/s100500050353} {\bibfield  {journal} {\bibinfo
  {journal} {Eur. Phys. J.}\ }\textbf {\bibinfo {volume} {A6}},\ \bibinfo
  {pages} {351} (\bibinfo {year} {1999})},\ \Eprint
  {http://arxiv.org/abs/nucl-th/9904016} {arXiv:nucl-th/9904016 [nucl-th]}
  \BibitemShut {NoStop}%
\bibitem [{\citenamefont {Mishra}\ \emph {et~al.}(2004)\citenamefont {Mishra},
  \citenamefont {Bratkovskaya}, \citenamefont {Schaffner-Bielich},
  \citenamefont {Schramm},\ and\ \citenamefont {St{\" o}cker}}]{Mishra:2003se}%
  \BibitemOpen
  \bibfield  {author} {\bibinfo {author} {\bibfnamefont {A.}~\bibnamefont
  {Mishra}}, \bibinfo {author} {\bibfnamefont {E.~L.}\ \bibnamefont
  {Bratkovskaya}}, \bibinfo {author} {\bibfnamefont {J.}~\bibnamefont
  {Schaffner-Bielich}}, \bibinfo {author} {\bibfnamefont {S.}~\bibnamefont
  {Schramm}}, \ and\ \bibinfo {author} {\bibfnamefont {H.}~\bibnamefont {St{\"
  o}cker}},\ }\href {\doibase 10.1103/PhysRevC.69.015202} {\bibfield  {journal}
  {\bibinfo  {journal} {Phys. Rev.}\ }\textbf {\bibinfo {volume} {C69}},\
  \bibinfo {pages} {015202} (\bibinfo {year} {2004})},\ \Eprint
  {http://arxiv.org/abs/nucl-th/0308082} {arXiv:nucl-th/0308082 [nucl-th]}
  \BibitemShut {NoStop}%
\bibitem [{\citenamefont {Mishra}\ and\ \citenamefont
  {Mazumdar}(2009)}]{Mishra:2008cd}%
  \BibitemOpen
  \bibfield  {author} {\bibinfo {author} {\bibfnamefont {A.}~\bibnamefont
  {Mishra}}\ and\ \bibinfo {author} {\bibfnamefont {A.}~\bibnamefont
  {Mazumdar}},\ }\href {\doibase 10.1103/PhysRevC.79.024908} {\bibfield
  {journal} {\bibinfo  {journal} {Phys. Rev.}\ }\textbf {\bibinfo {volume}
  {C79}},\ \bibinfo {pages} {024908} (\bibinfo {year} {2009})},\ \Eprint
  {http://arxiv.org/abs/0810.3067} {arXiv:0810.3067 [nucl-th]} \BibitemShut
  {NoStop}%
\bibitem [{\citenamefont {Kumar}\ and\ \citenamefont
  {Mishra}(2010)}]{Kumar:2010gb}%
  \BibitemOpen
  \bibfield  {author} {\bibinfo {author} {\bibfnamefont {A.}~\bibnamefont
  {Kumar}}\ and\ \bibinfo {author} {\bibfnamefont {A.}~\bibnamefont {Mishra}},\
  }\href {\doibase 10.1103/PhysRevC.81.065204} {\bibfield  {journal} {\bibinfo
  {journal} {Phys. Rev.}\ }\textbf {\bibinfo {volume} {C81}},\ \bibinfo {pages}
  {065204} (\bibinfo {year} {2010})},\ \Eprint {http://arxiv.org/abs/1005.5018}
  {arXiv:1005.5018 [nucl-th]} \BibitemShut {NoStop}%
\bibitem [{\citenamefont {Kumar}\ and\ \citenamefont
  {Mishra}(2011)}]{Kumar:2011ff}%
  \BibitemOpen
  \bibfield  {author} {\bibinfo {author} {\bibfnamefont {A.}~\bibnamefont
  {Kumar}}\ and\ \bibinfo {author} {\bibfnamefont {A.}~\bibnamefont {Mishra}},\
  }\href {\doibase 10.1140/epja/i2011-11164-6} {\bibfield  {journal} {\bibinfo
  {journal} {Eur. Phys. J.}\ }\textbf {\bibinfo {volume} {A47}},\ \bibinfo
  {pages} {164} (\bibinfo {year} {2011})},\ \Eprint
  {http://arxiv.org/abs/1102.4792} {arXiv:1102.4792 [nucl-th]} \BibitemShut
  {NoStop}%
\bibitem [{\citenamefont {Blaschke}\ \emph {et~al.}(2012)\citenamefont
  {Blaschke}, \citenamefont {Costa},\ and\ \citenamefont
  {Kalinovsky}}]{Blaschke:2011yv}%
  \BibitemOpen
  \bibfield  {author} {\bibinfo {author} {\bibfnamefont {D.}~\bibnamefont
  {Blaschke}}, \bibinfo {author} {\bibfnamefont {P.}~\bibnamefont {Costa}}, \
  and\ \bibinfo {author} {\bibfnamefont {{\relax Yu}.~L.}\ \bibnamefont
  {Kalinovsky}},\ }\href {\doibase 10.1103/PhysRevD.85.034005} {\bibfield
  {journal} {\bibinfo  {journal} {Phys. Rev.}\ }\textbf {\bibinfo {volume}
  {D85}},\ \bibinfo {pages} {034005} (\bibinfo {year} {2012})},\ \Eprint
  {http://arxiv.org/abs/1107.2913} {arXiv:1107.2913 [hep-ph]} \BibitemShut
  {NoStop}%
\bibitem [{\citenamefont {Suenaga}\ \emph {et~al.}(2014)\citenamefont
  {Suenaga}, \citenamefont {He}, \citenamefont {Ma},\ and\ \citenamefont
  {Harada}}]{Suenaga:2014dia}%
  \BibitemOpen
  \bibfield  {author} {\bibinfo {author} {\bibfnamefont {D.}~\bibnamefont
  {Suenaga}}, \bibinfo {author} {\bibfnamefont {B.-R.}\ \bibnamefont {He}},
  \bibinfo {author} {\bibfnamefont {Y.-L.}\ \bibnamefont {Ma}}, \ and\ \bibinfo
  {author} {\bibfnamefont {M.}~\bibnamefont {Harada}},\ }\href {\doibase
  10.1103/PhysRevC.89.068201} {\bibfield  {journal} {\bibinfo  {journal} {Phys.
  Rev.}\ }\textbf {\bibinfo {volume} {C89}},\ \bibinfo {pages} {068201}
  (\bibinfo {year} {2014})},\ \Eprint {http://arxiv.org/abs/1403.5140}
  {arXiv:1403.5140 [hep-ph]} \BibitemShut {NoStop}%
\bibitem [{\citenamefont {Sasaki}(2014)}]{Sasaki:2014asa}%
  \BibitemOpen
  \bibfield  {author} {\bibinfo {author} {\bibfnamefont {C.}~\bibnamefont
  {Sasaki}},\ }\href {\doibase 10.1103/PhysRevD.90.114007} {\bibfield
  {journal} {\bibinfo  {journal} {Phys. Rev.}\ }\textbf {\bibinfo {volume}
  {D90}},\ \bibinfo {pages} {114007} (\bibinfo {year} {2014})},\ \Eprint
  {http://arxiv.org/abs/1409.3420} {arXiv:1409.3420 [hep-ph]} \BibitemShut
  {NoStop}%
\bibitem [{\citenamefont {Suenaga}\ \emph {et~al.}(2015)\citenamefont
  {Suenaga}, \citenamefont {He}, \citenamefont {Ma},\ and\ \citenamefont
  {Harada}}]{Suenaga:2014sga}%
  \BibitemOpen
  \bibfield  {author} {\bibinfo {author} {\bibfnamefont {D.}~\bibnamefont
  {Suenaga}}, \bibinfo {author} {\bibfnamefont {B.-R.}\ \bibnamefont {He}},
  \bibinfo {author} {\bibfnamefont {Y.-L.}\ \bibnamefont {Ma}}, \ and\ \bibinfo
  {author} {\bibfnamefont {M.}~\bibnamefont {Harada}},\ }\href {\doibase
  10.1103/PhysRevD.91.036001} {\bibfield  {journal} {\bibinfo  {journal} {Phys.
  Rev.}\ }\textbf {\bibinfo {volume} {D91}},\ \bibinfo {pages} {036001}
  (\bibinfo {year} {2015})},\ \Eprint {http://arxiv.org/abs/1412.2462}
  {arXiv:1412.2462 [hep-ph]} \BibitemShut {NoStop}%
\bibitem [{\citenamefont {Sasaki}\ and\ \citenamefont
  {Redlich}(2015)}]{Sasaki:2014wma}%
  \BibitemOpen
  \bibfield  {author} {\bibinfo {author} {\bibfnamefont {C.}~\bibnamefont
  {Sasaki}}\ and\ \bibinfo {author} {\bibfnamefont {K.}~\bibnamefont
  {Redlich}},\ }\href {\doibase 10.1103/PhysRevD.91.074021} {\bibfield
  {journal} {\bibinfo  {journal} {Phys. Rev.}\ }\textbf {\bibinfo {volume}
  {D91}},\ \bibinfo {pages} {074021} (\bibinfo {year} {2015})},\ \Eprint
  {http://arxiv.org/abs/1412.7365} {arXiv:1412.7365 [hep-ph]} \BibitemShut
  {NoStop}%
\bibitem [{\citenamefont {Suenaga}\ and\ \citenamefont
  {Harada}(2016)}]{Suenaga:2015daa}%
  \BibitemOpen
  \bibfield  {author} {\bibinfo {author} {\bibfnamefont {D.}~\bibnamefont
  {Suenaga}}\ and\ \bibinfo {author} {\bibfnamefont {M.}~\bibnamefont
  {Harada}},\ }\href {\doibase 10.1103/PhysRevD.93.076005} {\bibfield
  {journal} {\bibinfo  {journal} {Phys. Rev.}\ }\textbf {\bibinfo {volume}
  {D93}},\ \bibinfo {pages} {076005} (\bibinfo {year} {2016})},\ \Eprint
  {http://arxiv.org/abs/1509.08578} {arXiv:1509.08578 [hep-ph]} \BibitemShut
  {NoStop}%
\bibitem [{\citenamefont {Park}\ \emph {et~al.}(2016)\citenamefont {Park},
  \citenamefont {Gubler}, \citenamefont {Harada}, \citenamefont {Lee},
  \citenamefont {Nonaka},\ and\ \citenamefont {Park}}]{Park:2016xrw}%
  \BibitemOpen
  \bibfield  {author} {\bibinfo {author} {\bibfnamefont {A.}~\bibnamefont
  {Park}}, \bibinfo {author} {\bibfnamefont {P.}~\bibnamefont {Gubler}},
  \bibinfo {author} {\bibfnamefont {M.}~\bibnamefont {Harada}}, \bibinfo
  {author} {\bibfnamefont {S.~H.}\ \bibnamefont {Lee}}, \bibinfo {author}
  {\bibfnamefont {C.}~\bibnamefont {Nonaka}}, \ and\ \bibinfo {author}
  {\bibfnamefont {W.}~\bibnamefont {Park}},\ }\href {\doibase
  10.1103/PhysRevD.93.054035} {\bibfield  {journal} {\bibinfo  {journal} {Phys.
  Rev.}\ }\textbf {\bibinfo {volume} {D93}},\ \bibinfo {pages} {054035}
  (\bibinfo {year} {2016})},\ \Eprint {http://arxiv.org/abs/1601.01250}
  {arXiv:1601.01250 [nucl-th]} \BibitemShut {NoStop}%
\bibitem [{\citenamefont {Harada}\ \emph {et~al.}(2016)\citenamefont {Harada},
  \citenamefont {Ma}, \citenamefont {Suenaga},\ and\ \citenamefont
  {Takeda}}]{Harada:2016uca}%
  \BibitemOpen
  \bibfield  {author} {\bibinfo {author} {\bibfnamefont {M.}~\bibnamefont
  {Harada}}, \bibinfo {author} {\bibfnamefont {Y.-L.}\ \bibnamefont {Ma}},
  \bibinfo {author} {\bibfnamefont {D.}~\bibnamefont {Suenaga}}, \ and\
  \bibinfo {author} {\bibfnamefont {Y.}~\bibnamefont {Takeda}},\ }\href
  {\doibase 10.1093/ptep/ptx140} {\bibfield  {journal} {\bibinfo  {journal}
  {PTEP}\ }\textbf {\bibinfo {volume} {2017}},\ \bibinfo {pages} {113D01}
  (\bibinfo {year} {2016})},\ \Eprint {http://arxiv.org/abs/1612.03496}
  {arXiv:1612.03496 [hep-ph]} \BibitemShut {NoStop}%
\bibitem [{\citenamefont {Suenaga}\ \emph {et~al.}(2017)\citenamefont
  {Suenaga}, \citenamefont {Yasui},\ and\ \citenamefont
  {Harada}}]{Suenaga:2017deu}%
  \BibitemOpen
  \bibfield  {author} {\bibinfo {author} {\bibfnamefont {D.}~\bibnamefont
  {Suenaga}}, \bibinfo {author} {\bibfnamefont {S.}~\bibnamefont {Yasui}}, \
  and\ \bibinfo {author} {\bibfnamefont {M.}~\bibnamefont {Harada}},\ }\href
  {\doibase 10.1103/PhysRevC.96.015204} {\bibfield  {journal} {\bibinfo
  {journal} {Phys. Rev.}\ }\textbf {\bibinfo {volume} {C96}},\ \bibinfo {pages}
  {015204} (\bibinfo {year} {2017})},\ \Eprint
  {http://arxiv.org/abs/1703.02762} {arXiv:1703.02762 [nucl-th]} \BibitemShut
  {NoStop}%
\bibitem [{\citenamefont {Suenaga}(2018)}]{Suenaga:2018kta}%
  \BibitemOpen
  \bibfield  {author} {\bibinfo {author} {\bibfnamefont {D.}~\bibnamefont
  {Suenaga}},\ }\href@noop {} {\  (\bibinfo {year} {2018})},\ \Eprint
  {http://arxiv.org/abs/1805.01709} {arXiv:1805.01709 [nucl-th]} \BibitemShut
  {NoStop}%
\bibitem [{\citenamefont {Sugiura}\ and\ \citenamefont
  {Hyodo}(2019)}]{Sugiura:2019ane}%
  \BibitemOpen
  \bibfield  {author} {\bibinfo {author} {\bibfnamefont {T.}~\bibnamefont
  {Sugiura}}\ and\ \bibinfo {author} {\bibfnamefont {T.}~\bibnamefont
  {Hyodo}},\ }\href {\doibase 10.1103/PhysRevC.99.065201} {\bibfield  {journal}
  {\bibinfo  {journal} {Phys. Rev.}\ }\textbf {\bibinfo {volume} {C99}},\
  \bibinfo {pages} {065201} (\bibinfo {year} {2019})},\ \Eprint
  {http://arxiv.org/abs/1904.00589} {arXiv:1904.00589 [hep-ph]} \BibitemShut
  {NoStop}%
\bibitem [{\citenamefont {Hosaka}\ \emph {et~al.}(2017)\citenamefont {Hosaka},
  \citenamefont {Hyodo}, \citenamefont {Sudoh}, \citenamefont {Yamaguchi},\
  and\ \citenamefont {Yasui}}]{Hosaka:2016ypm}%
  \BibitemOpen
  \bibfield  {author} {\bibinfo {author} {\bibfnamefont {A.}~\bibnamefont
  {Hosaka}}, \bibinfo {author} {\bibfnamefont {T.}~\bibnamefont {Hyodo}},
  \bibinfo {author} {\bibfnamefont {K.}~\bibnamefont {Sudoh}}, \bibinfo
  {author} {\bibfnamefont {Y.}~\bibnamefont {Yamaguchi}}, \ and\ \bibinfo
  {author} {\bibfnamefont {S.}~\bibnamefont {Yasui}},\ }\href {\doibase
  10.1016/j.ppnp.2017.04.003} {\bibfield  {journal} {\bibinfo  {journal} {Prog.
  Part. Nucl. Phys.}\ }\textbf {\bibinfo {volume} {96}},\ \bibinfo {pages} {88}
  (\bibinfo {year} {2017})},\ \Eprint {http://arxiv.org/abs/1606.08685}
  {arXiv:1606.08685 [hep-ph]} \BibitemShut {NoStop}%
\bibitem [{\citenamefont {Tol{\' o}s}\ \emph {et~al.}(2004)\citenamefont
  {Tol{\' o}s}, \citenamefont {Schaffner-Bielich},\ and\ \citenamefont
  {Mishra}}]{Tolos:2004yg}%
  \BibitemOpen
  \bibfield  {author} {\bibinfo {author} {\bibfnamefont {L.}~\bibnamefont
  {Tol{\' o}s}}, \bibinfo {author} {\bibfnamefont {J.}~\bibnamefont
  {Schaffner-Bielich}}, \ and\ \bibinfo {author} {\bibfnamefont
  {A.}~\bibnamefont {Mishra}},\ }\href {\doibase 10.1103/PhysRevC.70.025203}
  {\bibfield  {journal} {\bibinfo  {journal} {Phys. Rev.}\ }\textbf {\bibinfo
  {volume} {C70}},\ \bibinfo {pages} {025203} (\bibinfo {year} {2004})},\
  \Eprint {http://arxiv.org/abs/nucl-th/0404064} {arXiv:nucl-th/0404064
  [nucl-th]} \BibitemShut {NoStop}%
\bibitem [{\citenamefont {Lutz}\ and\ \citenamefont
  {Korpa}(2006)}]{Lutz:2005vx}%
  \BibitemOpen
  \bibfield  {author} {\bibinfo {author} {\bibfnamefont {M.~F.~M.}\
  \bibnamefont {Lutz}}\ and\ \bibinfo {author} {\bibfnamefont {C.~L.}\
  \bibnamefont {Korpa}},\ }\href {\doibase 10.1016/j.physletb.2005.11.046}
  {\bibfield  {journal} {\bibinfo  {journal} {Phys. Lett.}\ }\textbf {\bibinfo
  {volume} {B633}},\ \bibinfo {pages} {43} (\bibinfo {year} {2006})},\ \Eprint
  {http://arxiv.org/abs/nucl-th/0510006} {arXiv:nucl-th/0510006 [nucl-th]}
  \BibitemShut {NoStop}%
\bibitem [{\citenamefont {Mizutani}\ and\ \citenamefont
  {Ramos}(2006)}]{Mizutani:2006vq}%
  \BibitemOpen
  \bibfield  {author} {\bibinfo {author} {\bibfnamefont {T.}~\bibnamefont
  {Mizutani}}\ and\ \bibinfo {author} {\bibfnamefont {A.}~\bibnamefont
  {Ramos}},\ }\href {\doibase 10.1103/PhysRevC.74.065201} {\bibfield  {journal}
  {\bibinfo  {journal} {Phys. Rev.}\ }\textbf {\bibinfo {volume} {C74}},\
  \bibinfo {pages} {065201} (\bibinfo {year} {2006})},\ \Eprint
  {http://arxiv.org/abs/hep-ph/0607257} {arXiv:hep-ph/0607257 [hep-ph]}
  \BibitemShut {NoStop}%
\bibitem [{\citenamefont {Tol{\' o}s}\ \emph {et~al.}(2008)\citenamefont
  {Tol{\' o}s}, \citenamefont {Ramos},\ and\ \citenamefont
  {Mizutani}}]{Tolos:2007vh}%
  \BibitemOpen
  \bibfield  {author} {\bibinfo {author} {\bibfnamefont {L.}~\bibnamefont
  {Tol{\' o}s}}, \bibinfo {author} {\bibfnamefont {A.}~\bibnamefont {Ramos}}, \
  and\ \bibinfo {author} {\bibfnamefont {T.}~\bibnamefont {Mizutani}},\ }\href
  {\doibase 10.1103/PhysRevC.77.015207} {\bibfield  {journal} {\bibinfo
  {journal} {Phys. Rev.}\ }\textbf {\bibinfo {volume} {C77}},\ \bibinfo {pages}
  {015207} (\bibinfo {year} {2008})},\ \Eprint {http://arxiv.org/abs/0710.2684}
  {arXiv:0710.2684 [nucl-th]} \BibitemShut {NoStop}%
\bibitem [{\citenamefont {Molina}\ \emph {et~al.}(2009)\citenamefont {Molina},
  \citenamefont {Gamermann}, \citenamefont {Oset},\ and\ \citenamefont {Tol{\'
  o}s}}]{Molina:2008nh}%
  \BibitemOpen
  \bibfield  {author} {\bibinfo {author} {\bibfnamefont {R.}~\bibnamefont
  {Molina}}, \bibinfo {author} {\bibfnamefont {D.}~\bibnamefont {Gamermann}},
  \bibinfo {author} {\bibfnamefont {E.}~\bibnamefont {Oset}}, \ and\ \bibinfo
  {author} {\bibfnamefont {L.}~\bibnamefont {Tol{\' o}s}},\ }\href {\doibase
  10.1140/epja/i2009-10853-y} {\bibfield  {journal} {\bibinfo  {journal} {Eur.
  Phys. J.}\ }\textbf {\bibinfo {volume} {A42}},\ \bibinfo {pages} {31}
  (\bibinfo {year} {2009})},\ \Eprint {http://arxiv.org/abs/0806.3711}
  {arXiv:0806.3711 [nucl-th]} \BibitemShut {NoStop}%
\bibitem [{\citenamefont {Tol{\' o}s}\ \emph {et~al.}(2009)\citenamefont
  {Tol{\' o}s}, \citenamefont {Garc{\' i}a-Recio},\ and\ \citenamefont
  {Nieves}}]{Tolos:2009nn}%
  \BibitemOpen
  \bibfield  {author} {\bibinfo {author} {\bibfnamefont {L.}~\bibnamefont
  {Tol{\' o}s}}, \bibinfo {author} {\bibfnamefont {C.}~\bibnamefont {Garc{\'
  i}a-Recio}}, \ and\ \bibinfo {author} {\bibfnamefont {J.}~\bibnamefont
  {Nieves}},\ }\href {\doibase 10.1103/PhysRevC.80.065202} {\bibfield
  {journal} {\bibinfo  {journal} {Phys. Rev.}\ }\textbf {\bibinfo {volume}
  {C80}},\ \bibinfo {pages} {065202} (\bibinfo {year} {2009})},\ \Eprint
  {http://arxiv.org/abs/0905.4859} {arXiv:0905.4859 [nucl-th]} \BibitemShut
  {NoStop}%
\bibitem [{\citenamefont {Jim{\' e}nez-Tejero}\ \emph
  {et~al.}(2011)\citenamefont {Jim{\' e}nez-Tejero}, \citenamefont {Ramos},
  \citenamefont {Tol{\' o}s},\ and\ \citenamefont {Vida{\~
  n}a}}]{JimenezTejero:2011fc}%
  \BibitemOpen
  \bibfield  {author} {\bibinfo {author} {\bibfnamefont {C.~E.}\ \bibnamefont
  {Jim{\' e}nez-Tejero}}, \bibinfo {author} {\bibfnamefont {A.}~\bibnamefont
  {Ramos}}, \bibinfo {author} {\bibfnamefont {L.}~\bibnamefont {Tol{\' o}s}}, \
  and\ \bibinfo {author} {\bibfnamefont {I.}~\bibnamefont {Vida{\~ n}a}},\
  }\href {\doibase 10.1103/PhysRevC.84.015208} {\bibfield  {journal} {\bibinfo
  {journal} {Phys. Rev.}\ }\textbf {\bibinfo {volume} {C84}},\ \bibinfo {pages}
  {015208} (\bibinfo {year} {2011})},\ \Eprint {http://arxiv.org/abs/1102.4786}
  {arXiv:1102.4786 [hep-ph]} \BibitemShut {NoStop}%
\bibitem [{\citenamefont {Yasui}\ and\ \citenamefont
  {Sudoh}(2013)}]{Yasui:2012rw}%
  \BibitemOpen
  \bibfield  {author} {\bibinfo {author} {\bibfnamefont {S.}~\bibnamefont
  {Yasui}}\ and\ \bibinfo {author} {\bibfnamefont {K.}~\bibnamefont {Sudoh}},\
  }\href {\doibase 10.1103/PhysRevC.87.015202} {\bibfield  {journal} {\bibinfo
  {journal} {Phys. Rev.}\ }\textbf {\bibinfo {volume} {C87}},\ \bibinfo {pages}
  {015202} (\bibinfo {year} {2013})},\ \Eprint {http://arxiv.org/abs/1207.3134}
  {arXiv:1207.3134 [hep-ph]} \BibitemShut {NoStop}%
\bibitem [{\citenamefont {Fuchs}\ \emph {et~al.}(2006)\citenamefont {Fuchs},
  \citenamefont {Martemyanov}, \citenamefont {Faessler},\ and\ \citenamefont
  {Krivoruchenko}}]{Fuchs:2004fh}%
  \BibitemOpen
  \bibfield  {author} {\bibinfo {author} {\bibfnamefont {C.}~\bibnamefont
  {Fuchs}}, \bibinfo {author} {\bibfnamefont {B.~V.}\ \bibnamefont
  {Martemyanov}}, \bibinfo {author} {\bibfnamefont {A.}~\bibnamefont
  {Faessler}}, \ and\ \bibinfo {author} {\bibfnamefont {M.~I.}\ \bibnamefont
  {Krivoruchenko}},\ }\href {\doibase 10.1103/PhysRevC.73.035204} {\bibfield
  {journal} {\bibinfo  {journal} {Phys. Rev.}\ }\textbf {\bibinfo {volume}
  {C73}},\ \bibinfo {pages} {035204} (\bibinfo {year} {2006})},\ \Eprint
  {http://arxiv.org/abs/nucl-th/0410065} {arXiv:nucl-th/0410065 [nucl-th]}
  \BibitemShut {NoStop}%
\bibitem [{\citenamefont {He}\ \emph {et~al.}(2011)\citenamefont {He},
  \citenamefont {Fries},\ and\ \citenamefont {Rapp}}]{He:2011yi}%
  \BibitemOpen
  \bibfield  {author} {\bibinfo {author} {\bibfnamefont {M.}~\bibnamefont
  {He}}, \bibinfo {author} {\bibfnamefont {R.~J.}\ \bibnamefont {Fries}}, \
  and\ \bibinfo {author} {\bibfnamefont {R.}~\bibnamefont {Rapp}},\ }\href
  {\doibase 10.1016/j.physletb.2011.06.019} {\bibfield  {journal} {\bibinfo
  {journal} {Phys. Lett.}\ }\textbf {\bibinfo {volume} {B701}},\ \bibinfo
  {pages} {445} (\bibinfo {year} {2011})},\ \Eprint
  {http://arxiv.org/abs/1103.6279} {arXiv:1103.6279 [nucl-th]} \BibitemShut
  {NoStop}%
\bibitem [{\citenamefont {Ghosh}\ \emph {et~al.}(2013)\citenamefont {Ghosh},
  \citenamefont {Mitra},\ and\ \citenamefont {Sarkar}}]{Ghosh:2013xea}%
  \BibitemOpen
  \bibfield  {author} {\bibinfo {author} {\bibfnamefont {S.}~\bibnamefont
  {Ghosh}}, \bibinfo {author} {\bibfnamefont {S.}~\bibnamefont {Mitra}}, \ and\
  \bibinfo {author} {\bibfnamefont {S.}~\bibnamefont {Sarkar}},\ }\href
  {\doibase 10.1016/j.nuclphysa.2013.08.010} {\bibfield  {journal} {\bibinfo
  {journal} {Nucl. Phys.}\ }\textbf {\bibinfo {volume} {A917}},\ \bibinfo
  {pages} {71} (\bibinfo {year} {2013})},\ \Eprint
  {http://arxiv.org/abs/1309.0161} {arXiv:1309.0161 [nucl-th]} \BibitemShut
  {NoStop}%
\bibitem [{\citenamefont {Cleven}\ \emph {et~al.}(2017)\citenamefont {Cleven},
  \citenamefont {Magas},\ and\ \citenamefont {Ramos}}]{Cleven:2017fun}%
  \BibitemOpen
  \bibfield  {author} {\bibinfo {author} {\bibfnamefont {M.}~\bibnamefont
  {Cleven}}, \bibinfo {author} {\bibfnamefont {V.~K.}\ \bibnamefont {Magas}}, \
  and\ \bibinfo {author} {\bibfnamefont {A.}~\bibnamefont {Ramos}},\ }\href
  {\doibase 10.1103/PhysRevC.96.045201} {\bibfield  {journal} {\bibinfo
  {journal} {Phys. Rev.}\ }\textbf {\bibinfo {volume} {C96}},\ \bibinfo {pages}
  {045201} (\bibinfo {year} {2017})},\ \Eprint
  {http://arxiv.org/abs/1707.05728} {arXiv:1707.05728 [hep-ph]} \BibitemShut
  {NoStop}%
\bibitem [{\citenamefont {Machado}\ \emph {et~al.}(2014)\citenamefont
  {Machado}, \citenamefont {Matheus}, \citenamefont {Finazzo},\ and\
  \citenamefont {Noronha}}]{Machado:2013yaa}%
  \BibitemOpen
  \bibfield  {author} {\bibinfo {author} {\bibfnamefont {C.~S.}\ \bibnamefont
  {Machado}}, \bibinfo {author} {\bibfnamefont {R.~D.}\ \bibnamefont
  {Matheus}}, \bibinfo {author} {\bibfnamefont {S.~I.}\ \bibnamefont
  {Finazzo}}, \ and\ \bibinfo {author} {\bibfnamefont {J.}~\bibnamefont
  {Noronha}},\ }\href {\doibase 10.1103/PhysRevD.89.074027} {\bibfield
  {journal} {\bibinfo  {journal} {Phys. Rev.}\ }\textbf {\bibinfo {volume}
  {D89}},\ \bibinfo {pages} {074027} (\bibinfo {year} {2014})},\ \Eprint
  {http://arxiv.org/abs/1307.1797} {arXiv:1307.1797 [hep-ph]} \BibitemShut
  {NoStop}%
\bibitem [{\citenamefont {Gubler}\ \emph {et~al.}(2016)\citenamefont {Gubler},
  \citenamefont {Hattori}, \citenamefont {Lee}, \citenamefont {Oka},
  \citenamefont {Ozaki},\ and\ \citenamefont {Suzuki}}]{Gubler:2015qok}%
  \BibitemOpen
  \bibfield  {author} {\bibinfo {author} {\bibfnamefont {P.}~\bibnamefont
  {Gubler}}, \bibinfo {author} {\bibfnamefont {K.}~\bibnamefont {Hattori}},
  \bibinfo {author} {\bibfnamefont {S.~H.}\ \bibnamefont {Lee}}, \bibinfo
  {author} {\bibfnamefont {M.}~\bibnamefont {Oka}}, \bibinfo {author}
  {\bibfnamefont {S.}~\bibnamefont {Ozaki}}, \ and\ \bibinfo {author}
  {\bibfnamefont {K.}~\bibnamefont {Suzuki}},\ }\href {\doibase
  10.1103/PhysRevD.93.054026} {\bibfield  {journal} {\bibinfo  {journal} {Phys.
  Rev.}\ }\textbf {\bibinfo {volume} {D93}},\ \bibinfo {pages} {054026}
  (\bibinfo {year} {2016})},\ \Eprint {http://arxiv.org/abs/1512.08864}
  {arXiv:1512.08864 [hep-ph]} \BibitemShut {NoStop}%
\bibitem [{\citenamefont {Yoshida}\ and\ \citenamefont
  {Suzuki}(2016)}]{Yoshida:2016xgm}%
  \BibitemOpen
  \bibfield  {author} {\bibinfo {author} {\bibfnamefont {T.}~\bibnamefont
  {Yoshida}}\ and\ \bibinfo {author} {\bibfnamefont {K.}~\bibnamefont
  {Suzuki}},\ }\href {\doibase 10.1103/PhysRevD.94.074043} {\bibfield
  {journal} {\bibinfo  {journal} {Phys. Rev.}\ }\textbf {\bibinfo {volume}
  {D94}},\ \bibinfo {pages} {074043} (\bibinfo {year} {2016})},\ \Eprint
  {http://arxiv.org/abs/1607.04935} {arXiv:1607.04935 [hep-ph]} \BibitemShut
  {NoStop}%
\bibitem [{\citenamefont {Reddy~P.}\ \emph {et~al.}(2018)\citenamefont
  {Reddy~P.}, \citenamefont {Jahan C.~S.}, \citenamefont {Dhale}, \citenamefont
  {Mishra},\ and\ \citenamefont {Schaffner-Bielich}}]{Reddy:2017pqp}%
  \BibitemOpen
  \bibfield  {author} {\bibinfo {author} {\bibfnamefont {S.}~\bibnamefont
  {Reddy~P.}}, \bibinfo {author} {\bibfnamefont {A.}~\bibnamefont {Jahan
  C.~S.}}, \bibinfo {author} {\bibfnamefont {N.}~\bibnamefont {Dhale}},
  \bibinfo {author} {\bibfnamefont {A.}~\bibnamefont {Mishra}}, \ and\ \bibinfo
  {author} {\bibfnamefont {J.}~\bibnamefont {Schaffner-Bielich}},\ }\href
  {\doibase 10.1103/PhysRevC.97.065208} {\bibfield  {journal} {\bibinfo
  {journal} {Phys. Rev.}\ }\textbf {\bibinfo {volume} {C97}},\ \bibinfo {pages}
  {065208} (\bibinfo {year} {2018})},\ \Eprint
  {http://arxiv.org/abs/1712.07997} {arXiv:1712.07997 [nucl-th]} \BibitemShut
  {NoStop}%
\bibitem [{\citenamefont {Dhale}\ \emph {et~al.}(2018)\citenamefont {Dhale},
  \citenamefont {Reddy~P.}, \citenamefont {Jahan C.~S.},\ and\ \citenamefont
  {Mishra}}]{Dhale:2018plh}%
  \BibitemOpen
  \bibfield  {author} {\bibinfo {author} {\bibfnamefont {N.}~\bibnamefont
  {Dhale}}, \bibinfo {author} {\bibfnamefont {S.}~\bibnamefont {Reddy~P.}},
  \bibinfo {author} {\bibfnamefont {A.}~\bibnamefont {Jahan C.~S.}}, \ and\
  \bibinfo {author} {\bibfnamefont {A.}~\bibnamefont {Mishra}},\ }\href
  {\doibase 10.1103/PhysRevC.98.015202} {\bibfield  {journal} {\bibinfo
  {journal} {Phys. Rev.}\ }\textbf {\bibinfo {volume} {C98}},\ \bibinfo {pages}
  {015202} (\bibinfo {year} {2018})},\ \Eprint
  {http://arxiv.org/abs/1801.06405} {arXiv:1801.06405 [nucl-th]} \BibitemShut
  {NoStop}%
\bibitem [{\citenamefont {Casimir}(1948)}]{Casimir:1948dh}%
  \BibitemOpen
  \bibfield  {author} {\bibinfo {author} {\bibfnamefont {H.~B.~G.}\
  \bibnamefont {Casimir}},\ }\href@noop {} {\bibfield  {journal} {\bibinfo
  {journal} {Proc. Kon. Ned. Akad. Wetensch.}\ }\textbf {\bibinfo {volume}
  {51}},\ \bibinfo {pages} {793} (\bibinfo {year} {1948})}\BibitemShut
  {NoStop}%
\bibitem [{\citenamefont {Chernodub}\ \emph {et~al.}(2016)\citenamefont
  {Chernodub}, \citenamefont {Goy},\ and\ \citenamefont
  {Molochkov}}]{Chernodub:2016owp}%
  \BibitemOpen
  \bibfield  {author} {\bibinfo {author} {\bibfnamefont {M.~N.}\ \bibnamefont
  {Chernodub}}, \bibinfo {author} {\bibfnamefont {V.~A.}\ \bibnamefont {Goy}},
  \ and\ \bibinfo {author} {\bibfnamefont {A.~V.}\ \bibnamefont {Molochkov}},\
  }\href {\doibase 10.1103/PhysRevD.94.094504} {\bibfield  {journal} {\bibinfo
  {journal} {Phys. Rev.}\ }\textbf {\bibinfo {volume} {D94}},\ \bibinfo {pages}
  {094504} (\bibinfo {year} {2016})},\ \Eprint
  {http://arxiv.org/abs/1609.02323} {arXiv:1609.02323 [hep-lat]} \BibitemShut
  {NoStop}%
\bibitem [{\citenamefont {Chernodub}\ \emph
  {et~al.}(2017{\natexlab{a}})\citenamefont {Chernodub}, \citenamefont {Goy},\
  and\ \citenamefont {Molochkov}}]{Chernodub:2017mhi}%
  \BibitemOpen
  \bibfield  {author} {\bibinfo {author} {\bibfnamefont {M.~N.}\ \bibnamefont
  {Chernodub}}, \bibinfo {author} {\bibfnamefont {V.~A.}\ \bibnamefont {Goy}},
  \ and\ \bibinfo {author} {\bibfnamefont {A.~V.}\ \bibnamefont {Molochkov}},\
  }\href {\doibase 10.1103/PhysRevD.95.074511} {\bibfield  {journal} {\bibinfo
  {journal} {Phys. Rev.}\ }\textbf {\bibinfo {volume} {D95}},\ \bibinfo {pages}
  {074511} (\bibinfo {year} {2017}{\natexlab{a}})},\ \Eprint
  {http://arxiv.org/abs/1703.03439} {arXiv:1703.03439 [hep-lat]} \BibitemShut
  {NoStop}%
\bibitem [{\citenamefont {Chernodub}\ \emph
  {et~al.}(2017{\natexlab{b}})\citenamefont {Chernodub}, \citenamefont {Goy},\
  and\ \citenamefont {Molochkov}}]{Chernodub:2017gwe}%
  \BibitemOpen
  \bibfield  {author} {\bibinfo {author} {\bibfnamefont {M.~N.}\ \bibnamefont
  {Chernodub}}, \bibinfo {author} {\bibfnamefont {V.~A.}\ \bibnamefont {Goy}},
  \ and\ \bibinfo {author} {\bibfnamefont {A.~V.}\ \bibnamefont {Molochkov}},\
  }\href {\doibase 10.1103/PhysRevD.96.094507} {\bibfield  {journal} {\bibinfo
  {journal} {Phys. Rev.}\ }\textbf {\bibinfo {volume} {D96}},\ \bibinfo {pages}
  {094507} (\bibinfo {year} {2017}{\natexlab{b}})},\ \Eprint
  {http://arxiv.org/abs/1709.02262} {arXiv:1709.02262 [hep-lat]} \BibitemShut
  {NoStop}%
\bibitem [{\citenamefont {Chernodub}\ \emph {et~al.}(2018)\citenamefont
  {Chernodub}, \citenamefont {Goy}, \citenamefont {Molochkov},\ and\
  \citenamefont {Nguyen}}]{Chernodub:2018pmt}%
  \BibitemOpen
  \bibfield  {author} {\bibinfo {author} {\bibfnamefont {M.~N.}\ \bibnamefont
  {Chernodub}}, \bibinfo {author} {\bibfnamefont {V.~A.}\ \bibnamefont {Goy}},
  \bibinfo {author} {\bibfnamefont {A.~V.}\ \bibnamefont {Molochkov}}, \ and\
  \bibinfo {author} {\bibfnamefont {H.~H.}\ \bibnamefont {Nguyen}},\ }\href
  {\doibase 10.1103/PhysRevLett.121.191601} {\bibfield  {journal} {\bibinfo
  {journal} {Phys. Rev. Lett.}\ }\textbf {\bibinfo {volume} {121}},\ \bibinfo
  {pages} {191601} (\bibinfo {year} {2018})},\ \Eprint
  {http://arxiv.org/abs/1805.11887} {arXiv:1805.11887 [hep-lat]} \BibitemShut
  {NoStop}%
\bibitem [{\citenamefont {Chernodub}\ \emph {et~al.}(2019)\citenamefont
  {Chernodub}, \citenamefont {Goy},\ and\ \citenamefont
  {Molochkov}}]{Chernodub:2018aix}%
  \BibitemOpen
  \bibfield  {author} {\bibinfo {author} {\bibfnamefont {M.~N.}\ \bibnamefont
  {Chernodub}}, \bibinfo {author} {\bibfnamefont {V.~A.}\ \bibnamefont {Goy}},
  \ and\ \bibinfo {author} {\bibfnamefont {A.~V.}\ \bibnamefont {Molochkov}},\
  }\href {\doibase 10.1103/PhysRevD.99.074021} {\bibfield  {journal} {\bibinfo
  {journal} {Phys. Rev.}\ }\textbf {\bibinfo {volume} {D99}},\ \bibinfo {pages}
  {074021} (\bibinfo {year} {2019})},\ \Eprint
  {http://arxiv.org/abs/1811.01550} {arXiv:1811.01550 [hep-lat]} \BibitemShut
  {NoStop}%
\bibitem [{\citenamefont {Kitazawa}\ \emph {et~al.}(2019)\citenamefont
  {Kitazawa}, \citenamefont {Mogliacci}, \citenamefont {Kolbe},\ and\
  \citenamefont {Horowitz}}]{Kitazawa:2019otp}%
  \BibitemOpen
  \bibfield  {author} {\bibinfo {author} {\bibfnamefont {M.}~\bibnamefont
  {Kitazawa}}, \bibinfo {author} {\bibfnamefont {S.}~\bibnamefont {Mogliacci}},
  \bibinfo {author} {\bibfnamefont {I.}~\bibnamefont {Kolbe}}, \ and\ \bibinfo
  {author} {\bibfnamefont {W.~A.}\ \bibnamefont {Horowitz}},\ }\href {\doibase
  10.1103/PhysRevD.99.094507} {\bibfield  {journal} {\bibinfo  {journal} {Phys.
  Rev.}\ }\textbf {\bibinfo {volume} {D99}},\ \bibinfo {pages} {094507}
  (\bibinfo {year} {2019})},\ \Eprint {http://arxiv.org/abs/1904.00241}
  {arXiv:1904.00241 [hep-lat]} \BibitemShut {NoStop}%
\bibitem [{\citenamefont {Goity}(1990)}]{Goity:1990jb}%
  \BibitemOpen
  \bibfield  {author} {\bibinfo {author} {\bibfnamefont {J.~L.}\ \bibnamefont
  {Goity}},\ }\href {\doibase 10.1016/0370-2693(90)91023-5} {\bibfield
  {journal} {\bibinfo  {journal} {Phys. Lett.}\ }\textbf {\bibinfo {volume}
  {B249}},\ \bibinfo {pages} {495} (\bibinfo {year} {1990})}\BibitemShut
  {NoStop}%
\bibitem [{\citenamefont {Arndt}\ and\ \citenamefont
  {Lin}(2004)}]{Arndt:2004bg}%
  \BibitemOpen
  \bibfield  {author} {\bibinfo {author} {\bibfnamefont {D.}~\bibnamefont
  {Arndt}}\ and\ \bibinfo {author} {\bibfnamefont {C.~J.~D.}\ \bibnamefont
  {Lin}},\ }\href {\doibase 10.1103/PhysRevD.70.014503} {\bibfield  {journal}
  {\bibinfo  {journal} {Phys. Rev.}\ }\textbf {\bibinfo {volume} {D70}},\
  \bibinfo {pages} {014503} (\bibinfo {year} {2004})},\ \Eprint
  {http://arxiv.org/abs/hep-lat/0403012} {arXiv:hep-lat/0403012 [hep-lat]}
  \BibitemShut {NoStop}%
\bibitem [{\citenamefont {Bernardoni}\ \emph {et~al.}(2010)\citenamefont
  {Bernardoni}, \citenamefont {Hernandez},\ and\ \citenamefont
  {Necco}}]{Bernardoni:2009sx}%
  \BibitemOpen
  \bibfield  {author} {\bibinfo {author} {\bibfnamefont {F.}~\bibnamefont
  {Bernardoni}}, \bibinfo {author} {\bibfnamefont {P.}~\bibnamefont
  {Hernandez}}, \ and\ \bibinfo {author} {\bibfnamefont {S.}~\bibnamefont
  {Necco}},\ }\href {\doibase 10.1007/JHEP01(2010)070} {\bibfield  {journal}
  {\bibinfo  {journal} {JHEP}\ }\textbf {\bibinfo {volume} {01}},\ \bibinfo
  {pages} {070} (\bibinfo {year} {2010})},\ \Eprint
  {http://arxiv.org/abs/0910.2537} {arXiv:0910.2537 [hep-lat]} \BibitemShut
  {NoStop}%
\bibitem [{\citenamefont {Colangelo}\ \emph {et~al.}(2010)\citenamefont
  {Colangelo}, \citenamefont {Fuhrer},\ and\ \citenamefont
  {Lanz}}]{Colangelo:2010ba}%
  \BibitemOpen
  \bibfield  {author} {\bibinfo {author} {\bibfnamefont {G.}~\bibnamefont
  {Colangelo}}, \bibinfo {author} {\bibfnamefont {A.}~\bibnamefont {Fuhrer}}, \
  and\ \bibinfo {author} {\bibfnamefont {S.}~\bibnamefont {Lanz}},\ }\href
  {\doibase 10.1103/PhysRevD.82.034506} {\bibfield  {journal} {\bibinfo
  {journal} {Phys. Rev.}\ }\textbf {\bibinfo {volume} {D82}},\ \bibinfo {pages}
  {034506} (\bibinfo {year} {2010})},\ \Eprint {http://arxiv.org/abs/1005.1485}
  {arXiv:1005.1485 [hep-lat]} \BibitemShut {NoStop}%
\bibitem [{\citenamefont {Briceno}(2012)}]{Briceno:2011rz}%
  \BibitemOpen
  \bibfield  {author} {\bibinfo {author} {\bibfnamefont {R.~A.}\ \bibnamefont
  {Briceno}},\ }\href {\doibase 10.1103/PhysRevD.85.014508} {\bibfield
  {journal} {\bibinfo  {journal} {Phys. Rev.}\ }\textbf {\bibinfo {volume}
  {D85}},\ \bibinfo {pages} {014508} (\bibinfo {year} {2012})},\ \Eprint
  {http://arxiv.org/abs/1108.0120} {arXiv:1108.0120 [hep-lat]} \BibitemShut
  {NoStop}%
\bibitem [{\citenamefont {Scavenius}\ \emph {et~al.}(2001)\citenamefont
  {Scavenius}, \citenamefont {M{\' o}csy}, \citenamefont {Mishustin},\ and\
  \citenamefont {Rischke}}]{Scavenius:2000qd}%
  \BibitemOpen
  \bibfield  {author} {\bibinfo {author} {\bibfnamefont {O.}~\bibnamefont
  {Scavenius}}, \bibinfo {author} {\bibfnamefont {A.}~\bibnamefont {M{\'
  o}csy}}, \bibinfo {author} {\bibfnamefont {I.~N.}\ \bibnamefont {Mishustin}},
  \ and\ \bibinfo {author} {\bibfnamefont {D.~H.}\ \bibnamefont {Rischke}},\
  }\href {\doibase 10.1103/PhysRevC.64.045202} {\bibfield  {journal} {\bibinfo
  {journal} {Phys. Rev.}\ }\textbf {\bibinfo {volume} {C64}},\ \bibinfo {pages}
  {045202} (\bibinfo {year} {2001})},\ \Eprint
  {http://arxiv.org/abs/nucl-th/0007030} {arXiv:nucl-th/0007030 [nucl-th]}
  \BibitemShut {NoStop}%
\bibitem [{\citenamefont {Schaefer}\ and\ \citenamefont
  {Wagner}(2009)}]{Schaefer:2008hk}%
  \BibitemOpen
  \bibfield  {author} {\bibinfo {author} {\bibfnamefont {B.-J.}\ \bibnamefont
  {Schaefer}}\ and\ \bibinfo {author} {\bibfnamefont {M.}~\bibnamefont
  {Wagner}},\ }\href {\doibase 10.1103/PhysRevD.79.014018} {\bibfield
  {journal} {\bibinfo  {journal} {Phys. Rev.}\ }\textbf {\bibinfo {volume}
  {D79}},\ \bibinfo {pages} {014018} (\bibinfo {year} {2009})},\ \Eprint
  {http://arxiv.org/abs/0808.1491} {arXiv:0808.1491 [hep-ph]} \BibitemShut
  {NoStop}%
\bibitem [{\citenamefont {M{\' o}csy}\ \emph {et~al.}(2004)\citenamefont {M{\'
  o}csy}, \citenamefont {Mishustin},\ and\ \citenamefont
  {Ellis}}]{Mocsy:2004ab}%
  \BibitemOpen
  \bibfield  {author} {\bibinfo {author} {\bibfnamefont {A.}~\bibnamefont {M{\'
  o}csy}}, \bibinfo {author} {\bibfnamefont {I.~N.}\ \bibnamefont {Mishustin}},
  \ and\ \bibinfo {author} {\bibfnamefont {P.~J.}\ \bibnamefont {Ellis}},\
  }\href {\doibase 10.1103/PhysRevC.70.015204} {\bibfield  {journal} {\bibinfo
  {journal} {Phys. Rev.}\ }\textbf {\bibinfo {volume} {C70}},\ \bibinfo {pages}
  {015204} (\bibinfo {year} {2004})},\ \Eprint
  {http://arxiv.org/abs/nucl-th/0402070} {arXiv:nucl-th/0402070 [nucl-th]}
  \BibitemShut {NoStop}%
\bibitem [{\citenamefont {Schaefer}\ and\ \citenamefont
  {Wambach}(2005)}]{Schaefer:2004en}%
  \BibitemOpen
  \bibfield  {author} {\bibinfo {author} {\bibfnamefont {B.-J.}\ \bibnamefont
  {Schaefer}}\ and\ \bibinfo {author} {\bibfnamefont {J.}~\bibnamefont
  {Wambach}},\ }\href {\doibase 10.1016/j.nuclphysa.2005.04.012} {\bibfield
  {journal} {\bibinfo  {journal} {Nucl. Phys.}\ }\textbf {\bibinfo {volume}
  {A757}},\ \bibinfo {pages} {479} (\bibinfo {year} {2005})},\ \Eprint
  {http://arxiv.org/abs/nucl-th/0403039} {arXiv:nucl-th/0403039 [nucl-th]}
  \BibitemShut {NoStop}%
\bibitem [{\citenamefont {Bowman}\ and\ \citenamefont
  {Kapusta}(2009)}]{Bowman:2008kc}%
  \BibitemOpen
  \bibfield  {author} {\bibinfo {author} {\bibfnamefont {E.~S.}\ \bibnamefont
  {Bowman}}\ and\ \bibinfo {author} {\bibfnamefont {J.~I.}\ \bibnamefont
  {Kapusta}},\ }\href {\doibase 10.1103/PhysRevC.79.015202} {\bibfield
  {journal} {\bibinfo  {journal} {Phys. Rev.}\ }\textbf {\bibinfo {volume}
  {C79}},\ \bibinfo {pages} {015202} (\bibinfo {year} {2009})},\ \Eprint
  {http://arxiv.org/abs/0810.0042} {arXiv:0810.0042 [nucl-th]} \BibitemShut
  {NoStop}%
\bibitem [{\citenamefont {Mitter}\ and\ \citenamefont
  {Schaefer}(2014)}]{Mitter:2013fxa}%
  \BibitemOpen
  \bibfield  {author} {\bibinfo {author} {\bibfnamefont {M.}~\bibnamefont
  {Mitter}}\ and\ \bibinfo {author} {\bibfnamefont {B.-J.}\ \bibnamefont
  {Schaefer}},\ }\href {\doibase 10.1103/PhysRevD.89.054027} {\bibfield
  {journal} {\bibinfo  {journal} {Phys. Rev.}\ }\textbf {\bibinfo {volume}
  {D89}},\ \bibinfo {pages} {054027} (\bibinfo {year} {2014})},\ \Eprint
  {http://arxiv.org/abs/1308.3176} {arXiv:1308.3176 [hep-ph]} \BibitemShut
  {NoStop}%
\bibitem [{\citenamefont {Johnson}(1975)}]{Johnson:1975zp}%
  \BibitemOpen
  \bibfield  {author} {\bibinfo {author} {\bibfnamefont {K.}~\bibnamefont
  {Johnson}},\ }\href@noop {} {\bibfield  {journal} {\bibinfo  {journal} {Acta
  Phys. Polon.}\ }\textbf {\bibinfo {volume} {B6}},\ \bibinfo {pages} {865}
  (\bibinfo {year} {1975})}\BibitemShut {NoStop}%
\bibitem [{\citenamefont {Mamaev}\ and\ \citenamefont
  {Trunov}(1980)}]{Mamaev:1980jn}%
  \BibitemOpen
  \bibfield  {author} {\bibinfo {author} {\bibfnamefont {S.~G.}\ \bibnamefont
  {Mamaev}}\ and\ \bibinfo {author} {\bibfnamefont {N.~N.}\ \bibnamefont
  {Trunov}},\ }\href {\doibase 10.1007/BF00891938} {\bibfield  {journal}
  {\bibinfo  {journal} {Sov. Phys. J.}\ }\textbf {\bibinfo {volume} {23}},\
  \bibinfo {pages} {551} (\bibinfo {year} {1980})}\BibitemShut {NoStop}%
\bibitem [{\citenamefont {Cougo-Pinto}\ \emph {et~al.}(1996)\citenamefont
  {Cougo-Pinto}, \citenamefont {Farina},\ and\ \citenamefont
  {Tort}}]{CougoPinto:1996bh}%
  \BibitemOpen
  \bibfield  {author} {\bibinfo {author} {\bibfnamefont {M.~V.}\ \bibnamefont
  {Cougo-Pinto}}, \bibinfo {author} {\bibfnamefont {C.}~\bibnamefont {Farina}},
  \ and\ \bibinfo {author} {\bibfnamefont {A.}~\bibnamefont {Tort}},\ }\href
  {\doibase 10.1007/BF00398302} {\bibfield  {journal} {\bibinfo  {journal}
  {Lett. Math. Phys.}\ }\textbf {\bibinfo {volume} {38}},\ \bibinfo {pages}
  {97} (\bibinfo {year} {1996})}\BibitemShut {NoStop}%
\bibitem [{\citenamefont {Elizalde}\ \emph {et~al.}(2003)\citenamefont
  {Elizalde}, \citenamefont {Santos},\ and\ \citenamefont
  {Tort}}]{Elizalde:2002wg}%
  \BibitemOpen
  \bibfield  {author} {\bibinfo {author} {\bibfnamefont {E.}~\bibnamefont
  {Elizalde}}, \bibinfo {author} {\bibfnamefont {F.~C.}\ \bibnamefont
  {Santos}}, \ and\ \bibinfo {author} {\bibfnamefont {A.~C.}\ \bibnamefont
  {Tort}},\ }\href {\doibase 10.1142/S0217751X03014186} {\bibfield  {journal}
  {\bibinfo  {journal} {Int. J. Mod. Phys.}\ }\textbf {\bibinfo {volume}
  {A18}},\ \bibinfo {pages} {1761} (\bibinfo {year} {2003})},\ \Eprint
  {http://arxiv.org/abs/hep-th/0206114} {arXiv:hep-th/0206114 [hep-th]}
  \BibitemShut {NoStop}%
\bibitem [{\citenamefont {Ishikawa}\ \emph {et~al.}(2019)\citenamefont
  {Ishikawa}, \citenamefont {Nakayama},\ and\ \citenamefont
  {Suzuki}}]{Ishikawa:2018yey}%
  \BibitemOpen
  \bibfield  {author} {\bibinfo {author} {\bibfnamefont {T.}~\bibnamefont
  {Ishikawa}}, \bibinfo {author} {\bibfnamefont {K.}~\bibnamefont {Nakayama}},
  \ and\ \bibinfo {author} {\bibfnamefont {K.}~\bibnamefont {Suzuki}},\ }\href
  {\doibase 10.1103/PhysRevD.99.054010} {\bibfield  {journal} {\bibinfo
  {journal} {Phys. Rev.}\ }\textbf {\bibinfo {volume} {D99}},\ \bibinfo {pages}
  {054010} (\bibinfo {year} {2019})},\ \Eprint
  {http://arxiv.org/abs/1812.10964} {arXiv:1812.10964 [hep-ph]} \BibitemShut
  {NoStop}%
\bibitem [{\citenamefont {Kim}\ \emph {et~al.}(1987)\citenamefont {Kim},
  \citenamefont {Namgung}, \citenamefont {Soh},\ and\ \citenamefont
  {Yee}}]{Kim:1987db}%
  \BibitemOpen
  \bibfield  {author} {\bibinfo {author} {\bibfnamefont {S.~K.}\ \bibnamefont
  {Kim}}, \bibinfo {author} {\bibfnamefont {W.}~\bibnamefont {Namgung}},
  \bibinfo {author} {\bibfnamefont {K.~S.}\ \bibnamefont {Soh}}, \ and\
  \bibinfo {author} {\bibfnamefont {J.~H.}\ \bibnamefont {Yee}},\ }\href
  {\doibase 10.1103/PhysRevD.36.3172} {\bibfield  {journal} {\bibinfo
  {journal} {Phys. Rev.}\ }\textbf {\bibinfo {volume} {D36}},\ \bibinfo {pages}
  {3172} (\bibinfo {year} {1987})}\BibitemShut {NoStop}%
\bibitem [{\citenamefont {Song}\ and\ \citenamefont {Kim}(1990)}]{Song:1990dm}%
  \BibitemOpen
  \bibfield  {author} {\bibinfo {author} {\bibfnamefont {D.~Y.}\ \bibnamefont
  {Song}}\ and\ \bibinfo {author} {\bibfnamefont {J.~K.}\ \bibnamefont {Kim}},\
  }\href {\doibase 10.1103/PhysRevD.41.3165} {\bibfield  {journal} {\bibinfo
  {journal} {Phys. Rev.}\ }\textbf {\bibinfo {volume} {D41}},\ \bibinfo {pages}
  {3165} (\bibinfo {year} {1990})}\BibitemShut {NoStop}%
\bibitem [{\citenamefont {Song}(1993)}]{Song:1993da}%
  \BibitemOpen
  \bibfield  {author} {\bibinfo {author} {\bibfnamefont {D.~Y.}\ \bibnamefont
  {Song}},\ }\href {\doibase 10.1103/PhysRevD.48.3925} {\bibfield  {journal}
  {\bibinfo  {journal} {Phys. Rev.}\ }\textbf {\bibinfo {volume} {D48}},\
  \bibinfo {pages} {3925} (\bibinfo {year} {1993})}\BibitemShut {NoStop}%
\bibitem [{\citenamefont {Kim}\ \emph {et~al.}(1994)\citenamefont {Kim},
  \citenamefont {Han},\ and\ \citenamefont {Koh}}]{Kim:1994es}%
  \BibitemOpen
  \bibfield  {author} {\bibinfo {author} {\bibfnamefont {D.~K.}\ \bibnamefont
  {Kim}}, \bibinfo {author} {\bibfnamefont {Y.~D.}\ \bibnamefont {Han}}, \ and\
  \bibinfo {author} {\bibfnamefont {I.~G.}\ \bibnamefont {Koh}},\ }\href
  {\doibase 10.1103/PhysRevD.49.6943} {\bibfield  {journal} {\bibinfo
  {journal} {Phys. Rev.}\ }\textbf {\bibinfo {volume} {D49}},\ \bibinfo {pages}
  {6943} (\bibinfo {year} {1994})}\BibitemShut {NoStop}%
\bibitem [{\citenamefont {Vshivtsev}\ \emph {et~al.}(1995)\citenamefont
  {Vshivtsev}, \citenamefont {Magnitsky},\ and\ \citenamefont
  {Klimenko}}]{Vshivtsev:1995xx}%
  \BibitemOpen
  \bibfield  {author} {\bibinfo {author} {\bibfnamefont {A.~S.}\ \bibnamefont
  {Vshivtsev}}, \bibinfo {author} {\bibfnamefont {B.~V.}\ \bibnamefont
  {Magnitsky}}, \ and\ \bibinfo {author} {\bibfnamefont {K.~G.}\ \bibnamefont
  {Klimenko}},\ }\href@noop {} {\bibfield  {journal} {\bibinfo  {journal} {JETP
  Lett.}\ }\textbf {\bibinfo {volume} {61}},\ \bibinfo {pages} {871} (\bibinfo
  {year} {1995})},\ \bibinfo {note} {[Pisma Zh. Eksp. Teor.
  Fiz.61,847(1995)]}\BibitemShut {NoStop}%
\bibitem [{\citenamefont {Vdovichenko}\ and\ \citenamefont
  {Klimenko}(1998)}]{Vdovichenko:1998ev}%
  \BibitemOpen
  \bibfield  {author} {\bibinfo {author} {\bibfnamefont {M.~A.}\ \bibnamefont
  {Vdovichenko}}\ and\ \bibinfo {author} {\bibfnamefont {A.~K.}\ \bibnamefont
  {Klimenko}},\ }\href {\doibase 10.1134/1.567890} {\bibfield  {journal}
  {\bibinfo  {journal} {JETP Lett.}\ }\textbf {\bibinfo {volume} {68}},\
  \bibinfo {pages} {460} (\bibinfo {year} {1998})},\ \bibinfo {note} {[Pisma
  Zh. Eksp. Teor. Fiz.68,431(1998)]}\BibitemShut {NoStop}%
\bibitem [{\citenamefont {Vshivtsev}\ \emph {et~al.}(1998)\citenamefont
  {Vshivtsev}, \citenamefont {Vdovichenko},\ and\ \citenamefont
  {Klimenko}}]{Vshivtsev:1998fg}%
  \BibitemOpen
  \bibfield  {author} {\bibinfo {author} {\bibfnamefont {A.~S.}\ \bibnamefont
  {Vshivtsev}}, \bibinfo {author} {\bibfnamefont {M.~A.}\ \bibnamefont
  {Vdovichenko}}, \ and\ \bibinfo {author} {\bibfnamefont {K.~G.}\ \bibnamefont
  {Klimenko}},\ }\href {\doibase 10.1134/1.558650} {\bibfield  {journal}
  {\bibinfo  {journal} {J. Exp. Theor. Phys.}\ }\textbf {\bibinfo {volume}
  {87}},\ \bibinfo {pages} {229} (\bibinfo {year} {1998})},\ \bibinfo {note}
  {[Zh. Eksp. Teor. Fiz.114,418(1998)]}\BibitemShut {NoStop}%
\bibitem [{\citenamefont {Hayashi}\ and\ \citenamefont
  {Inagaki}(2010)}]{Hayashi:2010ru}%
  \BibitemOpen
  \bibfield  {author} {\bibinfo {author} {\bibfnamefont {M.}~\bibnamefont
  {Hayashi}}\ and\ \bibinfo {author} {\bibfnamefont {T.}~\bibnamefont
  {Inagaki}},\ }\href {\doibase 10.1142/S0217751X10049426} {\bibfield
  {journal} {\bibinfo  {journal} {Int. J. Mod. Phys.}\ }\textbf {\bibinfo
  {volume} {A25}},\ \bibinfo {pages} {3353} (\bibinfo {year} {2010})},\ \Eprint
  {http://arxiv.org/abs/1003.3163} {arXiv:1003.3163 [hep-ph]} \BibitemShut
  {NoStop}%
\bibitem [{\citenamefont {Ebert}\ and\ \citenamefont
  {Klimenko}(2010)}]{Ebert:2010eq}%
  \BibitemOpen
  \bibfield  {author} {\bibinfo {author} {\bibfnamefont {D.}~\bibnamefont
  {Ebert}}\ and\ \bibinfo {author} {\bibfnamefont {K.~G.}\ \bibnamefont
  {Klimenko}},\ }\href {\doibase 10.1103/PhysRevD.82.025018} {\bibfield
  {journal} {\bibinfo  {journal} {Phys. Rev.}\ }\textbf {\bibinfo {volume}
  {D82}},\ \bibinfo {pages} {025018} (\bibinfo {year} {2010})},\ \Eprint
  {http://arxiv.org/abs/1005.0699} {arXiv:1005.0699 [hep-ph]} \BibitemShut
  {NoStop}%
\bibitem [{\citenamefont {Wang}\ \emph {et~al.}(2018)\citenamefont {Wang},
  \citenamefont {Xia},\ and\ \citenamefont {Zong}}]{Wang:2018ovx}%
  \BibitemOpen
  \bibfield  {author} {\bibinfo {author} {\bibfnamefont {Q.-W.}\ \bibnamefont
  {Wang}}, \bibinfo {author} {\bibfnamefont {Y.}~\bibnamefont {Xia}}, \ and\
  \bibinfo {author} {\bibfnamefont {H.-S.}\ \bibnamefont {Zong}},\ }\href@noop
  {} {\  (\bibinfo {year} {2018})},\ \Eprint {http://arxiv.org/abs/1802.00258}
  {arXiv:1802.00258 [hep-ph]} \BibitemShut {NoStop}%
\bibitem [{\citenamefont {Inagaki}\ \emph {et~al.}(2019)\citenamefont
  {Inagaki}, \citenamefont {Matsuo},\ and\ \citenamefont
  {Shimoji}}]{Inagaki:2019kbc}%
  \BibitemOpen
  \bibfield  {author} {\bibinfo {author} {\bibfnamefont {T.}~\bibnamefont
  {Inagaki}}, \bibinfo {author} {\bibfnamefont {Y.}~\bibnamefont {Matsuo}}, \
  and\ \bibinfo {author} {\bibfnamefont {H.}~\bibnamefont {Shimoji}},\ }\href
  {\doibase 10.3390/sym11040451} {\bibfield  {journal} {\bibinfo  {journal}
  {Symmetry}\ }\textbf {\bibinfo {volume} {11}},\ \bibinfo {pages} {451}
  (\bibinfo {year} {2019})},\ \Eprint {http://arxiv.org/abs/1903.04244}
  {arXiv:1903.04244 [hep-th]} \BibitemShut {NoStop}%
\bibitem [{\citenamefont {Xu}\ and\ \citenamefont {Huang}(2019)}]{Xu:2019gia}%
  \BibitemOpen
  \bibfield  {author} {\bibinfo {author} {\bibfnamefont {K.}~\bibnamefont
  {Xu}}\ and\ \bibinfo {author} {\bibfnamefont {M.}~\bibnamefont {Huang}},\
  }\href@noop {} {\  (\bibinfo {year} {2019})},\ \Eprint
  {http://arxiv.org/abs/1903.08416} {arXiv:1903.08416 [hep-ph]} \BibitemShut
  {NoStop}%
\bibitem [{\citenamefont {Braun}\ \emph {et~al.}(2005)\citenamefont {Braun},
  \citenamefont {Klein},\ and\ \citenamefont {Pirner}}]{Braun:2005gy}%
  \BibitemOpen
  \bibfield  {author} {\bibinfo {author} {\bibfnamefont {J.}~\bibnamefont
  {Braun}}, \bibinfo {author} {\bibfnamefont {B.}~\bibnamefont {Klein}}, \ and\
  \bibinfo {author} {\bibfnamefont {H.~J.}\ \bibnamefont {Pirner}},\ }\href
  {\doibase 10.1103/PhysRevD.72.034017} {\bibfield  {journal} {\bibinfo
  {journal} {Phys. Rev.}\ }\textbf {\bibinfo {volume} {D72}},\ \bibinfo {pages}
  {034017} (\bibinfo {year} {2005})},\ \Eprint
  {http://arxiv.org/abs/hep-ph/0504127} {arXiv:hep-ph/0504127 [hep-ph]}
  \BibitemShut {NoStop}%
\bibitem [{\citenamefont {Braun}\ \emph {et~al.}(2006)\citenamefont {Braun},
  \citenamefont {Klein}, \citenamefont {Pirner},\ and\ \citenamefont
  {Rezaeian}}]{Braun:2005fj}%
  \BibitemOpen
  \bibfield  {author} {\bibinfo {author} {\bibfnamefont {J.}~\bibnamefont
  {Braun}}, \bibinfo {author} {\bibfnamefont {B.}~\bibnamefont {Klein}},
  \bibinfo {author} {\bibfnamefont {H.~J.}\ \bibnamefont {Pirner}}, \ and\
  \bibinfo {author} {\bibfnamefont {A.~H.}\ \bibnamefont {Rezaeian}},\ }\href
  {\doibase 10.1103/PhysRevD.73.074010} {\bibfield  {journal} {\bibinfo
  {journal} {Phys. Rev.}\ }\textbf {\bibinfo {volume} {D73}},\ \bibinfo {pages}
  {074010} (\bibinfo {year} {2006})},\ \Eprint
  {http://arxiv.org/abs/hep-ph/0512274} {arXiv:hep-ph/0512274 [hep-ph]}
  \BibitemShut {NoStop}%
\bibitem [{\citenamefont {Palhares}\ \emph {et~al.}(2011)\citenamefont
  {Palhares}, \citenamefont {Fraga},\ and\ \citenamefont
  {Kodama}}]{Palhares:2009tf}%
  \BibitemOpen
  \bibfield  {author} {\bibinfo {author} {\bibfnamefont {L.~F.}\ \bibnamefont
  {Palhares}}, \bibinfo {author} {\bibfnamefont {E.~S.}\ \bibnamefont {Fraga}},
  \ and\ \bibinfo {author} {\bibfnamefont {T.}~\bibnamefont {Kodama}},\ }\href
  {\doibase 10.1088/0954-3899/38/8/085101} {\bibfield  {journal} {\bibinfo
  {journal} {J. Phys.}\ }\textbf {\bibinfo {volume} {G38}},\ \bibinfo {pages}
  {085101} (\bibinfo {year} {2011})},\ \Eprint {http://arxiv.org/abs/0904.4830}
  {arXiv:0904.4830 [nucl-th]} \BibitemShut {NoStop}%
\bibitem [{\citenamefont {Braun}\ \emph {et~al.}(2011)\citenamefont {Braun},
  \citenamefont {Klein},\ and\ \citenamefont {Piasecki}}]{Braun:2010vd}%
  \BibitemOpen
  \bibfield  {author} {\bibinfo {author} {\bibfnamefont {J.}~\bibnamefont
  {Braun}}, \bibinfo {author} {\bibfnamefont {B.}~\bibnamefont {Klein}}, \ and\
  \bibinfo {author} {\bibfnamefont {P.}~\bibnamefont {Piasecki}},\ }\href
  {\doibase 10.1140/epjc/s10052-011-1576-7} {\bibfield  {journal} {\bibinfo
  {journal} {Eur. Phys. J.}\ }\textbf {\bibinfo {volume} {C71}},\ \bibinfo
  {pages} {1576} (\bibinfo {year} {2011})},\ \Eprint
  {http://arxiv.org/abs/1008.2155} {arXiv:1008.2155 [hep-ph]} \BibitemShut
  {NoStop}%
\bibitem [{\citenamefont {Tripolt}\ \emph {et~al.}(2014)\citenamefont
  {Tripolt}, \citenamefont {Braun}, \citenamefont {Klein},\ and\ \citenamefont
  {Schaefer}}]{Tripolt:2013zfa}%
  \BibitemOpen
  \bibfield  {author} {\bibinfo {author} {\bibfnamefont {R.-A.}\ \bibnamefont
  {Tripolt}}, \bibinfo {author} {\bibfnamefont {J.}~\bibnamefont {Braun}},
  \bibinfo {author} {\bibfnamefont {B.}~\bibnamefont {Klein}}, \ and\ \bibinfo
  {author} {\bibfnamefont {B.-J.}\ \bibnamefont {Schaefer}},\ }\href {\doibase
  10.1103/PhysRevD.90.054012} {\bibfield  {journal} {\bibinfo  {journal} {Phys.
  Rev.}\ }\textbf {\bibinfo {volume} {D90}},\ \bibinfo {pages} {054012}
  (\bibinfo {year} {2014})},\ \Eprint {http://arxiv.org/abs/1308.0164}
  {arXiv:1308.0164 [hep-ph]} \BibitemShut {NoStop}%
\bibitem [{\citenamefont {Phat}\ and\ \citenamefont
  {Thu}(2014)}]{Phat:2014asa}%
  \BibitemOpen
  \bibfield  {author} {\bibinfo {author} {\bibfnamefont {T.~H.}\ \bibnamefont
  {Phat}}\ and\ \bibinfo {author} {\bibfnamefont {N.~V.}\ \bibnamefont {Thu}},\
  }\href {\doibase 10.1142/S0217751X1450078X} {\bibfield  {journal} {\bibinfo
  {journal} {Int. J. Mod. Phys.}\ }\textbf {\bibinfo {volume} {A29}},\ \bibinfo
  {pages} {1450078} (\bibinfo {year} {2014})}\BibitemShut {NoStop}%
\bibitem [{\citenamefont {Almasi}\ \emph {et~al.}(2017)\citenamefont {Almasi},
  \citenamefont {Pisarski},\ and\ \citenamefont {Skokov}}]{Almasi:2016zqf}%
  \BibitemOpen
  \bibfield  {author} {\bibinfo {author} {\bibfnamefont {G.~A.}\ \bibnamefont
  {Almasi}}, \bibinfo {author} {\bibfnamefont {R.~D.}\ \bibnamefont
  {Pisarski}}, \ and\ \bibinfo {author} {\bibfnamefont {V.~V.}\ \bibnamefont
  {Skokov}},\ }\href {\doibase 10.1103/PhysRevD.95.056015} {\bibfield
  {journal} {\bibinfo  {journal} {Phys. Rev.}\ }\textbf {\bibinfo {volume}
  {D95}},\ \bibinfo {pages} {056015} (\bibinfo {year} {2017})},\ \Eprint
  {http://arxiv.org/abs/1612.04416} {arXiv:1612.04416 [hep-ph]} \BibitemShut
  {NoStop}%
\end{thebibliography}%

\end{document}